\documentclass[11pt]{article}
\usepackage[pdftex]{graphicx}     
\usepackage{enumerate}
\usepackage{natbib}
\usepackage{url} 
\usepackage{amsmath,amsfonts,enumerate,color,verbatim,booktabs,amssymb}
\usepackage{algorithm}
\usepackage{microtype}
\usepackage{xspace}
\usepackage{mathtools}
\usepackage{scalerel}
\usepackage[normalem]{ulem} 
\usepackage{times}
\usepackage[dvipsnames]{xcolor}
\usepackage{tikz}

\usepackage[english]{babel}
\graphicspath{{graphics/}}

\usepackage[figuresright]{rotating}

\newcommand{\blind}{0}

\renewcommand{\vec}[1]{\boldsymbol{#1}}

\newcommand{\thetavec}{\vec\theta}

\newcommand{\svec}{\vec{s}}

\newcommand{\xvec}{\vec{x}}

\newcommand{\rhovec}{\vec{\rho}}
\newcommand{\indep}{\overset{indep}{\sim }}

\newcommand{\N}{\text{N}}
\newcommand{\D}{\mathcal{D}}
\newcommand{\Ne}{{N^{\text{e}}}}

\newcommand{\tr}{{\text{r}}}
\newcommand{\tg}{{\text{g}}}

\newcommand{\phir}{{\phi_\text{r}}}
\newcommand{\phig}{{\phi_\text{g}}}
\newcommand{\phis}{{\phi_\text{s}}}
\newcommand{\phitheta}{\phi_\theta}
\newcommand{\INN}{{\mathbb{I}_{N}}}
\newcommand{\INGNG}{{\mathbb{I}_{\Ng}}}
\newcommand{\Ng}{{N^\text{g}}}

\newcommand{\lrvec}{{\vec{L}^\text{r}}}

\newcommand{\voneN}{\vec{1}_{\scaleto{N}{4pt}}}
\newcommand{\vzeroN}{\vec{0}_{\scaleto{N}{4pt}}}
\newcommand{\vzeroNg}{\vec{0}_{\scaleto{\Ng}{4pt}}}

\newcommand{\tmgtm}{\ensuremath{{t-1 \vert t-1}}}
\newcommand{\ttgtm}{\ensuremath{{tt \vert t-1}}}
\newcommand{\tgtm}{\ensuremath{{t \vert t-1}}}

\newcommand{\ltgtm}{\ensuremath{{\ell t \vert t-1}}}

\newcommand{\Cov}{{\text{Cov}}}

\newcommand{\p}{{\partial}}
\newcommand{\pp}{{\partial^2}}
\newcommand{\ppx}{{\partial x^2}}
\newcommand{\ppy}{{\partial y^2}}

\newcommand{\vxt}{{v_x^t}}
\newcommand{\vyt}{{v_y^t}}

\begin{document}

\def\spacingset#1{\renewcommand{\baselinestretch}%
{#1}\small\normalsize} \spacingset{1}


\if0\blind
{
  \title{\bf A Bayesian spatio-temporal model for high-resolution short-term forecasting of precipitation fields}
  \author{S. R. Johnson\thanks{This work was supported by NERC grant NE/P017134/1 (Flood-PREPARED: Predicting Rainfall Events by Physical Analytics of REaltime Data).}
\hspace{.2cm}\\
    School of Mathematics, Statistics \& Physics, Newcastle University,\\ Newcastle upon Tyne, NE1 7RU, UK\\
    S. E. Heaps\\
     School of Mathematics, Statistics \& Physics, Newcastle University,\\ Newcastle upon Tyne, NE1 7RU, UK\\
    K. J. Wilson\\
     School of Mathematics, Statistics \& Physics, Newcastle University,\\ Newcastle upon Tyne, NE1 7RU, UK
    and \\
    D. J. Wilkinson\\
     School of Mathematics, Statistics \& Physics, Newcastle University,\\ Newcastle upon Tyne, NE1 7RU, UK}
  \maketitle
  \newpage
} \fi

\if1\blind
{
  \bigskip
  \bigskip
  \bigskip
  \begin{center}
    {\LARGE\bf A Bayesian spatio-temporal model for high-resolution short-term forecasting of precipitation fields}
\end{center}
  \medskip
} \fi

\bigskip
\begin{abstract}
With extreme weather events becoming more common, the risk posed by surface water flooding is ever increasing.
In this work we propose a model, and associated
Bayesian inference scheme, for generating probabilistic (high-resolution short-term) forecasts of localised precipitation.
The parametrisation of our underlying hierarchical dynamic spatio-temporal model is motivated by 
a forward-time, centred-space finite difference solution to a collection of stochastic partial differential equations, where the main driving forces are advection and diffusion.
Observations from both weather radar and ground based rain gauges provide information from which we can learn about the likely values of the (latent) precipitation field in addition to other unknown model parameters.
Working in the Bayesian paradigm provides a coherent framework for capturing uncertainty both in the underlying model
parameters and also in our forecasts. Further, appealing to simulation based (MCMC) sampling yields a straightforward solution to handling zeros, treated as censored observations, via data augmentation.
Both the underlying state and the observations are of moderately large dimension ($\mathcal{O}(10^4)$ and $\mathcal{O}(10^3)$ respectively) and this renders standard inference approaches computationally infeasible.
Our solution is to embed the ensemble Kalman smoother within a Gibbs sampling scheme
to facilitate approximate Bayesian inference in reasonable time.
Both the methodology and the effectiveness of our
posterior sampling scheme are demonstrated via simulation studies and also by a case study of
real data from the Urban Observatory project based in Newcastle upon Tyne, UK.
\end{abstract}

\noindent%
\vfill

\spacingset{1.45} 
 \section{Introduction}
\label{sec:introduction}

Increased urbanisation coupled with the effects of global climate change means that cities are becoming increasingly susceptible to localised surface water flooding as a result of intense rainfall events \citep{kendon2014heavier,barr2020flood}.
Such events can result in substantial damage to private properties, businesses and also cause severe disruption to inner city transport systems and infrastructure.
Recent advances in high performance computing have enabled the development of high-performance integrated hydrodynamic modelling systems for rainfall-induced surface water flows within urban areas \citep{liang2015high,xia2018new,xing2019city}.
Such models rely on high-dimensional spatio-temporal inputs that characterise the amount of rainfall observed at ground level.
Ground-based rain gauges are perhaps the natural choice for obtaining such information.
However, it is often not practical to deploy these at sufficient spatial resolution.
Thus, in practice, inputs are typically obtained from weather radar systems, but these do not necessarily provide an accurate description of the rainfall rates on the ground.

In what follows we propose a physically motivated spatio-temporal statistical model that is capable of jointly modelling observations stemming from a (potentially sparse) rain gauge network in addition to a weather radar, allowing the synthesis of these two different data sources.
The basic premise is to leverage information from the weather radar, for example, where the rain clouds are, and where they are moving, and use the ground-based rain gauge information to calibrate the spatio-temporal field of interest.
We apply our model to real data obtained from Newcastle's Urban Observatory project \citep{james2014urban} that collects data from a large range of sensors distributed around the North East of England.
This work forms part of a wider consortium of analytics and models developed as part of the NERC funded Flood-PREPARED (Predicting Rainfall Events by Physical Analytics of REaltime Data) project; further details are given by \cite{barr2020flood}.

The analysis of spatio-temporal data has a rich history in the statistical literature and a wealth of approaches have been proposed for modelling these types of data; we refer the reader to \cite{liu2021statistical} who provide an excellent overview of this field.
Here we appeal to the hierarchical dynamical spatio-temporal modelling framework \citep{wikle1998hierarchical, wikle1999dimension,berliner2003physical,stroud2010ensemble,katzfuss2020ensemble} which we find naturally appealing for several reasons.
In particular the modelling procedure can be decomposed into a sequence of conditional model specifications of which the main components are a \textit{system model} and  an \textit{observation model}.
The system model describes how the (latent) process of interest evolves over space and time, and the observations are then modelled conditionally on this process, via the observation model, which provides a straightforward framework for handling observations from numerous data sources.
The direct specification of a system model can be a rather daunting task due to the complex spatial and temporal dependencies that need to be captured.
Given the physical nature of our problem we choose to motivate the dynamics of the system model via a physical advection-diffusion-reaction process.
This approach avoids the need for direct specification of space-time covariance functions (which can be difficult in practice) and, when coupled with an appropriate discretisation scheme, gives rise to a tractable vector auto-regressive (VAR) model \citep{cressiewikle2015,krainski2018advanced}.

The dimension of both the observations and also the spatio-temporal field of interest are moderately large and so solving the inference (inverse) problem is challenging, especially within the Bayesian paradigm.
We consider an efficient Markov chain Monte Carlo (MCMC) scheme for sampling from the posterior distribution of interest.
By opting for a simulation based inference approach we are able to straightforwardly handle both missing and zero (censored) observations by appealing to data augmentation \citep{tanner1987calculation}.
More specifically we introduce an appropriate collection of latent variables \citep{sanso2000nonstationary,heaps2015bayesian} that when coupled with an appropriate prior distribution gives rise to a tractable dynamic linear model (DLM) for the complete data.
For these types of models state inference is typically achieved via the forwards filtering backwards sampling (FFBS) algorithm \citep{carter1994gibbs,fruhwirth1994data}, however the dimension of our state-space renders (exact) simulation from the joint full conditional distribution of the dynamic states computationally infeasible.
We therefore consider an MCMC algorithm in which an approximate sample from the joint full conditional distribution of the dynamic states is obtained via the ensemble Kalman smoother (EnKS) proposed by \cite{evensen2000ensemble}.
The remaining unknown quantities can be sampled from their respective full conditional distributions and so the resulting MCMC algorithm is akin to the Gibbs ensemble Kalman smoother (GEnKS) outlined in \cite{katzfuss2020ensemble}.

The remainder of the paper is structured as follows.
In Section~\ref{sec:system_model} we derive our physically motivated system model by taking an appropriate discretisation of a collection of stochastic partial differential equations (SPDEs).
The corresponding observation model is outlined in Section~\ref{sec:obs_model}.
In Section~\ref{sec:Bayes_model} we define the complete Bayesian model together with our prior specification and also discuss parameter identifiability.
Our Bayesian approach to inference is discussed in Section~\ref{sec:posterior_comp}.
Simulation studies illustrating the effectiveness of our posterior sampling scheme are given in
Section~3  of the supplementary material and are also summarised briefly in Section~\ref{sec:posterior_comp} of this paper.
Section~\ref{sec:real_data} illustrates the use of our proposed model in a real data setting where we consider observations obtained from the Urban Observatory project based in Newcastle upon Tyne, UK.
Section~\ref{sec:conc} offers some conclusions.

\section{System model}
\label{sec:system_model}

In what follows we suppose that precipitation intensity at ground level is represented by a two-dimensional latent field that takes values on the real line.
For ease of understanding, here positive values can be thought of as a (possibly transformed) rainfall rate with negative values indicating no rainfall; this notion is formalised in our observation model in Section~\ref{sec:obs_model}.
More specifically, let $\theta(\svec,t)\in \mathbb{R}$ denote the precipitation intensity field at location~$\svec \in \mathcal{S}$ and time $t \in \mathcal{T}$, where $\mathcal{S} \subset \mathbb{R}^2$ defines a finite two-dimensional field and $\mathcal{T} \subset \mathbb{R}$ a temporal domain of interest.
We suppose the main features that determine the evolution of this field are governed by advection, diffusion and reaction and so we let the evolution be described by the corresponding SPDE, namely
\begin{equation}
\frac{\p \theta(\svec,t)}{\p t} = \nabla \cdot \Sigma \nabla \theta(\svec,t) - \nabla  \cdot v(\svec,t)  \theta(\svec,t) - a\, \theta(\svec,t) + S(\svec,t) + \epsilon(\svec,t),
\label{eqn:theta_pde}
\end{equation}
where $\nabla = (\p/\p x, \p/\p y)$ and $\cdot$ denotes the dot product. Here $\Sigma$ is a matrix controlling the degree of anisotropy in the diffusion term, $v(\svec,t)$ is the velocity field and the reaction is a simple decay function with rate~$a>0$. Note that if we assume isotropic diffusion, that is, $\Sigma = b \, \mathbb{I}_2$ for some scalar $b \in \mathbb{R}_{\geq 0}$, then $\nabla \cdot \Sigma \nabla \theta(\svec,t) = b\, \Delta \theta(\svec,t)$ where $\Delta$ denotes the standard Laplace operator and $\mathbb{I}_q$ is the~$q$ dimensional identity matrix.
The auxiliary processes~$S(\svec,t)$ and~$\epsilon(\svec,t)$ in~\eqref{eqn:theta_pde} are defined over the same spatial and temporal domains as~$\theta(\svec,t)$; the former denotes a source-sink process that allows for instantaneous injections into (or decay from) the system and~$\epsilon(\svec,t)$ represents a noise process.

Incorporating both the source-sink and error processes is a modelling choice as they each allow for instantaneous injections/decay to and from the system.
Indeed some simplification of~\eqref{eqn:theta_pde} is possible if both the source-sink and error processes are assumed to be temporally white. In this case $S(\svec,t)$ can be absorbed into the error process ($\epsilon(\svec,t) \to S(\svec,t) + \epsilon(\svec,t)$) and, if desired, spatially correlated injections/decay can be achieved by imposing a spatially coloured covariance kernel on the error process; see, for example, \cite{sigrist2015stochastic}.
In our application it seems sensible to assume that any injection to the system, for example, the formation of a rain cloud, is likely to be both spatially and temporally correlated.
Put another way, if the rainfall rate is increasing at location~$\vec{s}$ at time~$t$ ($S(\svec,t)>0$) then we think it is more likely that it is increasing within the neighbourhood of~$\vec{s}$ and also at time $t + dt$.
This assumption can be embedded within the dynamics in one of two ways.
First, the source-sink process could be temporally coloured and spatially white, with the spatial dependence instead embedded within the error process; this approach was taken by \cite{stroud2010ensemble} in the context of the advection-diffusion model.
Alternatively, and the approach we consider here, 
the source-sink process can be both spatially and temporally coloured and coupled with a simple (spatially and temporally white) error process.
We favour this approach because it keeps the form of the error process simple, thereby avoiding direct specification of a covariance kernel, and also allows us to motivate the spatial and temporal interactions via a physical process.
More specifically, we assume the evolution of $S(\svec,t)$ is governed by the SPDE corresponding to the (isotropic) diffusion-reaction process, namely
\begin{equation}
\frac{\p S(\svec,t)}{\p t} = b^* \Delta S(\svec,t) - a^* S(\svec,t)+ \tilde{\epsilon}(\svec,t),
\label{eqn:source_sink_pde}
\end{equation}
where, as in~\eqref{eqn:theta_pde}, the reaction is taken to be a simple decay function (with rate~$a^*>0$).
Note that in~\eqref{eqn:source_sink_pde} the additional source-sink term has been absorbed into the error process and we assume that this process is both spatially and temporally white.

The SPDEs \eqref{eqn:theta_pde} and \eqref{eqn:source_sink_pde} define the evolution of the system and, in general, we suppose that the processes, the constants, and the initial conditions of the system are unknown.
We now appeal to a numerical finite-difference scheme that will give rise to a natural model parameterisation; this is the topic of the next section.

\subsection{Discretisation}
\label{sec:discretisation}
To discretise the spatial field we impose a (regular) lattice of dimension $N = n \times n$ over the domain~$\mathcal{S}$, that is, we consider a finite collection of equally spaced locations $\svec_{ij} = (x_i,y_j) \in \mathcal{S}$ for $i,j=1,\dots,n$.
The temporal domain~$\mathcal{T}$ is also discretised by considering equally spaced time points and, without loss of generality, we consider $t\in \{1,2,\dots,T\}$.
Let $\theta_{ij}^t \equiv \theta(\svec_{ij},t)$ be the state of the system at location~$\svec_{ij}$ and time~$t$, with $S_{ij}^t, \epsilon_{ij}^t$ and $\tilde{\epsilon}^{\,t}_{ij}$ similarly defined.
We also assume isotropic diffusion in~\eqref{eqn:theta_pde} and so $\Sigma = b\,\mathbb{I}_2$.
Further, for reasons of parsimony, we assume the velocity field is spatially homogeneous and so $v(\svec,t) = \vec{v}(t) \in \mathbb{R}^2$.
The velocity/drift vector~$\vec{v}_t = (\vxt,\vyt)' \equiv (v_1(t),v_2(t))'$ contains the velocities in the~$x$ and~$y$ directions at time~$t$.

The SPDEs given by \eqref{eqn:theta_pde} and \eqref{eqn:source_sink_pde} can now be written as
\begin{align}
\frac{\p \theta}{\p t} &= b \left[\frac{\pp \theta}{\ppx} + \frac{\pp \theta}{\ppy}\right]-  \left[ \vxt \frac{\p \theta}{\p x} + \vyt \frac{\p \theta}{\p y}  \right] - a \theta + S + \epsilon, \label{eqn:theta_disc_pde} \\
\frac{\p S}{\p t} &= b^* \left[\frac{\pp S}{\ppx} + \frac{\pp S}{\ppy}\right]- a^* S + \tilde{\epsilon}, \label{eqn:source_disc_pde}
\end{align}
and several possible methods for approximating the partial derivatives can now be taken.
Recall that ultimately our interest lies in using these physical processes to motivate the parametrisation of a vector auto-regressive model.
It follows that here we seek an explicit solution that is also first-order in time.
We also stress that it is not our intention to obtain the exact solutions to the SPDEs and so we have no interest in studying the \textit{consistency} (and thus \textit{convergence}) or even accuracy of our approximation.

If we consider the advection-diffusion-reaction equation~\eqref{eqn:theta_disc_pde} then it is straightforward to obtain a solution that is both explicit and first-order in time via the forward-time centred-space (FTCS) method.
That is, a solution can be obtained by approximating the temporal and spatial derivatives by $\p \theta / \p t \simeq (\theta_{ij}^{t+1} - \theta_{ij}^t)/\delta t$, $\p \theta / \p x \simeq (\theta_{i+1,j}^t - \theta_{i-1,j}^t)/2\delta x$, $\pp \theta / \ppx \simeq (\theta_{i+1,j}^t + \theta_{i-1,j}^t - 2\theta_{ij}^t)/(\delta x)^2$, with obvious generalisations to derivatives taken w.r.t~$y$, and then solving for~$\theta_{ij}^{t+1}$.
Whilst such a solution is valid, in the context of VAR models, this approach yields an atypical parametrisation.
A more natural parametrisation can be obtained by appealing to \textit{operator splitting} methods. 
Here we consider sequential splitting
in which the original equation is split into two parts over a single time step.
More specifically, in the context of~\eqref{eqn:theta_disc_pde}, we first consider the advection-diffusion equation ($\p\theta/\p t = b \Delta\theta - \vec{v} \nabla \cdot \theta$) in isolation and, as in \cite{stroud2010ensemble} and \cite{xu2007estimation}, we consider a FTCS solution; let the solution be given by $\tilde{\theta}_{ij}^{t+1}$.
The remaining reaction (decay) equation given by the SDE $d\theta/d t = -a \theta + S + \epsilon$ is solved using the first-order forward time approximation ($d\theta/d t \simeq (\theta_{ij}^{t+1} - \theta_{ij}^t)/\delta t$), however here $\theta_{ij}^t =   \tilde{\theta}_{ij}^{t+1}$ is initialised from the solution to the advection-diffusion equation.
The diffusion-reaction equation~\eqref{eqn:source_disc_pde} is solved using the same approach by considering $\vec{v}_t = \vec{0}$ fixed throughout; full details are provided in Section~1 of the supplementary material.
We note in passing that the FTCS solution to the advection-diffusion equation is only appropriate if the diffusion dominates the solution and we assume this here.
In settings where this is not the case an \textit{upwind} solution
could instead be sought, although we note that such solutions are rather inconvenient as the evolution equations change depending on (the sign of) the velocity components~$\vec{v}_t$, which are unknown.
For the interested reader, \cite{hoffman2018numerical} and \cite{leveque2007finite} provide a comprehensive discussion of the numerical methods considered here.

Applying the approach above yields the evolution equations
\begin{align}
\theta_{ij}^{t+1} &= \alpha \left[ (1-4\beta)\theta_{ij}^t + (\beta - \nu_x^t)\theta_{i+1,j}^t + (\beta +\nu_x^t)\theta_{i-1,j}^t \right. \notag\\
&\hspace{3cm} \left. + (\beta - \nu_y^t)\theta_{i,j+1}^t + (\beta + \nu_y^t)\theta_{i,j+1}^t \right] + S_{ij}^t  + \epsilon_{ij}^t, \label{eqn:theta_evolution}  \\
S_{ij}^{t+1} &= \alpha^* \left[ (1-4\beta^*)S_{ij}^t + \beta^* \left(S_{i+1,j}^t +  S_{i-1,j}^t +  S_{i,j+1}^t +  S_{i,j+1}^t \right) \right] + \tilde{\epsilon}^{\,t}_{ij}. \label{eqn:source_evolution} 
\end{align}
Note that here the parameters do not directly correspond to those given in~\eqref{eqn:theta_disc_pde} and~\eqref{eqn:source_disc_pde}; the discretisation constants ($\delta x,\delta y,\delta t$) have been absorbed into other quantities, now denoted by $\alpha,\beta,\alpha^*,\beta^*$ and $\nu$, and these quantities will be considered unknown. In principle this solution could be viewed as a dimensionless parametrisation,
although we do not find this interpretation particularly useful; ultimately the parameters in \eqref{eqn:theta_evolution} and \eqref{eqn:source_evolution} now simply correspond to the (assumed unknown) quantities that result from the finite difference solution and no longer have a strictly physical interpretation.

The evolution equations \eqref{eqn:theta_evolution} and \eqref{eqn:source_evolution} are well defined for the majority of the points within the spatial domain, however issues arise along the edge of the system as not all neighbouring locations exist.
Thus we need to enforce an appropriate boundary condition on the spatial domain.
We consider periodic boundary conditions and, whilst this may provoke concerns about spurious periodicity,
 we note that this can be alleviated by constructing the problem such that the forecast domain of interest, say~$\mathcal{S}^p$, is relatively small in comparison to the model domain~$\mathcal{S}$ and so the effect of any spurious periodicity is likely to be negligible within~$\mathcal{S}^p$ \citep{cressie1991stat, lawson1999edge,liu2021statistical}.

\subsubsection{A note on spatial dependence}
\label{sec:spatial_dep}
For a particular location~$i,j$ the relationship between the value of the processes at time~$t+1$ and that at the previous time point, that is, the evolution defined within the square braces ([]) of~\eqref{eqn:theta_evolution} and~\eqref{eqn:source_evolution}, can be conveniently expressed by the five-point stencils \citep{milne1953numerical}
{\small
\begin{equation}
\begin{bmatrix}
- & \beta-\nu_y^t & -\\
\beta+\nu_x^t & 1-4\beta  & \beta-\nu_x^t\\
- & \beta+\nu_y^t & -\\
\end{bmatrix} \hspace{1cm} \text{\normalsize and} \hspace{1cm}
\begin{bmatrix}
- & \beta^* & -\\
\beta^* & 1 - 4\beta^*  & \beta^*\\
- & \beta^* & -\\
\end{bmatrix},
\label{eqn:stencils}
\end{equation}}
respectively.
These stencils highlight the first-order spatial neighbourhood structure; this evolution is akin to that defined by a (first-order) STAR($1_1$) model \citep{pfeifer1980three,cressie1991stat,garside2003dynamic}.
Of course, higher order neighbourhoods could also be considered.
For example, assuming anisotropic diffusion ($\Sigma$ non-diagonal in \eqref{eqn:theta_pde}) would introduce dependence on the second-order (corner) neighbours, at the cost of additional unknown parameters.
Further, an arbitrary number of spatial neighbours could be introduced by considering higher order (spatial) approximations in the finite difference solution.

A natural approach to determining the spatial order would be to pose this as a model selection problem, but this approach is not straightforward. 
Solving the inference (inverse) problem is challenging, especially within the Bayesian paradigm, and requires a bespoke approach for each different model.
Thus, rather than consider a collection of models, we instead propose to address the issue indirectly by considering the system model given by~\eqref{eqn:theta_evolution}-\eqref{eqn:source_evolution} and ``imputing time-steps'' between the observation times. 
Put another way, the latent processes (potentially) operate on a different temporal scale to the observed process.
The basic premise is that increasing the number of imputed time-steps is akin to considering a larger neighbourhood when viewed from the observation times.
For example, with observations at time~$t$ and~$t+1$, from~\eqref{eqn:theta_evolution} we have that the process at time~$t+1$ is spatially dependent on the first order neighbourhood at time~$t$.
However, if we introduce an intermediate time~$t + \frac{1}{2}$, then propagate $\theta_t \to \theta_{t+\frac{1}{2}} \to \theta_{t+1}$ via~\eqref{eqn:theta_evolution} then information from higher order spatial neighbours is passed from time~$t$ to time~$t+1$ via the intermediate time~$t + \frac{1}{2}$.
As a result, increasing the number of imputed time-steps also allows for the process to transition across numerous grid squares between the observation times; without imputation the first-order neighbourhood structure~\eqref{eqn:stencils} would only allow for transitions across a single grid square per time-step, which is perhaps physically questionable.
The benefit of considering imputing time-steps as opposed to a collection of different models/evolution stencils is that this approach has minimal impact on our solution to the Bayesian inference problem.
It is therefore straightforward to consider varying levels of imputation and use standard model selection methods to determine which model is best; this is discussed further in Section~\ref{sec:real_data}.

\subsection{State space representation}
\label{sec:state_space_rep}
The evolution equations~\eqref{eqn:theta_evolution} and~\eqref{eqn:source_evolution} can be straightforwardly rewritten as vector autoregressive relationships.
Let $\thetavec_t = (\theta_{11}^t,\dots,\theta_{n1}^t,\theta_{12}^t,\dots,\theta_{n2}^t,\dots,\theta_{1n}^t,\dots,\theta_{nn}^t)'$ denote the state of the process at time~$t$, with $\vec{S}_t, \vec{\epsilon}_t$ and $\vec{\epsilon}^*_t$ similarly defined.
We can then construct \textit{evolution} matrices $G(\vec\nu_{t})$ and~$G^*$, where row~$i$ contains the neighbour weights (for location~$i$) as prescribed by the stencils~\eqref{eqn:stencils}, left and right respectively.
We also impose temporal dependence within each component of the velocity vector~$\vec{\nu}$; more specifically a first-order (vector) autoregressive relationship.
Note that in the continuous context of Section~\ref{sec:system_model} this is equivalent to assuming that the advection terms evolve according to the SDE governing a simple decay function, that is, $d \vec{\nu}(t)/d t = -a_\nu \vec{\nu}(t) + \epsilon^\nu(t)$.
Finally, subject to certain restrictions on the parameter values (e.g. $\alpha \in (0,1)$) that we impose via the prior distribution, the system has a \textit{steady state} solution.
It follows that as $t \to \infty$ the system tends to its marginal mean, which is zero.
The assumption of a zero-mean process is not physically justifiable  and so we also introduce an additional (spatially and temporally homogeneous) mean parameter~$\mu$.

The dynamics of the system can then be expressed as
\begin{align}
\label{eqn:system_model}
\thetavec_t - \mu \voneN &= G(\vec\nu_{t-1})(\thetavec_{t-1}- \mu \voneN) + \vec{S}_{t-1}+ \vec{\epsilon}_t, \notag\\
\vec{S}_t &= G^* \vec{S}_{t-1} + \vec{\epsilon}_t^*, \\
\vec{\nu}_{t} &= \alpha_v\vec{\nu}_{t-1} +\vec\epsilon^\nu_t, \notag
\end{align}
for $t=1,\dots,T$, where~$\voneN$ is an $N$-length column vector containing ones.
The initial value/condition of the system~$\thetavec_0,\vec{S}_0$ and $\vec{\nu}_0$ are assumed to be unknown and assigned a prior distribution; full details are given in Section~\ref{sec:Bayes_model}.

\section{The observation model}
\label{sec:obs_model}
In general, we suppose that all quantities in~\eqref{eqn:system_model} are unknown and attempt to infer the likely values given some observed data.
More specifically we suppose that both the radar and rain gauges provide noisy and potentially biased observations of the latent rain field.
Further, both of these observation sources provide readings in units of mm/h and are therefore bounded below at zero.
We therefore suppose that these data provide censored (at zero) observations of the latent precipitation intensity field (that exists on the real line).
The relationship between the observations and the latent process is modelled via a (Type~I) Tobit model \citep{tobin1958estimation} where, for a zero-censored variable, if $Y\sim \text{Tobit}(\mu,\phi)$ then~$Y$ has a mixed discrete-continuous distribution with likelihood contribution $\pi(Y|\mu,\phi) = \left[ \sqrt{\phi} \; \varphi \left\{ \sqrt{\phi}(y - \mu) \right\} \right]^{\mathbb{I}(Y>0)} \left[ 1 -  \Phi \left\{ \sqrt{\phi} (\mu-0) \right\} \right]^{\mathbb{I}(Y=0)}$.
Here $\varphi$ denotes the standard normal probability density function and $\Phi$ the corresponding cumulative distribution function.
In what follows we parameterise the relationship between the observations and the latent field of interest~($\theta$) in a general framework.
Full details about the transformations and how these data are acquired for our application are given in Section~\ref{sec:real_data}.

\subsection{Information provided by the gauges}
\label{sec:gauge_obs_model}
We suppose that we have observations from~$\Ng$ rain gauges where gauge~$g$ provides a single observation~$G_{tg} \geq 0$ at each time~$t$. 
Let~$\mathcal{D}^\tg = \{G_{tg};  g=1,\dots,\Ng, t=1,\dots,T\}$ denote the collection of observed gauge data.
These observations are censored at zero and typically exhibit a long right tail; let $\tilde{Y}^\tg_{tg} = \mathcal{M}(G_{tg})$ denote a suitably transformed observation where $\mathcal{M} \in \mathcal{F}$ is from the set of functions $\mathcal{F}  =\{f : \mathbb{R}_{\geq 0} \to\mathbb{R}_{\geq 0} \; \text{s.t} \; f(0) = 0  \}$.
In our application we take the transformation proposed by \cite{yeo2000new} with parameter~$\lambda=0$ and so $\mathcal{M}(x) = \log(x+1)$.

Each gauge is located within a single cell of our regular lattice grid.
Let $\ell_g \in \{1,\dots,N\}$ denote the location of gauge~$g$ for $g=1,\dots,\Ng$ and note that several gauges may reside at the same location, that is, we may have $\ell_g = \ell_{g'}$ for $g \neq g'$.
Conditional on the quantities in~\eqref{eqn:system_model}, we assume
\begin{equation}
\tilde{Y}^\tg_{tg}|\cdot \sim \text{Tobit}(\theta_{t \ell_g},\phig)
\label{eqn:gauge_obs_eqn}
\end{equation}
for $g=1,\dots,\Ng$ and $t=1,\dots,T$, where $\phig = \sigma_g^{-2}$ is a (potentially) unknown precision parameter governing the observation error.

\subsection{Information provided by the radar}

For the purpose of model specification we suppose that the radar provides observations of rainfall intensity (mm/h) at each location within our spatial domain; further details about how these data are obtained are provided in our case study in Section~\ref{sec:real_data}.
Following Section~\ref{sec:gauge_obs_model} we let $\mathcal{D}^\tr = \{R_{ti};  i=1,\dots,N, t=1,\dots,T\}$ denote the collection of observed radar data where~$R_{ti} \geq 0$ is the radar observation from location~$i$ at time~$t$ and $\tilde{Y}^\tr_{ti} = \mathcal{M}(R_{ti})$ denotes the corresponding transformed observation. 
Similarly, conditional on the quantities in~\eqref{eqn:system_model}, we assume
\begin{equation}
\tilde{Y}^\tr_{ti}|\cdot \sim \text{Tobit}(\theta_{t i} + \mu_r,\phir)
\label{eqn:radar_obs_eqn}
\end{equation}
for $i=1,\dots,N$ and $t=1,\dots,T$.
Here an additional bias parameter~$\mu_r \in \mathbb{R}$ is introduced to allow for potential miscalibration of the radar and $\phir = \sigma_r^{-2}$ is a (potentially) unknown precision parameter governing the observation error.

\section{The Bayesian model}
\label{sec:Bayes_model}
We are now in a position to define our complete Bayesian model that combines the observation models~\eqref{eqn:gauge_obs_eqn} and~\eqref{eqn:radar_obs_eqn} together with the system model~\eqref{eqn:system_model}. 
More specifically
\begin{align}
\label{eqn:full_bayes_model}
\hspace{3cm}\tilde{Y}^\tg_{tg}|\cdot &\sim \text{Tobit}(\theta_{t\ell_g} , \phig), & \hspace{-5cm} g&=1,\dots,\Ng, \notag\\
\tilde{Y}^\tr_{ti}|\cdot &\sim \text{Tobit}(\theta_{ti} + \mu_r, \phir), &  i&=1,\dots,N, \notag\\
\thetavec_t - \mu \voneN &= G(\vec\nu_{t-1})(\thetavec_{t-1}- \mu \voneN), + \vec{S}_{t-1}+ \vec{\epsilon}_t, \\
\vec{S}_t &= G^* \vec{S}_{t-1} + \vec{\epsilon}_t^*, \notag\\
\vec{\nu}_{t} &= \alpha_v\vec{\nu}_{t-1} +\vec\epsilon^\nu_t, \notag
\end{align}
for $t=1,\dots,T$.
As noted in Section~\ref{sec:system_model}, given that both spatial and temporal dependence are allowed for in the source-sink process~($S$) we propose to keep the form of the errors simple. In particular we assume (i.i.d) zero-mean Gaussian innovation errors and so $\vec\epsilon_t \indep \N_N(\vec{0}, \phi_\theta^{-1} \INN)$ and $\vec\epsilon^*_t \indep \N_N(\vec{0}, \phi_s^{-1} \INN)$ for $t=1,\dots,T$.
Similarly we take the innovation error on the ``velocities'' to be $\vec{\epsilon}^\nu_t \indep \N_2(\vec{0}, \phi^{-1}_\nu \mathbb{I}_2)$ for $t=1,\dots,T$.
These assumptions, coupled with an appropriate prior distribution for the initial values of the dynamic states ($\thetavec_0,\vec{S}_0,\vec{\nu}_0$) and also the unknown static quantities of interest, will allow an appropriate collection of latent variables to be introduced that in turn facilitate the full conditional distributions for all unknown quantities to be written down in closed form.

\subsection{Prior specification}
\label{sec:prior_spec}

Specifying a suitable prior distribution is a problem well discussed within the Bayesian literature~\citep{BernardoS94}.
Given the Gaussian form of our observation model and, with tractability in mind, we choose to place Gaussian distributions over the initial values of the system.
More specifically we take $\thetavec_0|\mu \sim \N_N(\vec{m}_\theta + \mu \voneN, C_\theta), \vec{S}_0 \sim \N_N(\vec{m}_s, C_s)$ and $\vec{\nu}_0 \sim \N_2(\vec{m}_\nu, C_\nu)$, where the mean vectors~($\vec{m}_\cdot$) and covariance matrices~($C_\cdot$) are specified \textit{a priori}.
The static mean and bias parameters are also assumed to follow Gaussian distributions  $\mu \sim \N(m_{\mu},v_{\mu})$ and $\mu_r \sim \N(m_{\mu_r},v_{\mu_r})$.
Our prior specification is completed by choosing appropriate distributions for the static parameters within the evolution matrix~$G_t$; we take $\beta \sim \N(m_{\beta},v_{\beta})$ where, to encourage positive evolution coefficients in~\eqref{eqn:theta_evolution}, $m_{\beta}$ and $v_{\beta}$ are to be chosen so that $\Pr(\beta \in (0,0.25)) \simeq 1$ \textit{a priori}.
To ensure a stationary solution we take $\alpha \sim \text{TN}(m_{\alpha},v_{\alpha},0,1)$, that is, a truncated Gaussian distribution so that $\alpha \in (0,1)$.

The remaining static quantities in~\eqref{eqn:full_bayes_model} are considered to be fixed (known) constants, although we note that this is not a limitation of our approach.
For example, appropriate (Gamma) prior distributions could also be placed on the innovation precision parameters ($\phitheta,\phis,\phi_\nu$), though we found that these parameters are only weakly identified in practice.
Similar issues arise for the observation precision parameters~$\phir$ and~$\phig$.
This is further complicated given that, in our application, the radar provides a much larger number of observations (and therefore information) in comparison to the rain gauges ($N \gg \Ng$).
It follows that, with~$\phig$ unknown, there is a potential risk of the gauge observations being explained by large observation error ($\phig \simeq 0$) and this is not desirable given that the gauges provide the most accurate information about the precipitation intensity field at ground level; this issue can be alleviated by an appropriate choice of $\phig \gg \phir$.
Of course, choosing suitable values for these static quantities poses its own challenge, although such choices can be guided by inspection of the prior predictive distribution.
The choice of state innovation error~$\phitheta$ from a finite set of possible values is posed as a model selection problem.

\section{Posterior computation}
\label{sec:posterior_comp}

The Tobit likelihood(s) together with the prior distribution given in Section~\ref{sec:prior_spec} leads to a posterior distribution that is not available in closed form.
We therefore propose a simulation based inference approach, specifically MCMC, to obtain draws from the posterior distribution of interest.
The design of the sampling algorithm can be simplified by the introduction of appropriate latent variables that give rise to tractable full conditional distributions (FCDs) for all unknown quantities of interest.
However, the dimension of our state-space renders (exact) simulation from the joint full conditional distribution of the dynamic states computationally infeasible.
A potential, relatively straightforward, solution to this issue would be to sample each of the latent states one-at-a-time from their respective full conditional distribution.
Unfortunately this approach is likely to lead to poor mixing within the MCMC scheme and therefore require an infeasible number of iterations to obtain a reasonable number of near uncorrelated posterior draws.
We therefore propose to use the ensemble Kalman smoother (EnKS) to obtain an approximate sample from the joint full conditional distribution of the dynamic states.
The EnKS sampling step is embedded within a Gibbs sampling algorithm (in place of its exact counterpart) to facilitate approximate posterior sampling in reasonable time.
Section~\ref{sec:latent_vars} introduces an appropriate collection of latent variables and the resulting complete data model is given in Section~\ref{sec:complete_data_model}. The EnKS algorithm is outlined in Section~\ref{sec:enks} and a discussion of our sampling algorithm is given in Section~\ref{sec:mcmc_sampling}.

\subsection{Latent variables}
\label{sec:latent_vars}

The form of the Tobit likelihood(s) precludes use of conjugate Bayesian inference for the unknown quantities in the model~\eqref{eqn:full_bayes_model}.
Of course, samples from the posterior distribution could be obtained via Metropolis-Hastings (MH) steps within an MCMC scheme.
However, whilst it is fairly straightforward to conceive of a sensible proposal distribution for the static parameters, it is less clear how to construct a suitable joint proposal for the dynamic states.
In particular, the high-dimension of the state-space is likely to lead to poor acceptance rates, which is not appealing.
One-at-a-time MH steps are also unlikely to yield an effective sampling scheme as discussed above.

Fortunately the inference problem can be made more tractable by appealing to data augmentation \citep{tanner1987calculation}.
The basic premise is to introduce an appropriate collection of \textit{latent observations} such that the complete data likelihood, that is, the joint distribution of the latent observations and the observed data, is of a convenient form.
Indeed, an equivalent way to arrive at the observation model~\eqref{eqn:full_bayes_model} is to directly define (partially) latent vectors that denote the complete data,~$\vec{Y}^\tg_t$ and~$\vec{Y}^\tr_t$, and assume $Y^\tg_{tg}|\cdot \sim N(\theta_{t\ell_g} , \phig)$ and $Y^\tr_{ti}|\cdot \sim N(\theta_{ti} + \mu_r, \phir)$ for $g=1,\dots,\Ng$, $i=1,\dots,N$ and $t=1,\dots,T$.
Then, conditional on the complete data (and all other unknown quantities), we can let the observed data $\tilde{Y}^\tg_{tg} = Y^\tg_{ti} \mathbb{I}\left( Y^\tg_{ti} > 0\right)$ and $\tilde{Y}^\tr_{ti} = Y^\tr_{ti} \mathbb{I}\left( Y^\tr_{ti} > 0\right)$ from which it follows that the observed data likelihood contributions, for example, $\pi (\tilde{Y}^\tg_{tg}|\cdot ) = \int \pi(\tilde{Y}^\tg_{tg}|Y^\tg_{tg})\pi(Y^\tg_{tg}|\cdot) dY^\tg_{tg}$ are as outlined in Section~\ref{sec:obs_model}.
Note that, given a collection of observed radar and gauge observations $\D = \{ \D^\tg,\D^\tr\}$, the full conditional distributions of the quantities within the (partially) latent vectors are straightforward to obtain and are 
$Y^\tg_{tg}|\D^\tg, \cdot \sim \text{TN}(\theta_{t \ell_g},\phig,-\infty,0)$ if $\tilde{y}^\tg_{tg} = 0$ and $Y^\tg_{tg} = \tilde{y}^\tg_{tg}$ otherwise; similarly $Y^\tr_{ti}|\D^\tr, \cdot \sim \text{TN}(\theta_{t i} + \mu_r,\phir,-\infty,0)$ if $\tilde{y}^\tr_{tr} = 0$ and $Y^\tr_{ti} = \tilde{y}^\tr_{ti}$ otherwise.

Let~$\vec{Y}_t = \left(\vec{Y}'^\tr_t, \vec{Y}'^\tg_t \right)'$ be the $(N+\Ng)$-length complete data vector that contains the positive (radar and gauge) observations when available, and the latent (truncated Gaussian) variables otherwise.
Further, for notational simplicity, let $\rhovec = (\mu, \mu_r,\alpha,\beta)$ denote the unknown static quantities and
$\vec{q}_{1:t} \equiv \{\vec{q}_1,\dots,\vec{q}_t\}$ for any vector valued quantity~$\vec{q}$.
From above it follows that the complete data likelihood $\pi(\vec{Y}_{1:T}|\thetavec_{0:T},\vec{S}_{0:T},\vec{\nu}_{0:T},\vec{\rho})$ is a $T \times (N+\Ng)$-dimensional multivariate Gaussian density.
Crucially this allows us to write down a dynamic linear model for the complete data from which the full conditional distributions of all remaining unknown quantities can be straightforwardly derived; this is the topic of the next section.

\subsection{Complete data model}
\label{sec:complete_data_model}
The \textit{complete data model} is given by our original state and observation model coupled with that for the latent observations~$\vec{Y}_{1:T}$.
With notational simplicity in mind, let $\xvec_t = (\thetavec_t',\vec{S}'_t)'$ denote the joint vector containing both of the dynamic processes at time~$t$.
It follows that the unknown quantities of interest are $\vec{Y}_{1:T}$, $\xvec_{0:T}$, $\vec{\nu}_{0:T}$ and $\vec{\rho}$ and the density of all stochastic quantities is 
$\pi(\D,\vec{Y}_{1:T},\xvec_{0:T},\vec{\nu}_{0:T},\vec{\rho}) = \pi(\D|\vec{Y}_{1:T})\pi(\vec{Y}_{1:T}|\xvec_{0:T},\vec{\nu}_{0:T},\vec{\rho})\pi(\xvec_{1:T}|\xvec_{0},\vec{\nu}_{0:T},\vec{\rho}) \linebreak[1] \pi(\xvec_{0}|\vec{\rho})\pi(\vec{\nu}_{0:T})\pi(\vec{\rho})$, from which the full conditional distributions can be obtained in closed form.
Note that the full conditional distributions for the latent variables~$\vec{Y}_{1:T}$ are those outlined in Section~\ref{sec:latent_vars}.
Now, the upshot of introducing the latent observations is that zeros within the observed data no longer make a contribution to the complete data likelihood.
So, conditional on the latent observations (and the observed data), we can construct the conditional distribution
$\pi(\xvec_{0:T},\vec{\nu}_{0:T},\vec{\rho}|\vec{Y}_{1:T},\D) \propto  \pi(\vec{Y}_{1:T}|\xvec_{0:T},\vec{\nu}_{0:T},\vec{\rho})\pi(\xvec_{1:T}|\vec{\nu}_{0:T},\vec{\rho})\pi(\xvec_{0}|\vec{\rho})\pi(\vec{\nu}_{0:T})\pi(\vec{\rho}).$
Crucially, conditional to the prior specification outlined in Section~\ref{sec:prior_spec}, this is simply the density corresponding to the Dynamic Linear Model
\begin{align}
\hspace{3cm}  \vec{Y}_t&= F \xvec_t + \mu_r \lrvec + \vec{v}_t, & \vec{v}_t &\sim \N_{N+\Ng}(\vec0,V ), \notag \\
(\xvec_t - \mu\vec{L})&= \tilde{G}_t\left(\xvec_{t-1}- \mu\vec{L}\right) + \vec{w}_t, & \vec{w}_t &\sim \N_{2N}(\vec0,W), \label{eqn:complete_data_model} \\
\vec{\nu}_{t} &= \alpha_v\vec{\nu}_{t-1} +\vec\epsilon^\nu_t, \notag
\end{align}
for $t=1,\dots,T$.
Here $\vec{L} = (\voneN',\vzeroN')'$ and $\lrvec = (\voneN',\vzeroNg')'$ are indicator vectors that map the mean and bias parameters to the respective process and observation layer.
The matrices
{\small
\begin{equation*}
F = 
\begin{bmatrix}
\INN & 0_{N\times N}\\
\mathcal{I}^\tg & 0_{\Ng\times N}
\end{bmatrix}, 
\tilde{G}_t = 
\begin{bmatrix}
G(\vec{\nu}_{t-1}) & \INN\\
0_{N \times N} & G^*
\end{bmatrix}, 
V = 
\begin{bmatrix}
\phi_\tr^{-1} \INN & 0_{N\times\Ng}\\
0_{\Ng\times N} & \phi_\tg^{-1}\INGNG
\end{bmatrix} \text{\normalsize and} \;
W = 
\begin{bmatrix}
\phitheta^{-1} \INN & 0_{N\times N}\\
0_{N\times N} & \phi_s^{-1} \INN
\end{bmatrix},
\end{equation*}}%
where $\mathcal{I}^\tg$ maps the latent rain field ($\theta$) at location~$\ell_g$ to the observation from gauge~$g$; specifically an ($\Ng \times N$) dimensional matrix where row~$g$ contains a 1 in column~$\ell_g$ and zeros otherwise.

The full conditional distributions for $\xvec_{0:T}$, $\vec{\nu}_{0:T}$ and $\vec{\rho}$ can now straightforwardly be derived from $\pi(\xvec_{0:T},\vec{\nu}_{0:T},\vec{\rho}|\vec{Y}_{1:T},\D)$.
In particular, given this density corresponds to a dynamic linear model, it follows that a joint sample from $\pi(\xvec_{0:T}|\cdot)$ can be obtained via the FFBS algorithm.
Thus, in principle, it is straightforward to construct a Gibbs sampling algorithm in which we draw a joint sample of the dynamic states (via the FFBS algorithm) and then proceed to draw from the full conditional distributions of the remaining parameters (including the latent observations) in turn.
Whilst this strategy is naturally appealing, unfortunately the FFBS algorithm becomes computationally infeasible in large state dimension.
In our application~$\xvec_t$ has length $2N = 2 \times 72^2 = 10368$ and so analytically computing the moments of the required (Gaussian) distributions is challenging; in particular the required covariance matrices are of dimension $10368 \times 10368$.
Our solution to this problem is to instead use the Ensemble Kalman Smoother to obtain an approximate draw from $\pi(\xvec_{0:T}|\cdot)$; this is the topic of the next section.

\subsection{State sampling via the Ensemble Kalman Smoother}
\label{sec:enks}
The ensemble Kalman smoother (EnKS) is a method for obtaining approximate draws from~$\pi(\xvec_{0:T}|\cdot)$.
The basic premise is that, rather than analytically computing the moments of the required (Gaussian) distributions (as is done in the FFBS algorithm), to instead consider an ensemble of~$\Ne$ (equally weighted) particles that are draws from the distributions of interest. 
These particles are propagated and sequentially updated as observations ``arrive'', where the (cross-)covariance matrices required to compute the update are approximated from the ensemble.
For dynamic linear models this method provides an exact sample from~$\pi(\xvec_{0:T}|\cdot)$ in the limit of $\Ne \to \infty$ \citep{katzfuss2020ensemble}.
The EnKS is a generalisation of the more well known ensemble Kalman filter (EnKF) and numerous variants of the smoother can be found within the literature,  see,  for example, \cite{evensen2000ensemble, khare2008investigation} and \cite{bocquet2014iterative}.
Here we consider the ensemble Kalman smoother of \cite{evensen2000ensemble}; the basic idea is that the smoothed states can be obtained by applying the standard EnKF update to the augmented state that includes all the previous history.
It follows that, at each time~$t$, the entire state history $(\xvec_{0},\dots,\xvec_{t})$ is updated to incorporate the information in the observations~$\vec{y}_t$; see \cite{katzfuss2020ensemble} for further details.
In what follows, the notation $\xvec_{n|m}$ represents a draw from the distribution~$\pi(\xvec_n|\vec{Y}_{1:m},\cdot)$, that is, a sample of the state vector at time $n$ given observations up to and including at time $m$.
For the model given by~\eqref{eqn:complete_data_model} an approximate sample~$\xvec_{0:T|T}$ from~$\pi(\xvec_{0:T}|\cdot)$ is obtained as follows.

Initialise: draw $\xvec_{0|0}^j \indep \N(\vec{m}_0, C_0)$ for $j=1,\dots,\Ne$. Then, for $t=1,\dots,T$,
\begin{enumerate}
\item Forecast step: for $j=1,\dots,\Ne$,
\begin{itemize}
\item Let $\xvec_\tgtm^j = \tilde{G}_t (\xvec_\tmgtm^j-\mu\vec{L}) + \mu\vec{L} + \vec{w}_t^j$ where  $\vec{w}^j_t \indep \N(\vec{0},W)$. 
\end{itemize}
\item Smoothing step: for $j=1,\dots,\Ne$,
\begin{itemize}
\item Generate pseudo-observations $\hat{\vec{y}}^j_\tgtm = F \xvec_\tgtm^j + \mu_r \lrvec + \vec{v}_t^j$ where $\vec{v}_t^j \indep \N(\vec{0}, V)$.
\item For $\ell=0,\dots,t$, compute (the smoothed state) $\xvec_{\ell|t}^j = \xvec_{\ell|t-1}^j + \hat{K}_{\ell t}(\vec{y}_t - \hat{\vec{y}}^j_\tgtm)$.
\end{itemize}
\end{enumerate}
A sample of the state vector is obtained by drawing $j \in \{1,\dots,\Ne \}$ uniformly at random and letting $\xvec_{0:T|T} = \xvec_{0:T|T}^j$.
Here $\hat{K}_{\ell t} = \Sigma^{x y}_{\ell t|t-1} \left(\Sigma^{yy}_{tt|t-1}\right)^{-1}$ is an approximation of the \textit{optimal} Kalman gain computed from the ensemble.
More specifically $\Sigma^{x y}_{\ell t|t-1} = \Cov(\mathcal{X}_{\ell|t-1},\mathcal{Y}_\tgtm)$ is the sample cross-covariance of the ensemble of states $\mathcal{X}_{\ell|t-1} = (\xvec_{\ell|t-1}^1,\dots,\xvec_{\ell|t-1}^\Ne)$ and pseudo observations $\mathcal{Y}_\tgtm = (\hat{\vec{y}}_\tgtm^1,\dots,\hat{\vec{y}}_\tgtm^\Ne)$; similarly $\Sigma^{y y}_{t t|t-1}$ is the sample covariance of the ensemble of pseudo observations $\mathcal{Y}_\tgtm$.

From a practical standpoint, updating (smoothing) the entire state history in Step~2 is computationally burdensome, even for modest length time-series.
Thus in practice it is typically only feasible to consider a ``fixed-lag smoother'' in which only the states at the previous~$\tau$ time points are updated given the observations at each time~$t$.
More formally to only consider $\ell=t^*,\dots,t$ where $t^* = \text{max}(0,t-\tau)$ in Step~2 of the algorithm above; the EnKF is a special case and is recovered by taking $\tau = 0$.
Finally, we note that if there are no observations available at time~$t$, for example, if~$\vec{x}_t$ is the state vector at an imputed time-step, then the smoothing step (Step~2) becomes trivial and we simply let $\xvec_{\ell|t}^j = \xvec_{\ell|t-1}^j$ for $\ell=0,\dots,t$ and $j=1,\dots,\Ne$.

\subsubsection{A note on covariance estimation}
\label{sec:covar_est_inf}
Key to the successful implementation of the EnKS algorithm is the effective estimation of the (cross-)covariance matrices required to compute~$\hat{K}_{\ell t}$, especially when the ensemble size ($\Ne$) is small, relative to the dimension of the states and/or observations.
Several methods have been proposed for improving the accuracy of the covariance approximations in this setting, most commonly,
\textit{covariance tapering} \citep{houtekamer2001sequential} and \textit{covariance inflation} \citep{anderson1999monte}.
However there is little guidance on how to choose an appropriate tapering function and/or inflation factor.
The approach we favour, and also that which we find works well in practice, relies on approximating the matrices from the so-called \textit{deterministic ensemble} and then using analytic results to obtain the required matrix.
More specifically we introduce $\tilde\xvec_\tgtm^j = \tilde{G}_t\xvec_\tmgtm^j$ in Step~1 of the EnKS algorithm and let $\xvec_\tgtm^j = \tilde\xvec_\tgtm^j + \vec{w}_t^j$ for $j=1,\dots,\Ne$.
The Kalman gain matrix can then be written as
\begin{equation}
\hat{K}_{\ell t} =
\begin{cases}
\tilde\Sigma_\ltgtm  F'\left( F\tilde\Sigma_\ttgtm F' +FWF' + V\right)^{-1}, &\text{for} \; \ell < t, \\
(\tilde\Sigma_\ttgtm + W) F_t'\left( F\tilde\Sigma_\ttgtm F' +FWF' + V\right)^{-1}, &\text{for} \; \ell = t.
\end{cases}
\end{equation}
Here $\tilde{\Sigma}_{\ttgtm}$ is the sample covariance of the deterministic ensemble $\tilde{\mathcal{X}}_{t|t-1} = (\tilde{\xvec}_{t|t-1}^1,\dots,\tilde{\xvec}_{t|t-1}^\Ne)$ and $\tilde{\Sigma}_{\ltgtm}$ is the cross-covariance matrix  computed from the ensembles $\mathcal{X}_{\ell|t-1}$ and $\tilde{\mathcal{X}}_{t|t-1}$.
Additional discussion is provided in Section~2 of the supplementary material and further guidance on the practical implementation of these algorithms is given by \cite{evensen2003ensemble}.

\subsection{MCMC sampling algorithm}
\label{sec:mcmc_sampling}
With all full conditional distributions available in closed form we can straightforwardly implement a Gibbs sampling scheme to obtain draws from the desired posterior distribution \citep{geman1987stochastic,smith1993bayesian,tierney1994markov}.
Our MCMC algorithm is akin to the Gibbs ensemble Kalman smoother (GEnKS) algorithm outlined in \cite{katzfuss2020ensemble}.
In particular we draw a joint sample of the dynamic states $\vec{x}_{0:T}$ via the EnKS algorithm outlined in Section~\ref{sec:enks}, then proceed to draw a sample of each quantity in $\vec{\rho}$ is from its corresponding (univariate) FCD.
Samples of the latent observations~($Z$) are drawn from their FCDs given in Section~\ref{sec:latent_vars} and finally the ``velocity'' parameters are sampled from $\pi(\vec{\nu}_t|\cdot)$ for $t=0,\dots,T$.
The latent variable method of \cite{damien2001sampling} is used to facilitate numerically stable sampling from truncated (Gaussian) distributions.
We note in passing that a joint sample of~$\vec{\nu}_{0:T}$ could be obtained via the FFBS algorithm that may promote better mixing of the MCMC chain.
However, in general, we find that our chains mix well and so do not consider this here.

Although Gibbs samplers are inherently serial algorithms the total execution time can often be reduced by appealing to parallel computing within each of the sampling steps.
In particular many of the computations required within the EnKS algorithm can be effectively performed in parallel; see, for example, \cite{houtekamer2014parallel}.
Further, the large matrix operations, including those required to compute the moments of the full conditional distributions, can be straightforwardly computed in parallel, for example, via routines within the Eigen library \citep{eigenwebv3}.
C++ code to run our algorithm can be found at the GitHub repository \url{https://github.com/
srjresearch/BayesianRainfallModelling}.

\subsection{Simulation study}
\label{sec:sim_study_paper}
To investigate the effectiveness of the posterior sampling scheme outlined in Section~\ref{sec:mcmc_sampling} we analyse several synthetic datasets for which the values of the underlying parameters are known; the results are given in Section~3 of the supplementary material.
These synthetic datasets closely resemble that within our real data application which follows in Section~\ref{sec:real_data} and consider an $N=72\times 72$ spatial grid with $\Ng=15$ rain gauges over $T=72$ observation times.
The choice of ensemble size and the length of the smoothing window to consider are primarily limited by computational resource.
From a practical standpoint we would like to be able to make inferences in a reasonable amount of time (hours as opposed to days) and to this extent we let $\Ne =100$ and back smooth the latent states up to and including the previous $\tau = 3$ observation times within the EnKS step.
These simulation studies reveal that, in general, we are able to effectively recover the unknown quantities of interest given these practical choices; considering a larger smoothing window of $\tau = 6$ increases the total execution time of the algorithm by (approximately) a factor of~$3$ and does not have a notable effect on parameter inference.

Our model~\eqref{eqn:full_bayes_model} is governed by numerous parameters in addition to multiple layers of latent variables and this may give rise to parameter identifiability issues.
However, the simulation studies suggest that the underlying parameters are identifiable and are generally well recovered despite the bias introduced by the approximation in the EnKS step.
For example, the mean~$\mu$ and bias~$\mu_r$ parameters are well identified, with the synthetic values obtaining reasonable support under their respective marginal posterior distributions.
Perhaps unsurprisingly inferences for the static parameters~$\alpha$ and~$\beta$ that govern the system evolution are those most affected by EnKS approximation; this may also be an artefact of the information loss as these data are only partially observed.
That said, it is pleasing to see that the velocity components~$\vec{\nu}_t$ are generally well inferred, with the majority of the synthetic values looking plausible under their marginal posterior distributions.
Crucially, these simulation studies suggest that we are able to make plausible inferences about the latent process of interest~$\thetavec_t$, with spatial clusters of positive (and negative) precipitation intensities being well recovered.
Finally we note that the observed data look plausible under the respective posterior predictive distributions and so, based on these simulation studies, we suggest that we are able to make reasonable inferences given the choice of $\Ne =100$ and $\tau = 3$.

\section{Real data application}
\label{sec:real_data}

The Urban Observatory \citep{james2014urban} collects data from a large number of sensors distributed around the North East of England.
Here we consider observations obtained from a WR--10X radar situated on Claremont Tower, Newcastle upon Tyne, UK (Lat 54.9804, Long -1.6146) in addition to a collection of tipping bucket rain gauges distributed around the city centre.
We consider a 12 hour period from the 12th August 2018 when moderately large rainfall rates were observed and also for which a reasonable amount of rain gauge data is available.
Our modelling domain~$\mathcal{S}$ is taken to be an 18Km$^2$ area centered at the location of the radar (Claremont Tower).
This domain is discretised as outlined in Section~\ref{sec:discretisation} by imposing a regular lattice grid of dimension $N=72\times 72$ and so each grid cell is of dimension $500$m$^2$.

The radar provides, at~$10$ minute time intervals, $240 \times 360$ raw observations~$z$ that are given in units of dBZ (decibel relative to~$Z$).
The corresponding rainfall rates~$R$ (mm/h) are obtained via the \cite{marshall1948distribution} relationship.
Additionally we suppose that values of $z<0$ should correspond to zero rainfall and so we let the $z$-$R$ relationship be $R(z) = \left( 10^{(z/10)}/200 \right)^{0.625} \mathbb{I}(z>0)$.
Note that $(1/200)^{0.625} = 0.04$ and so only rainfall rates less than $0.04$ mm/h are truncated to zero.
The radar records observations at $150$m intervals along ``beams'' that are $36$Km in length; $360$ beams cover azimuth angles of $1$-$360$ degrees (at an angular resolution of $1$ degree).
It follows that these observations are irregularly spaced over the modelling domain~$\mathcal{S}$ and so we take $R_{ti}$ to be the average rainfall rate when numerous observations fall within the same grid square; observations that fall outside of our modelling domain are discarded.
Additionally we also obtain observations from~$\Ng = 15$ tipping bucket rain gauges that reside within our modelling domain~$\mathcal{S}$.
Each gauge is located within a single grid square and provides rainfall accumulations over a fixed time domain and, where appropriate, we sum over the appropriate time domains so as to obtain the~$10$ minute accumulation before then converting into units of mm/h (multiplying by a factor of~$6$).
It follows that the length of the time series is $T = 72$ and the number of the observations (at each time-step) is $N+\Ng = 5199$. 
These data (in units of mm/h) are reproduced within the associated GitHub repository.

Our ultimate interest lies in inferring the (latent) precipitation intensity field at ground level and to this extent we suppose that the rain gauges provide reasonably accurate information about this process (in comparison to the radar observations).
To reflect this we take $\phi_g = 100$ and $\phi_r=2$ which allows the gauge observations to more strongly influence the state inference; these values were also used in the simulation studies discussed in Section~\ref{sec:sim_study_paper}, which suggests our sampling scheme works well in this setting.
We consider the prior distribution outlined in Section~\ref{sec:prior_spec} with
$\thetavec_0|\mu \sim \N_N( \mu \voneN, 2^2 \INN)$, $\vec{S}_0 \sim \N_N(\vec{0}, 0.5^2 \INN)$, $\vec{\nu}_0 \sim \N_2(\vec{0}, 0.1^2 \mathbb{I}_2)$,   $\mu,\mu_r \sim \N(0,1)$, $\alpha \sim \text{TN}(0.8,250^{-1},0,1)$ and $\beta \sim \N(0.1,500^{-1})$ \textit{a priori}.
The parameters governing the evolution of the source-sink process, $\alpha^*=0.85$, $\beta^*=0.15$ and $\phi_s=20$, are considered fixed and were chosen via visual inspection of the prior predictive distribution.
We suppose that the components of the velocity vector~$\vec{\nu}$ should exhibit large temporal dependence and to this extent we let $\alpha_\nu = 0.95$ and $\phi_\nu = 2000$ so that each component is modelled via a stationary AR(1) process with stationary standard deviation of~$0.07$.
Finally we pose the choice of the innovation precision parameter~$\phi_\theta$ as a model selection problem.
More specifically we consider multiple values of~$\phi_\theta$ and select the value which gives rise to the lowest deviance information criterion (DIC) proposed by \cite{spiegelhalter2002bayesian}; the effective number of parameters is approximated by (two times) the sample variance of the log observed data likelihood  \citep{gelman2014bayesian}.
Of course, other model selection methods could also be considered, however the DIC is particularly appealing as it is computationally cheap to evaluate and does not require saving the MCMC draws of all unknown quantities, which is prohibitive in large scale applications.

Unfortunately attempts to directly fit the model~\eqref{eqn:full_bayes_model} to these data (with various values of $\phi_\theta$) gives rise to poor model fit and predictive performance (results not reported here).
Inspection of the radar observations reveals that rain clouds move fairly quickly across the spatial domain, that is, they transition across numerous grid squares between a single time-step.
Thus, following the discussion in Section~\ref{sec:spatial_dep}, we insert~$\tilde{t}$ time-steps between the observation times to allow our model to capture this behaviour.
Note that the dynamic processes of interest are~$\vec{x}_{0:\tilde{T}}$ and~$\vec{\nu}_{0:\tilde{T}}$ where the length of the augmented time series is $\tilde{T} = \tilde{t}(T-1) + T$.
Further, so as to only introduce similar levels of stochastic noise in to the latent process between the observation times, we scale the innovation (precision) matrix and let $W^{-1} \to W^{-1} \times (\tilde{t} + 1)$.
We consider $\phi_\theta \in \{20,40,60 \}$ and $\tilde{t} \in \{0,1,\dots,10\}$.
Posterior draws are obtained via the MCMC algorithm outlined in Section~\ref{sec:mcmc_sampling} where we perform $2000$ iterations with the first $1000$ discarded as burn-in.
Convergence is assessed by monitoring the trace of the (log) complete data likelihood in addition to the static parameters and randomly selected dynamic states.

These analyses reveal that the optimal model (from those considered) is that with $\phi_\theta = 40$ and $\tilde{t} = 7$ imputed time-steps.
Inspection of the marginal posterior distributions of the static parameters reveals that there is high correlation between the latent states at successive time points with E$(\alpha|\D) = 0.99$ (SD$(\alpha|\D) =3.87 \times 10^{-4}$); this is perhaps not surprising given the modestly large number of imputed time-steps.
Interestingly we find E$(\beta|\D) = 0.24$ (SD$(\beta|\D) =6.50\times 10^{-4}$) that, when coupled with the velocities~$\nu_x,\nu_y$, suggests that the (first-order) neighbours have a large influence with relatively little weight ($1-4\beta$) assigned to the same location at the previous time-step.
This analysis also reveals that the radar typically provides lower rainfall rates than those observed on the ground with E$(\mu_r|\D) = -1.09$ (SD$(\mu_r|\D) = 0.03$); if the radar was calibrated with the gauge observations we would expect to see $\mu_r \simeq 0$.

\begin{figure}[t]
\begin{center}
\begin{minipage}[b]{0.49\linewidth}
        \centering
   	    \includegraphics[width=\linewidth, clip, trim=0 20 30 60]{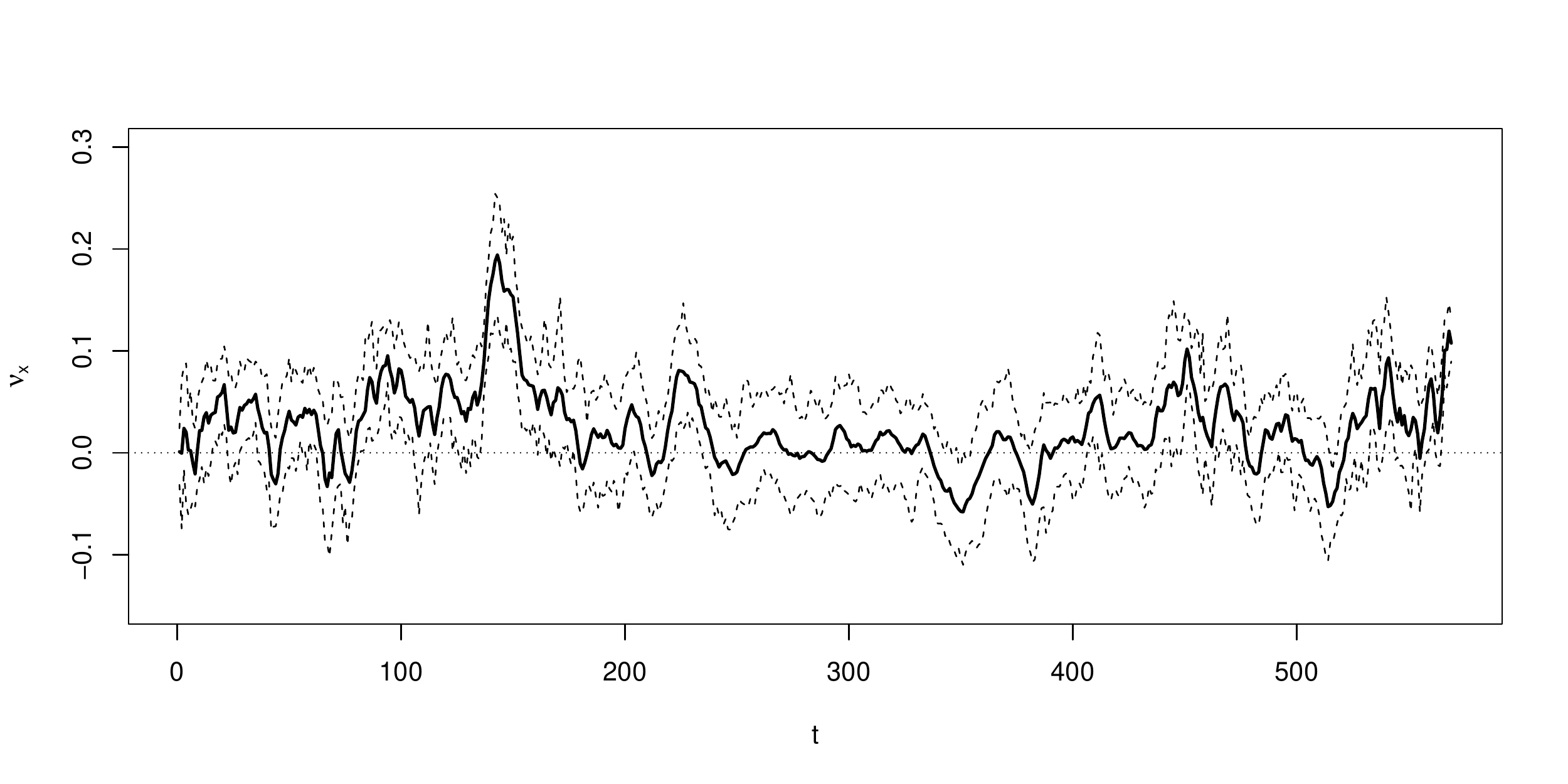}
\end{minipage} 
\begin{minipage}[b]{0.49\linewidth}
        \centering
        \includegraphics[width=\linewidth, clip, trim=0 20 30 60]{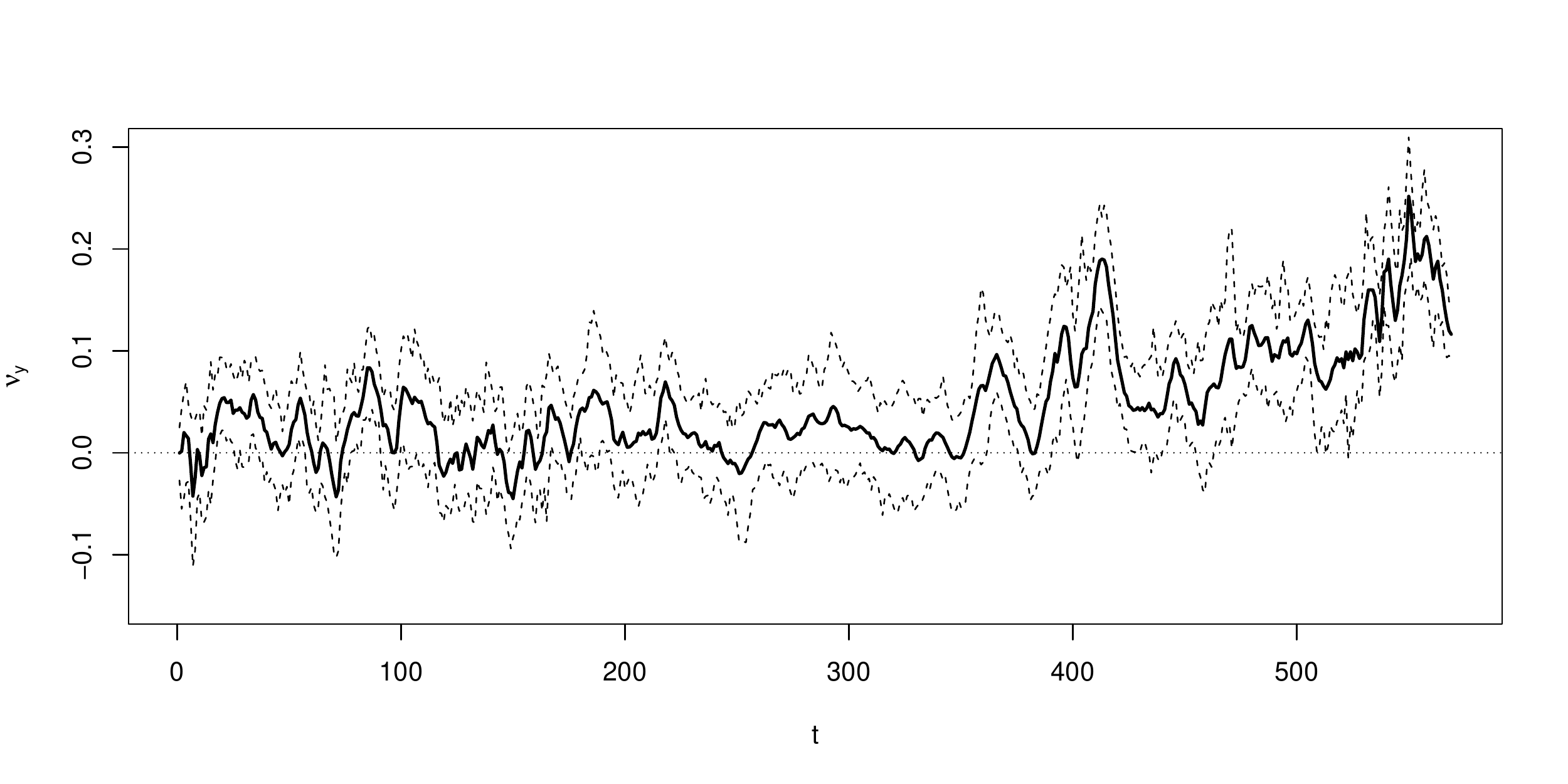}
\end{minipage}
\caption{Marginal posterior means (--) and the $95\%$ credible regions (- -) for the velocity components~$\nu_x$ (left) and $\nu_y$ (right) for $t=1,\dots,\tilde{T}$. Horizontal line (- $\cdot$ -) drawn at~$\nu_x,\nu_y = 0$. }
\label{fig:real_velocities}
\end{center}
\end{figure}

\begin{figure}[t]
\hspace{1cm} $t=12$ \hspace{1.05cm} $t=24$ \hspace{1.05cm} $t=36$ \hspace{1.05cm} $t=48$ \hspace{1.05cm} $t=60$ \hspace{1.05cm} $t=72$
\vspace{-0.25cm}
\begin{center}
        \centering
\includegraphics[width=0.15\linewidth, clip, trim= 55 70 30 55]{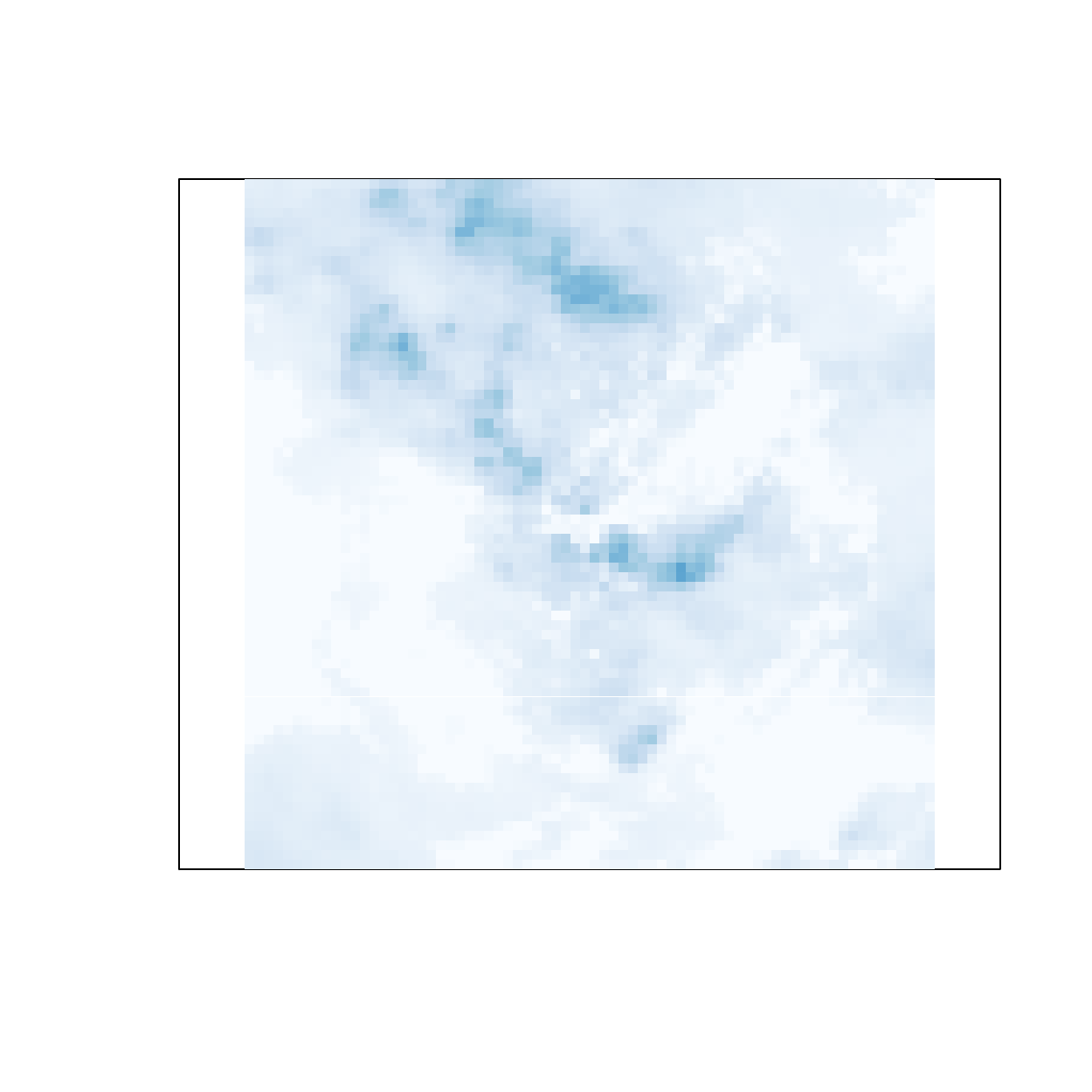}%
\includegraphics[width=0.15\linewidth, clip, trim= 55 70 30 55]{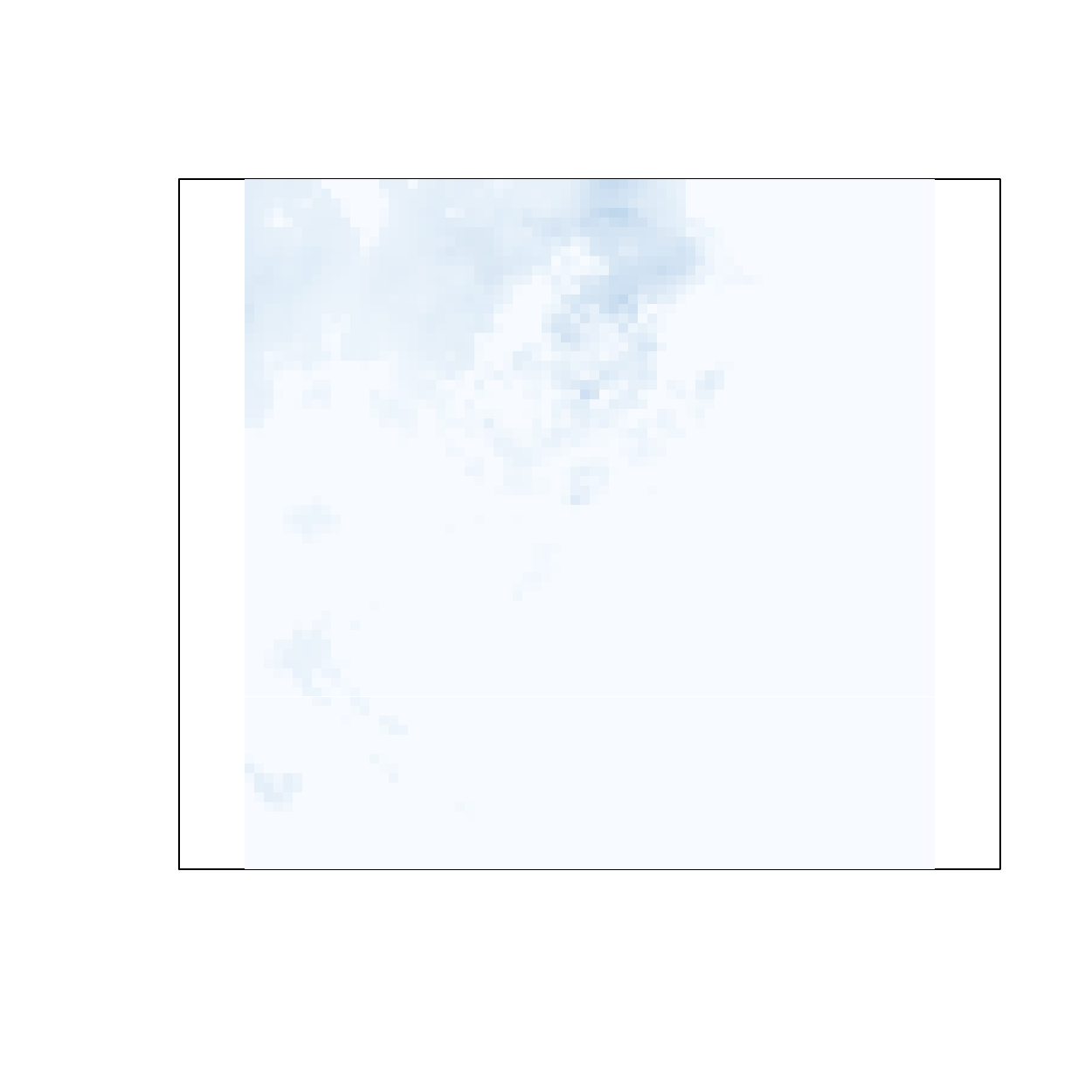}%
\includegraphics[width=0.15\linewidth, clip, trim= 55 70 30 55]{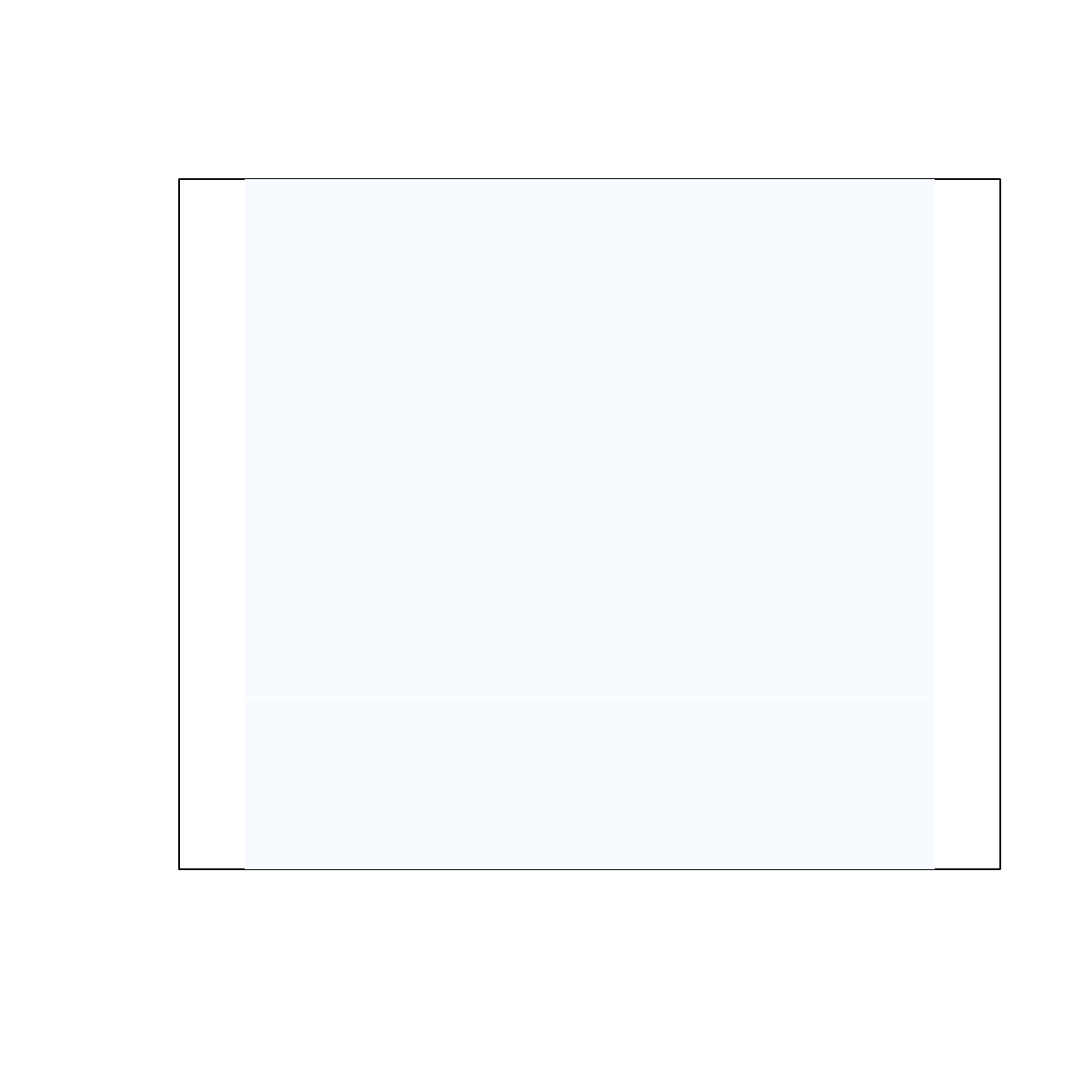}%
\includegraphics[width=0.15\linewidth, clip, trim= 55 70 30 55]{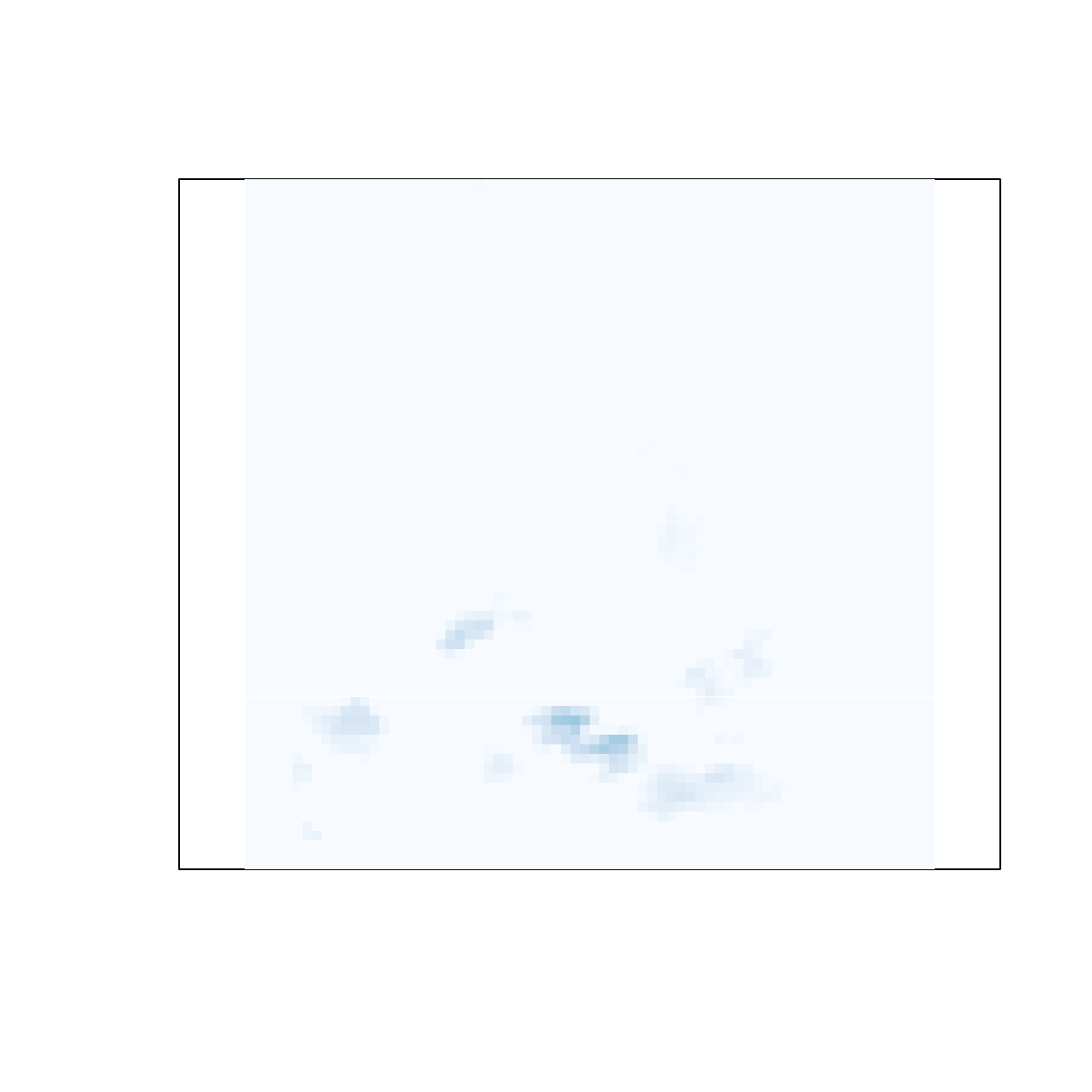}%
\includegraphics[width=0.15\linewidth, clip, trim= 55 70 30 55]{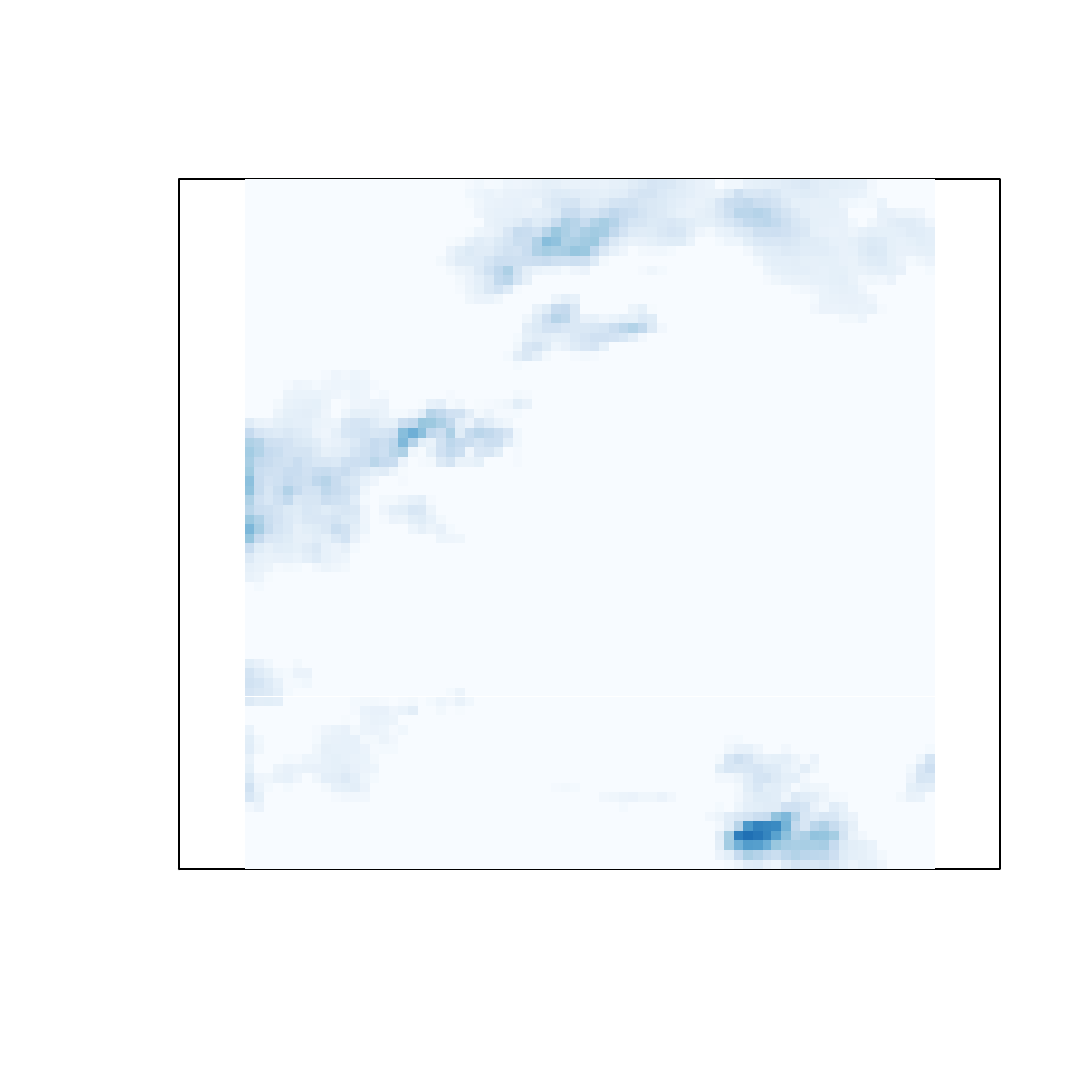}%
\includegraphics[width=0.15\linewidth, clip, trim= 55 70 30 55]{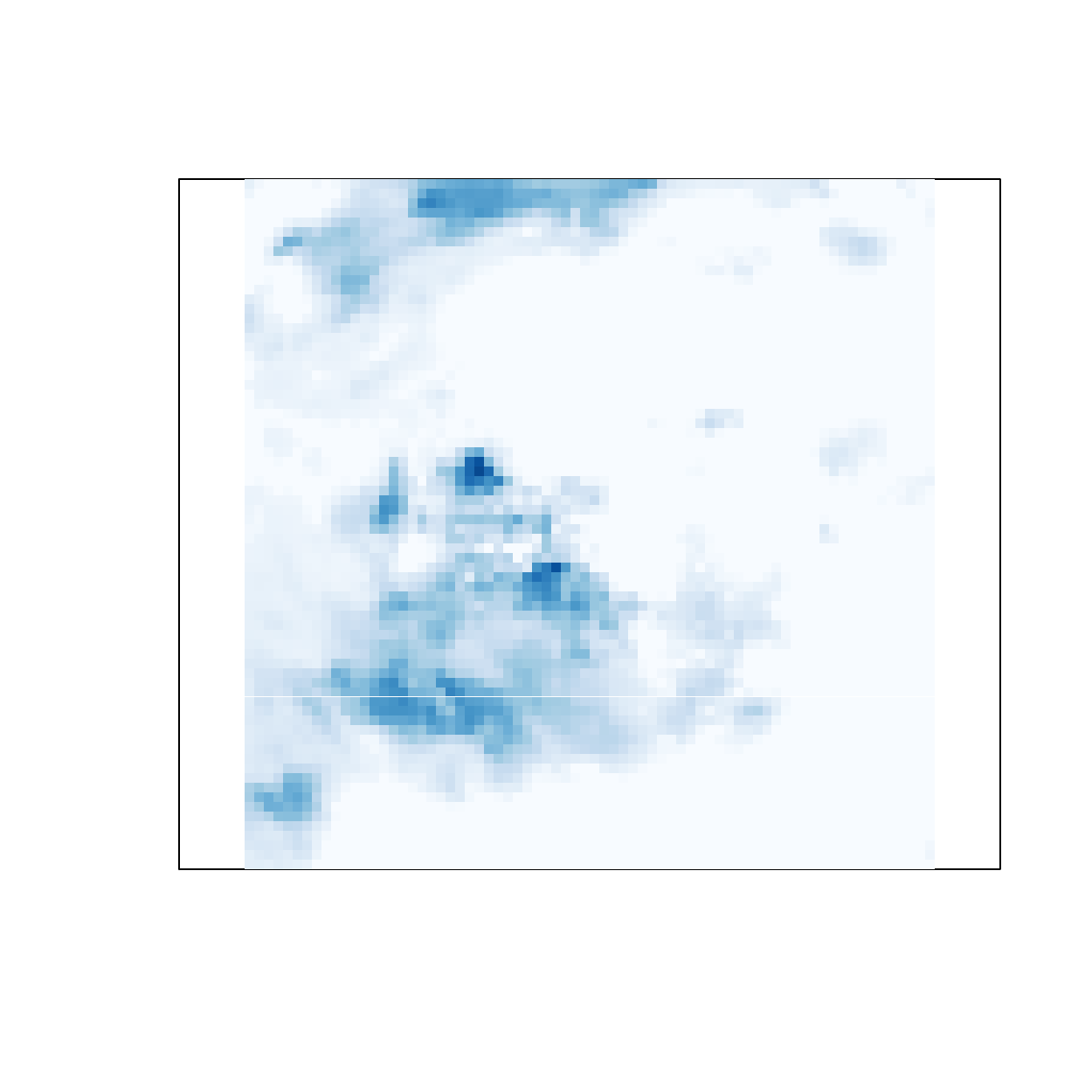}%
\includegraphics[width=0.015\linewidth, clip, trim= 660 120 0 100 ]{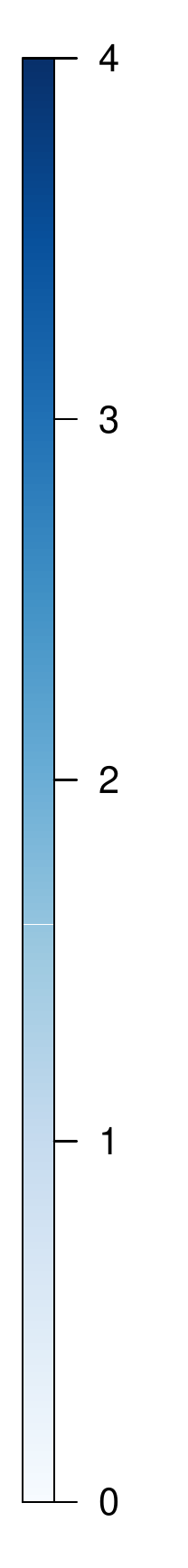}
\includegraphics[width=0.15\linewidth, clip, trim= 55 70 30 55]{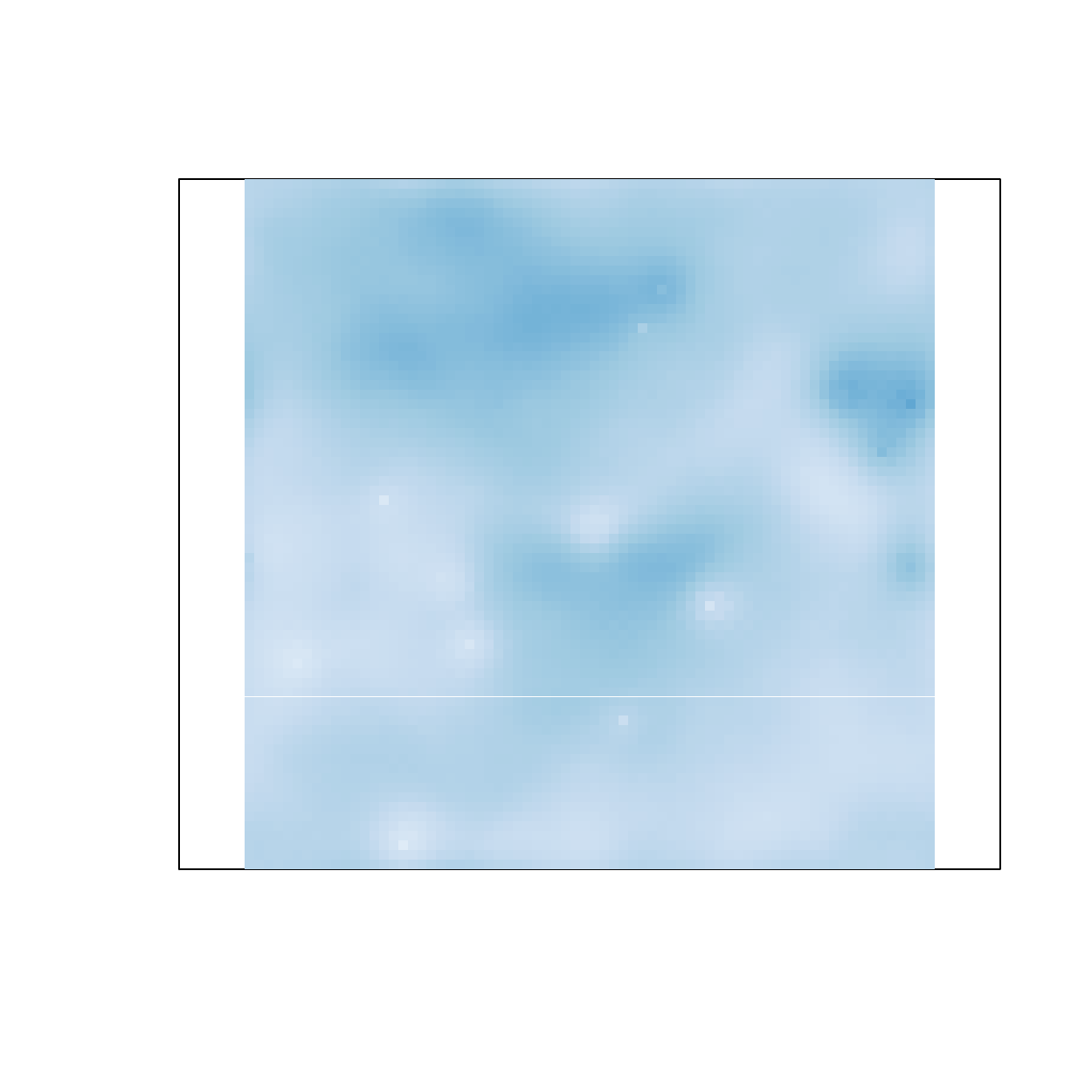}%
\includegraphics[width=0.15\linewidth, clip, trim= 55 70 30 55]{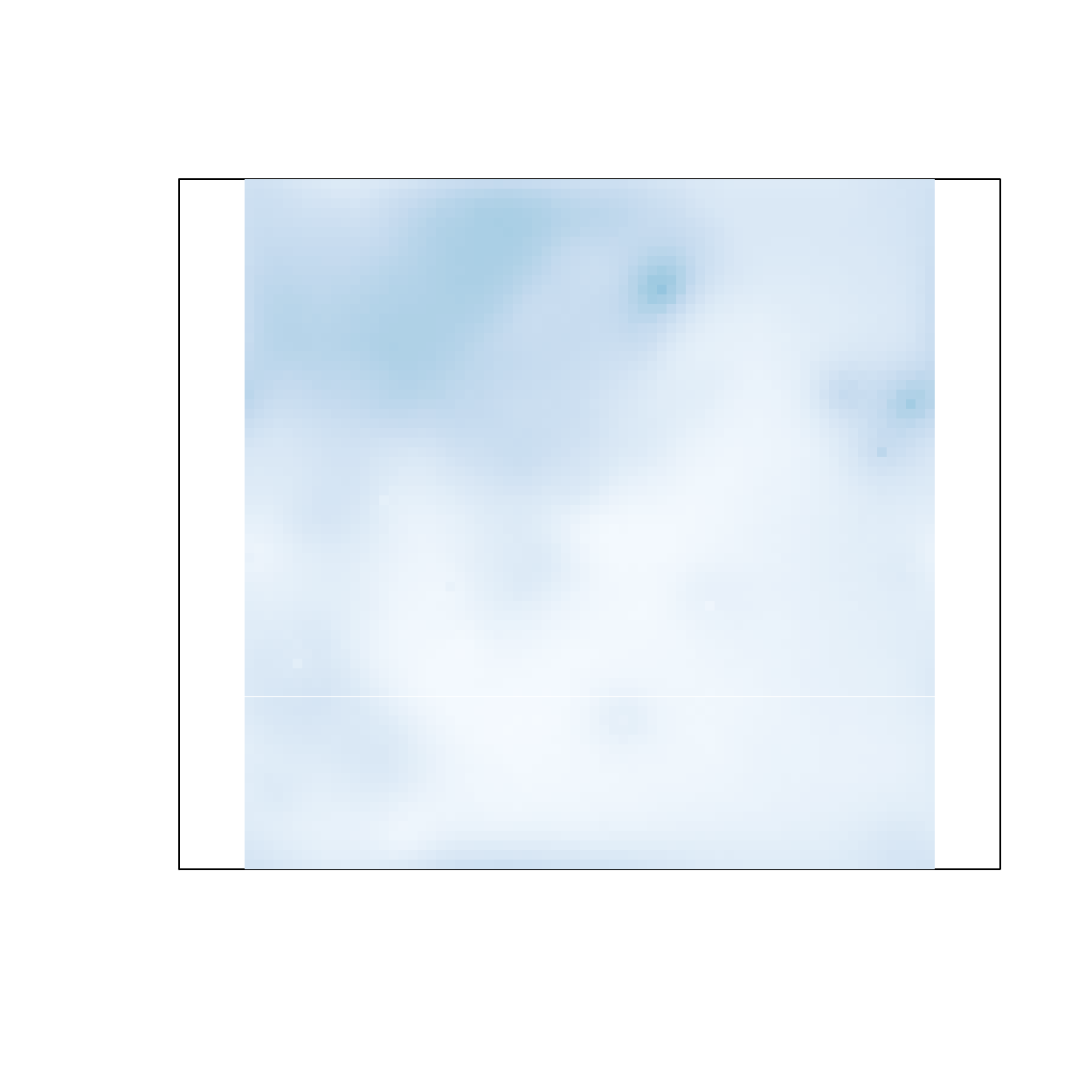}%
\includegraphics[width=0.15\linewidth, clip, trim= 55 70 30 55]{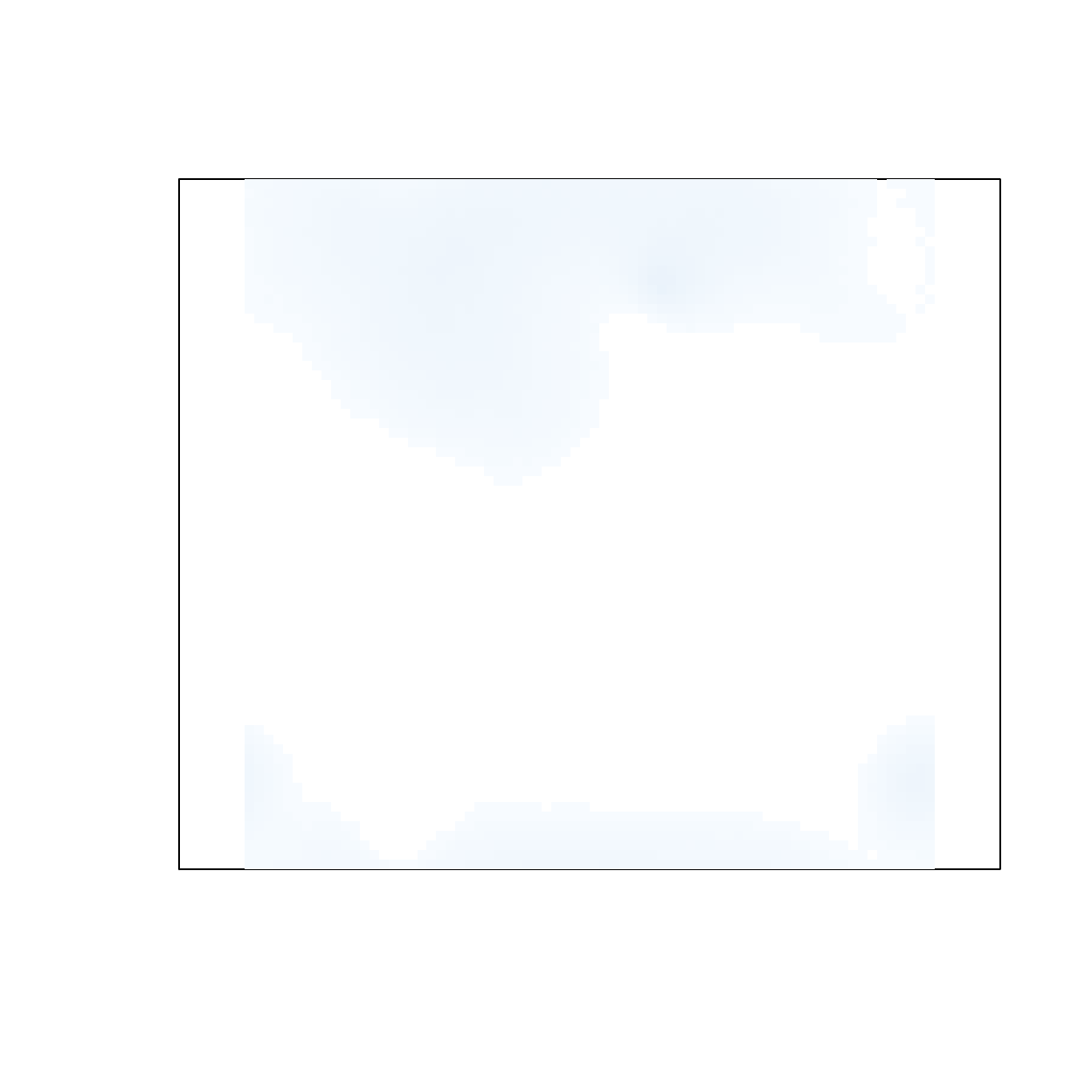}%
\includegraphics[width=0.15\linewidth, clip, trim= 55 70 30 55]{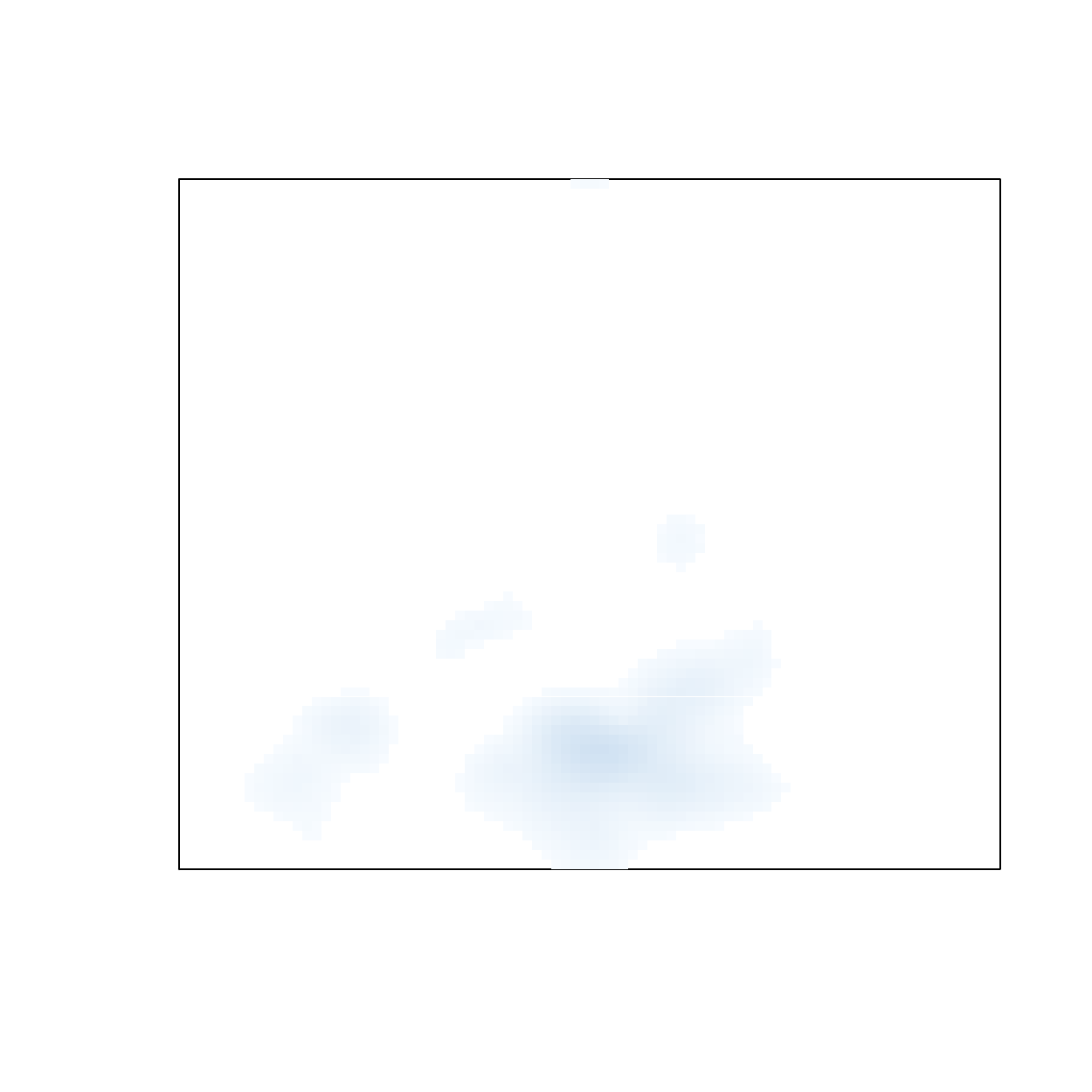}%
\includegraphics[width=0.15\linewidth, clip, trim= 55 70 30 55]{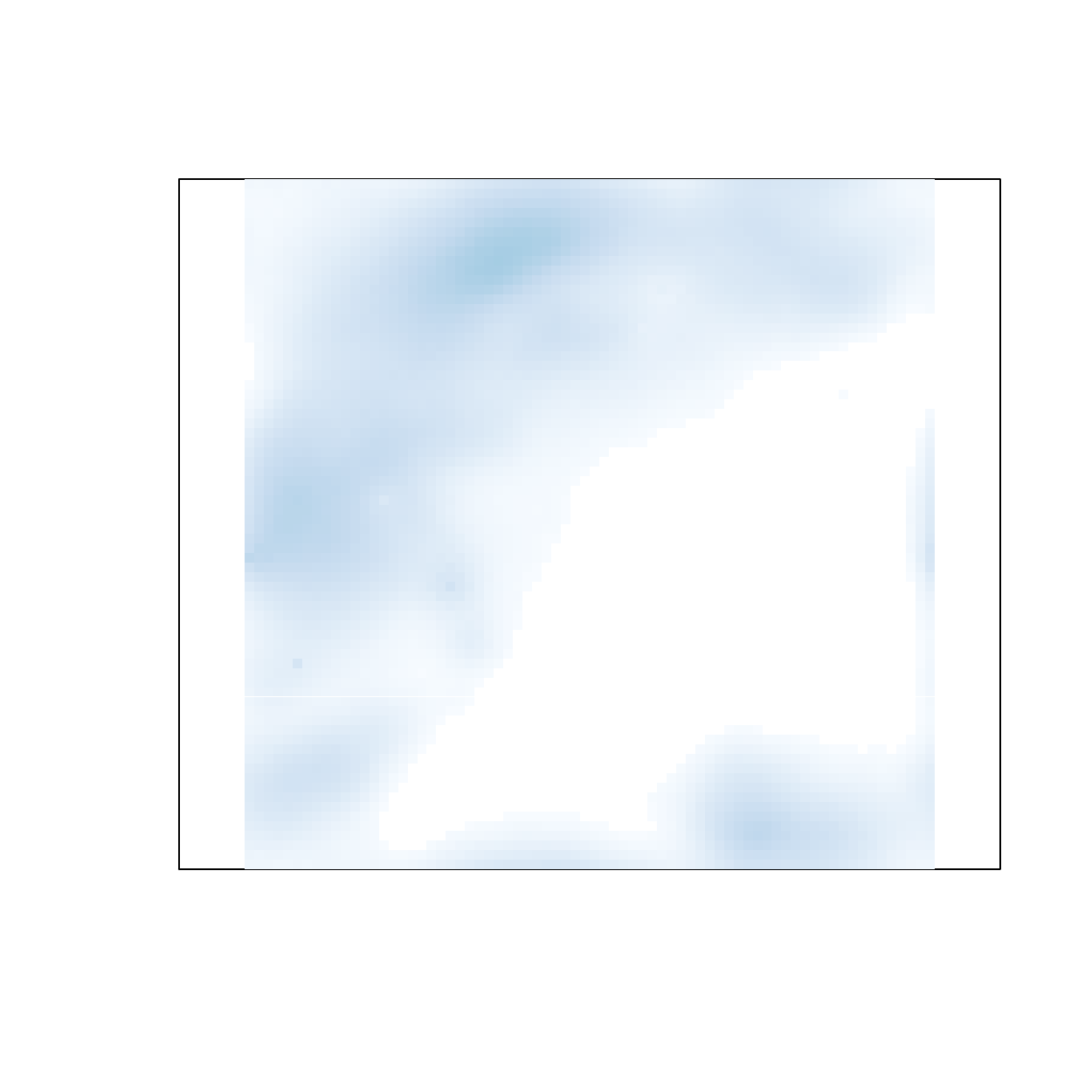}%
\includegraphics[width=0.15\linewidth, clip, trim= 55 70 30 55]{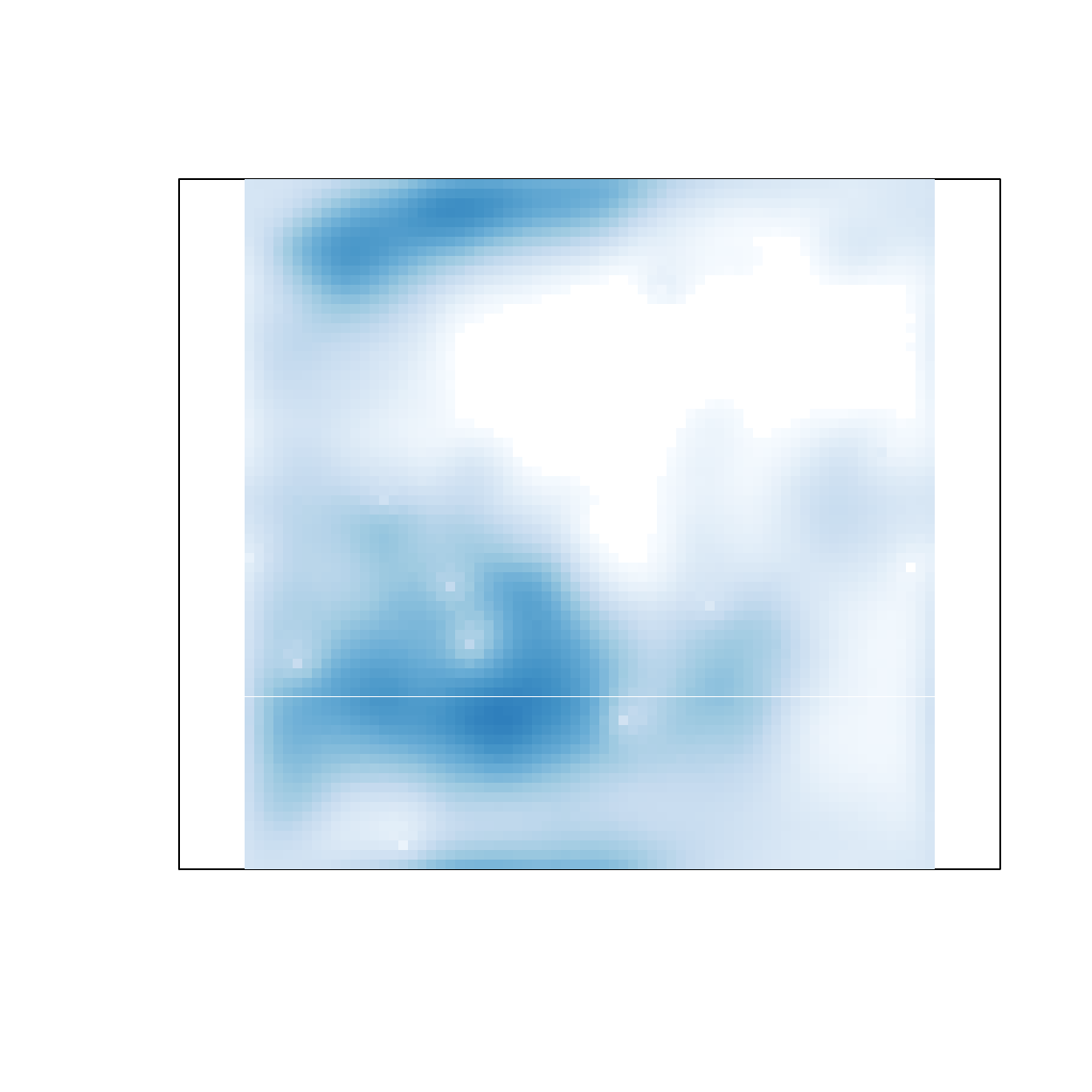}%
\includegraphics[width=0.015\linewidth, clip, trim= 660 120 0 100 ]{legend_4.pdf}
\includegraphics[width=0.15\linewidth, clip, trim= 55 70 30 55]{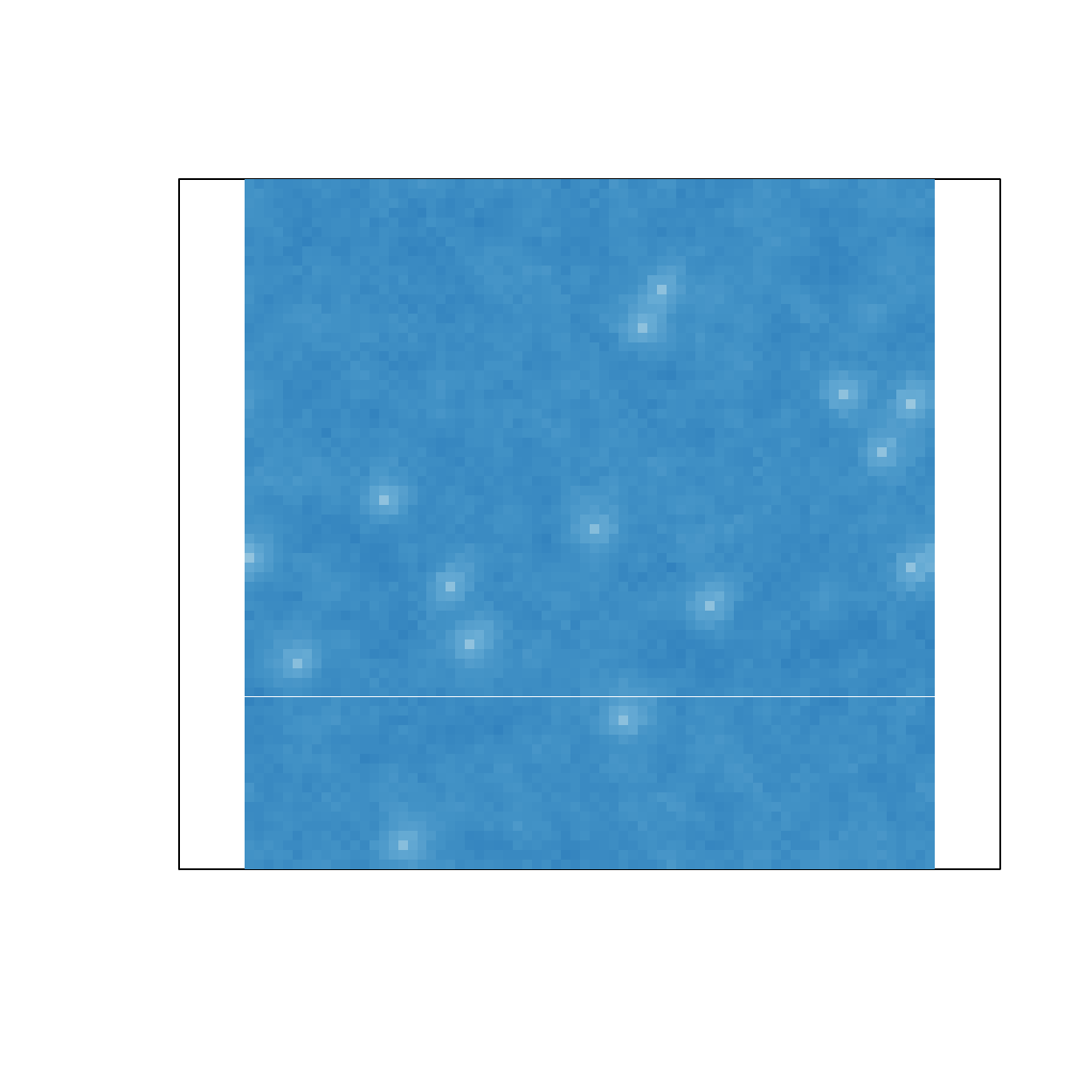}%
\includegraphics[width=0.15\linewidth, clip, trim= 55 70 30 55]{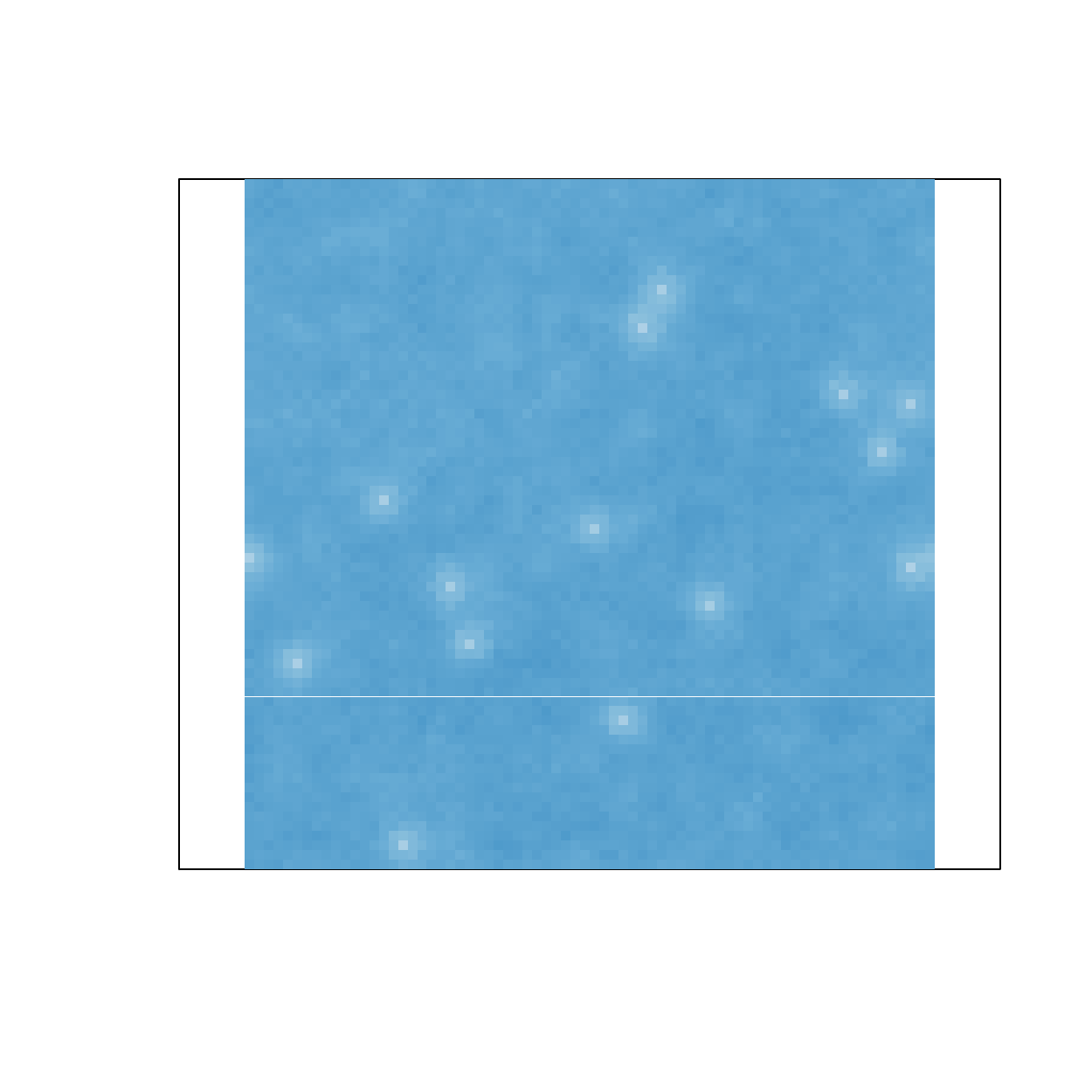}%
\includegraphics[width=0.15\linewidth, clip, trim= 55 70 30 55]{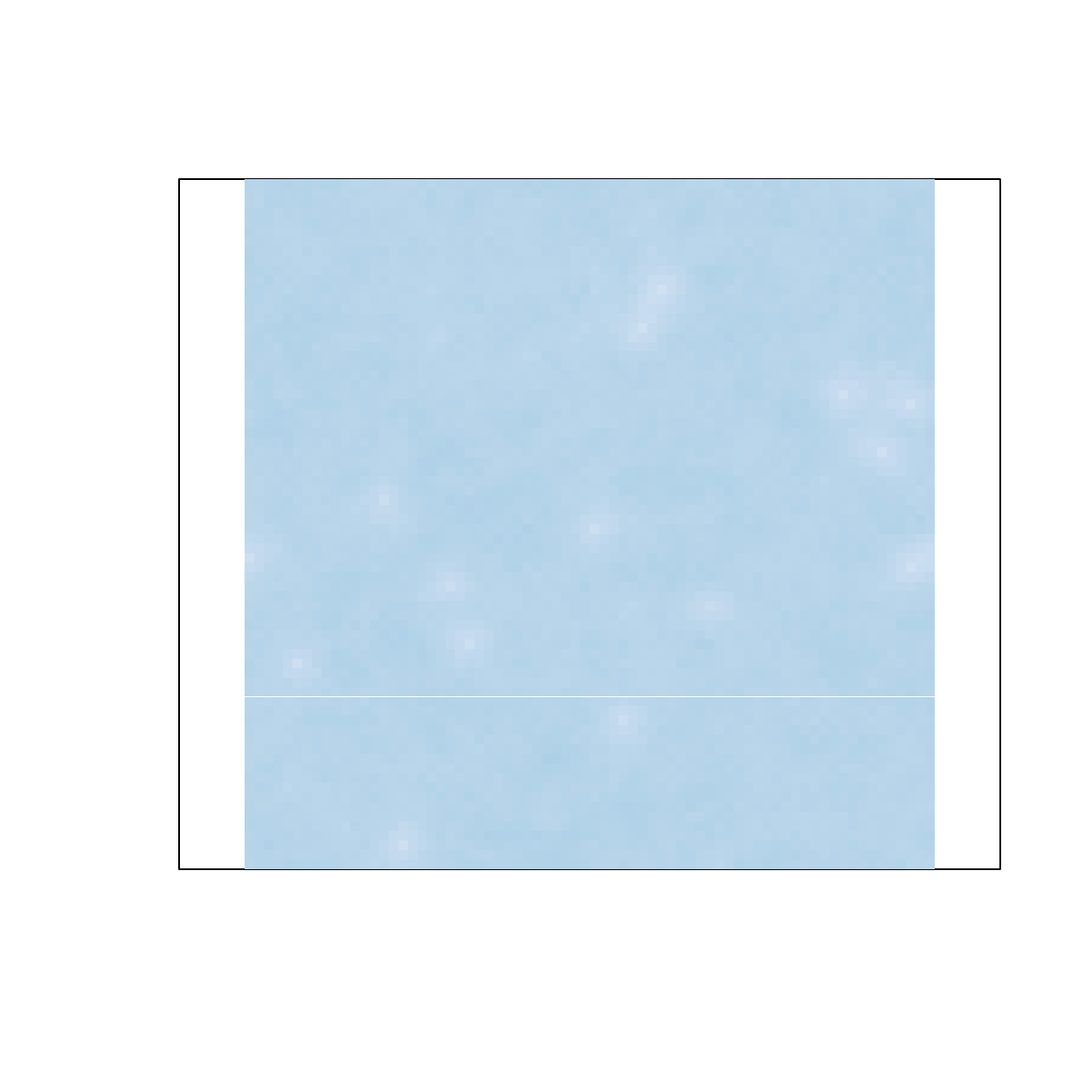}%
\includegraphics[width=0.15\linewidth, clip, trim= 55 70 30 55]{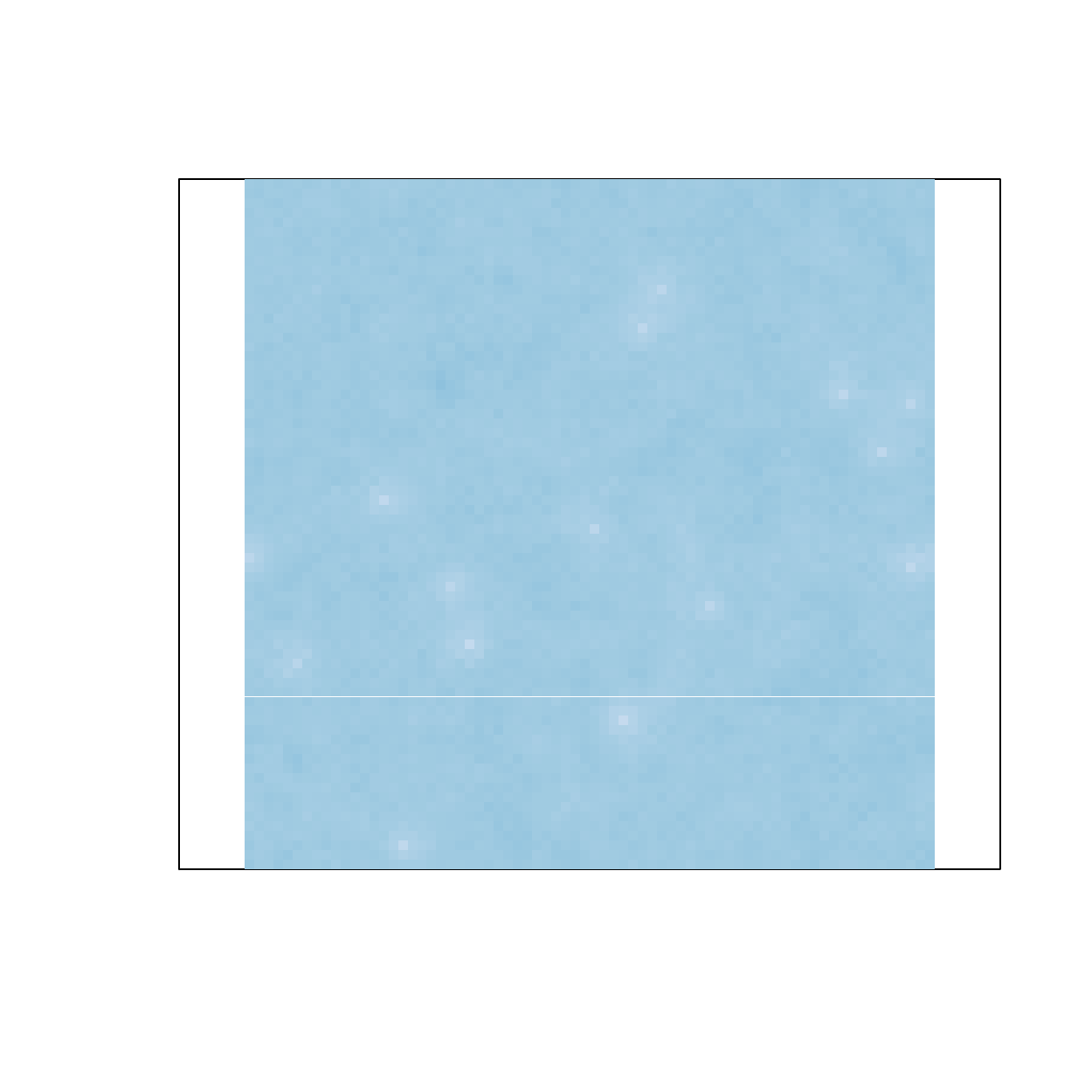}%
\includegraphics[width=0.15\linewidth, clip, trim= 55 70 30 55]{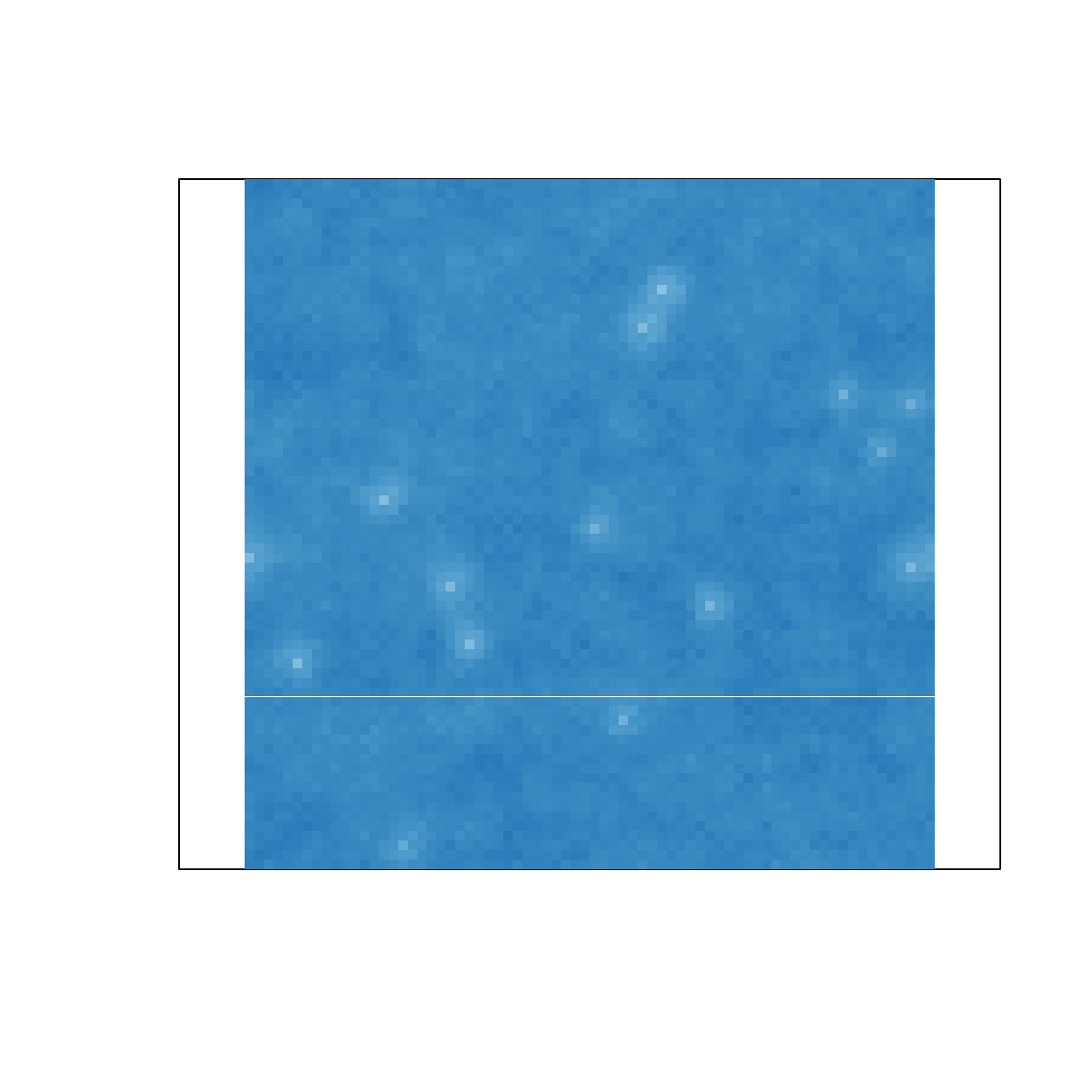}%
\includegraphics[width=0.15\linewidth, clip, trim= 55 70 30 55]{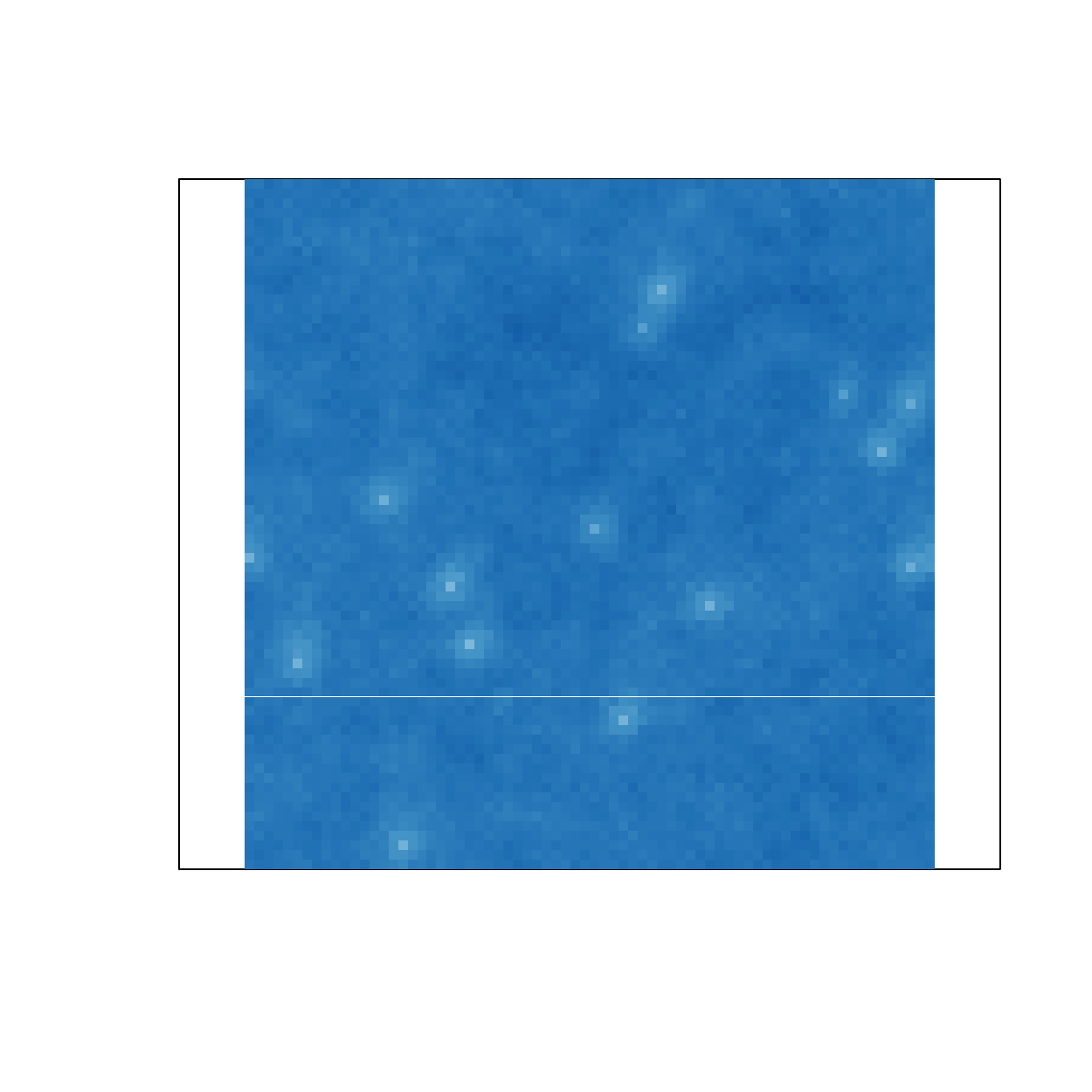}%
\includegraphics[width=0.015\linewidth, clip, trim= 660 120 0 100 ]{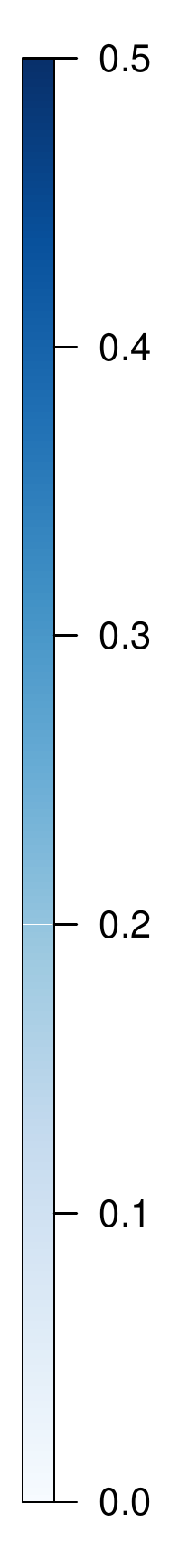}
\includegraphics[width=0.15\linewidth, clip, trim= 55 70 30 55]{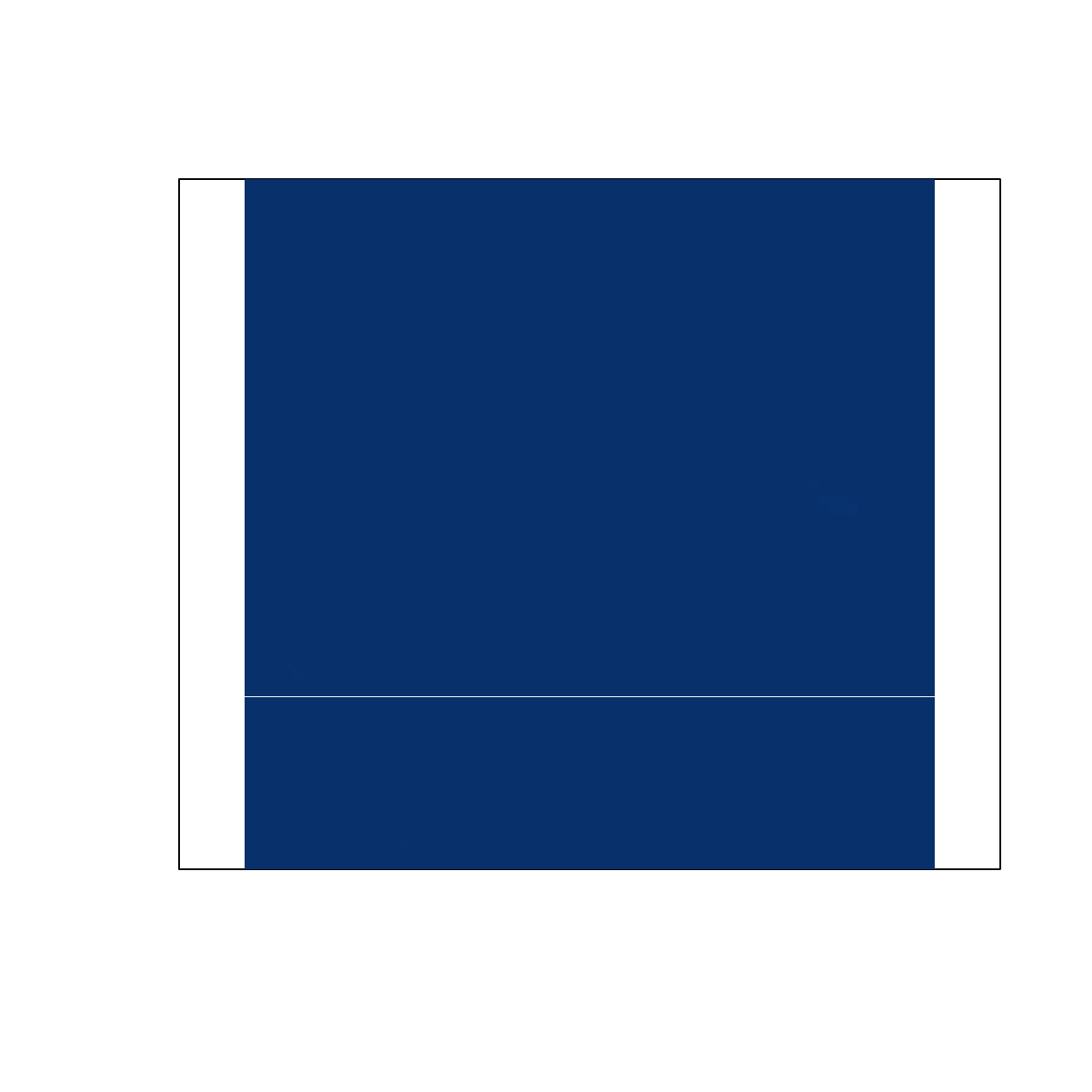}%
\includegraphics[width=0.15\linewidth, clip, trim= 55 70 30 55]{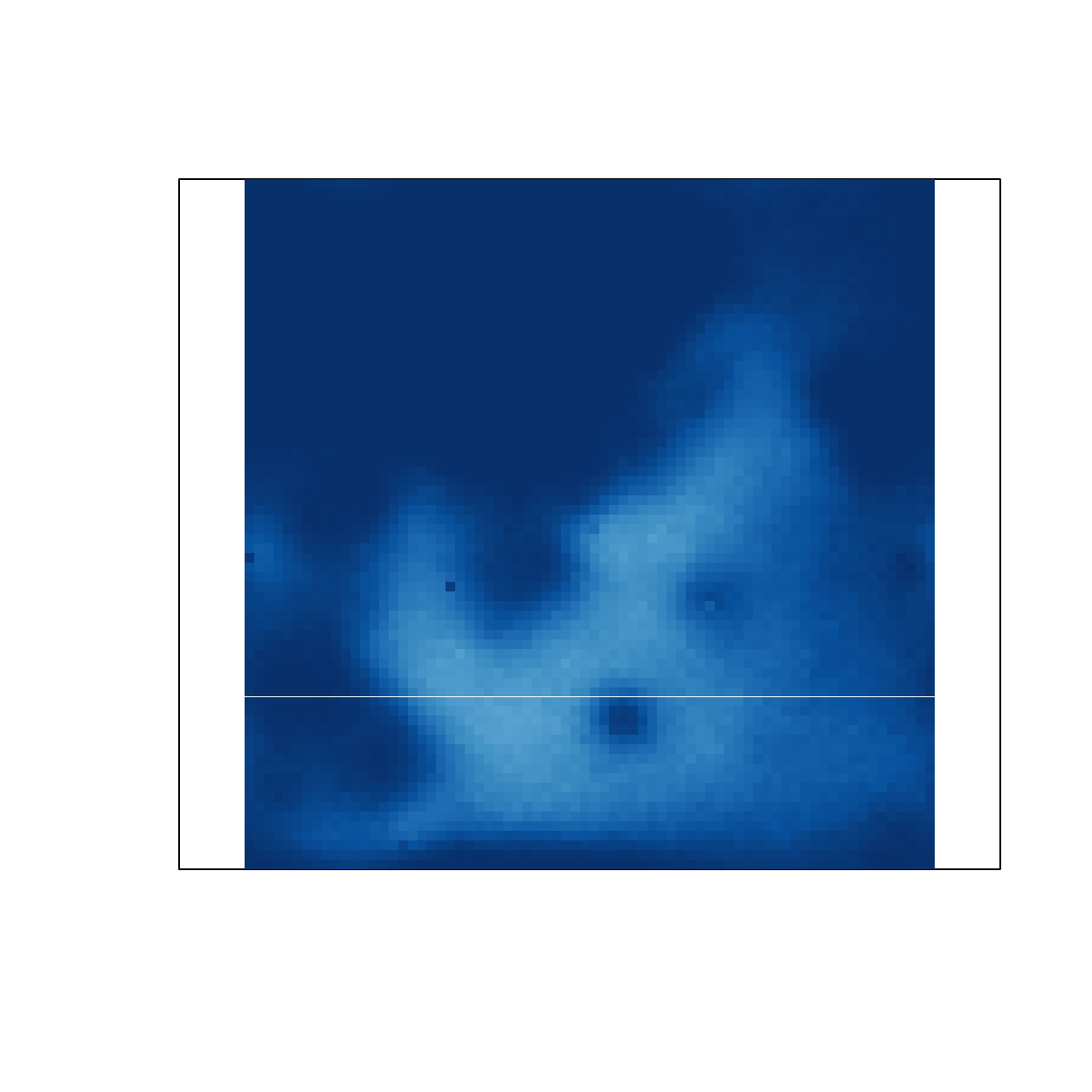}%
\includegraphics[width=0.15\linewidth, clip, trim= 55 70 30 55]{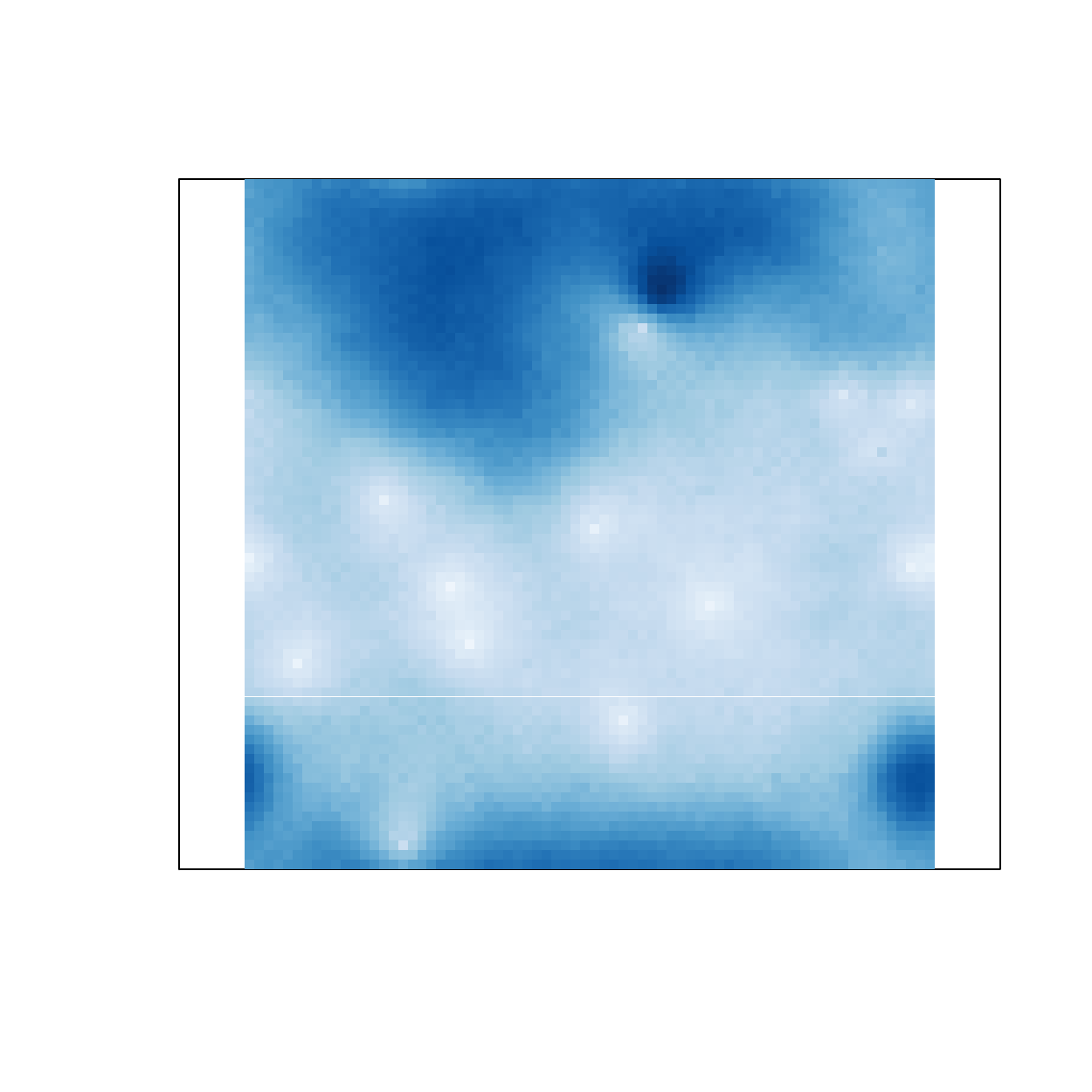}%
\includegraphics[width=0.15\linewidth, clip, trim= 55 70 30 55]{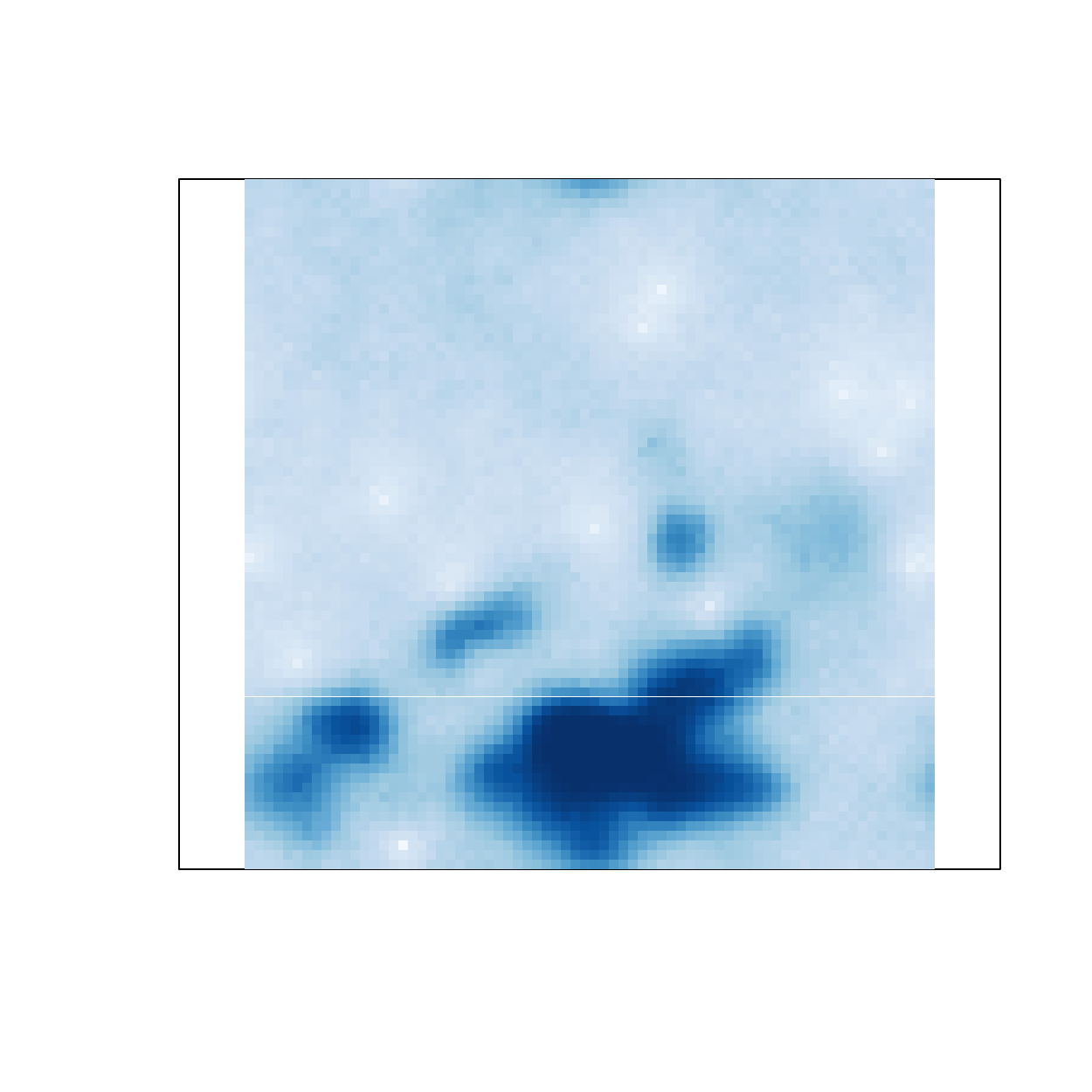}%
\includegraphics[width=0.15\linewidth, clip, trim= 55 70 30 55]{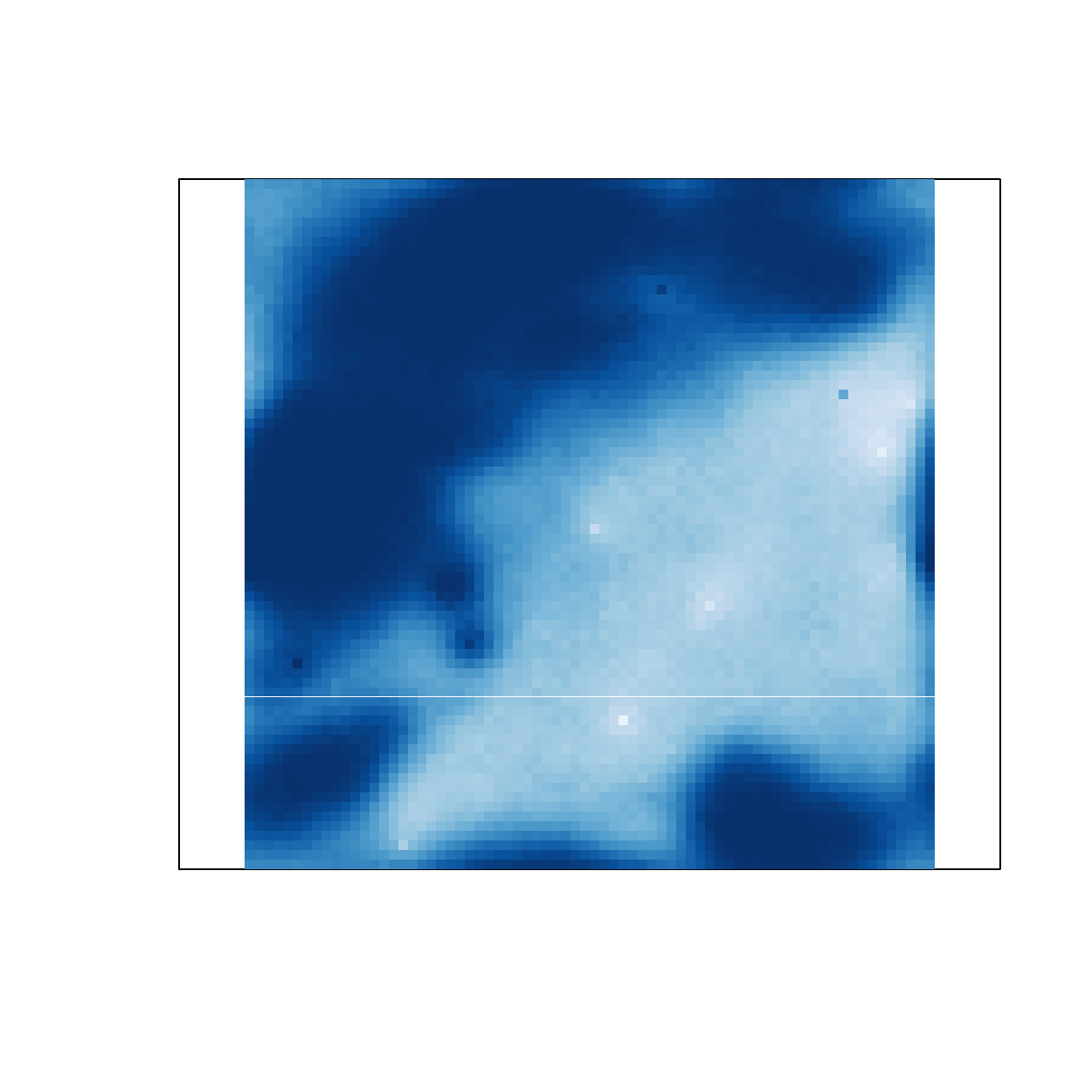}%
\includegraphics[width=0.15\linewidth, clip, trim= 55 70 30 55]{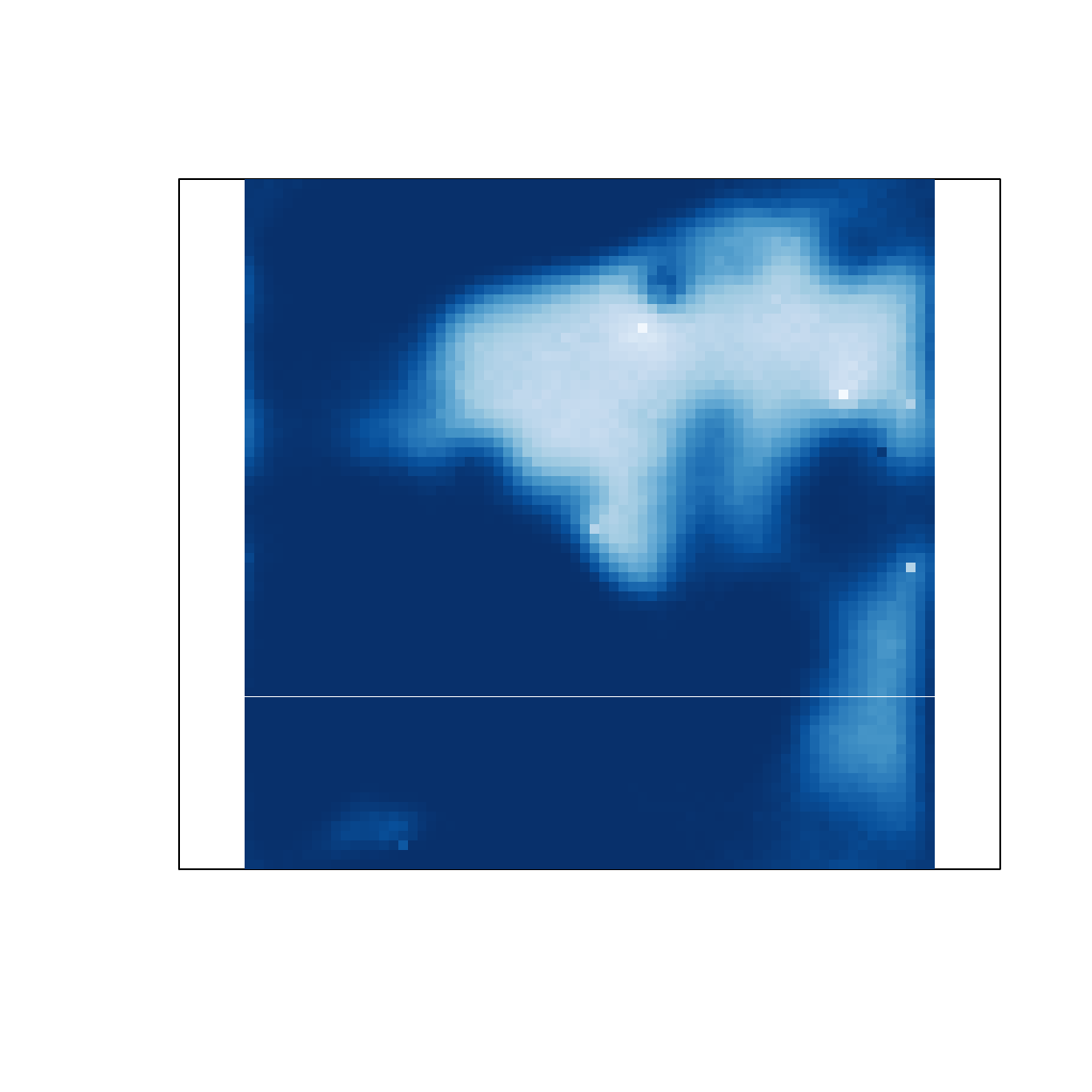}%
\includegraphics[width=0.015\linewidth, clip, trim= 660 120 0 100 ]{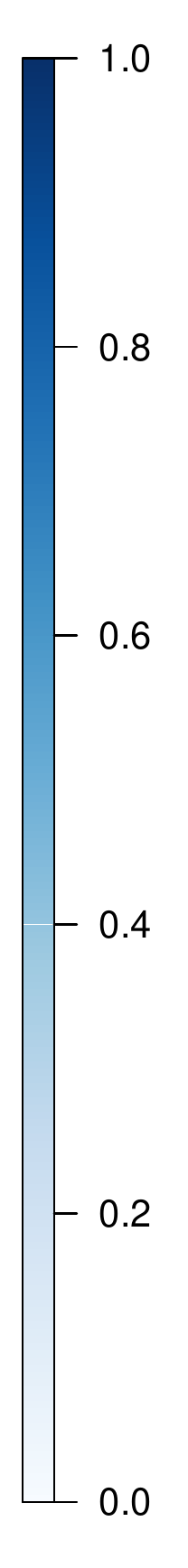}
\includegraphics[width=0.15\linewidth, clip, trim= 35 35 20 50]{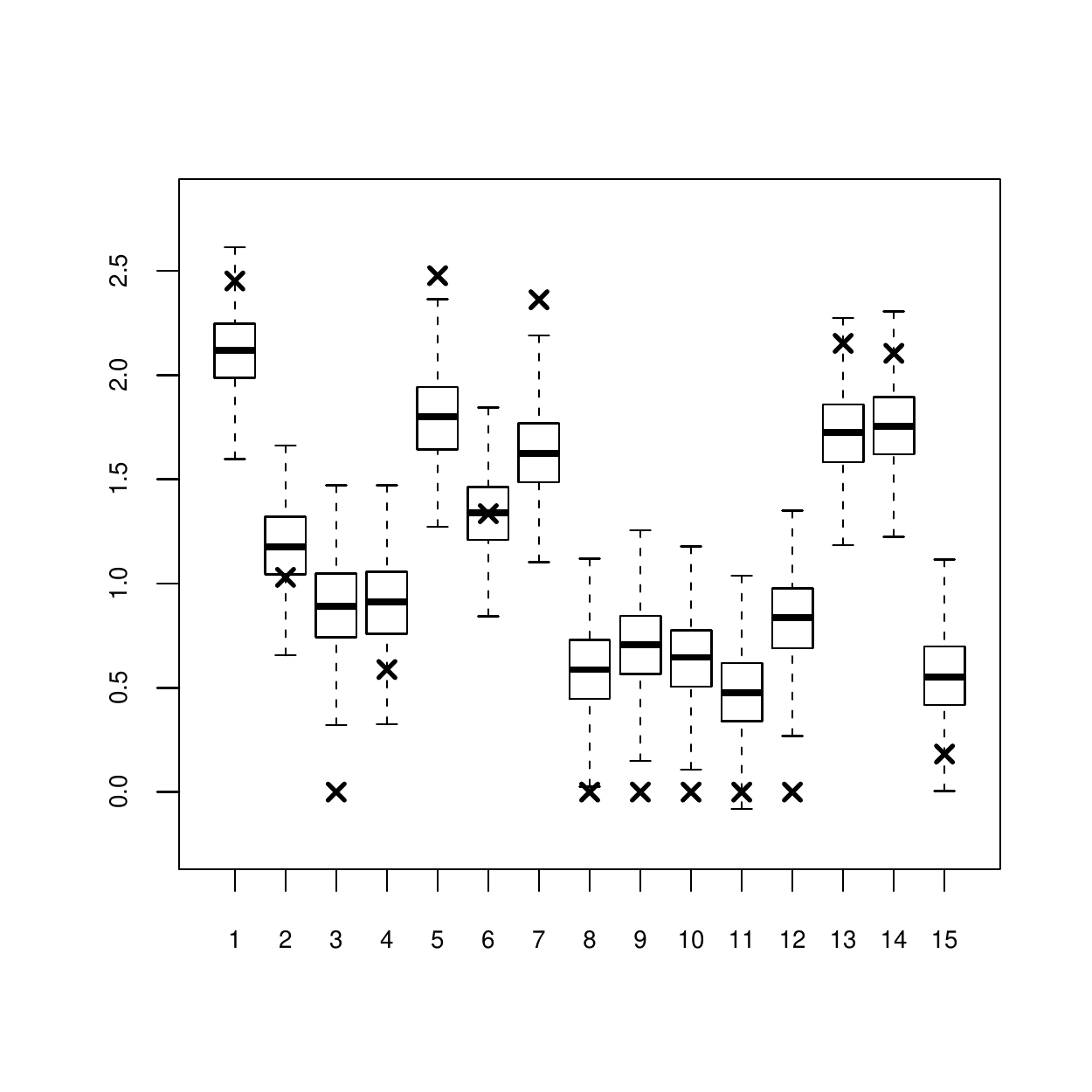}%
\includegraphics[width=0.15\linewidth, clip, trim= 35 35 20 50]{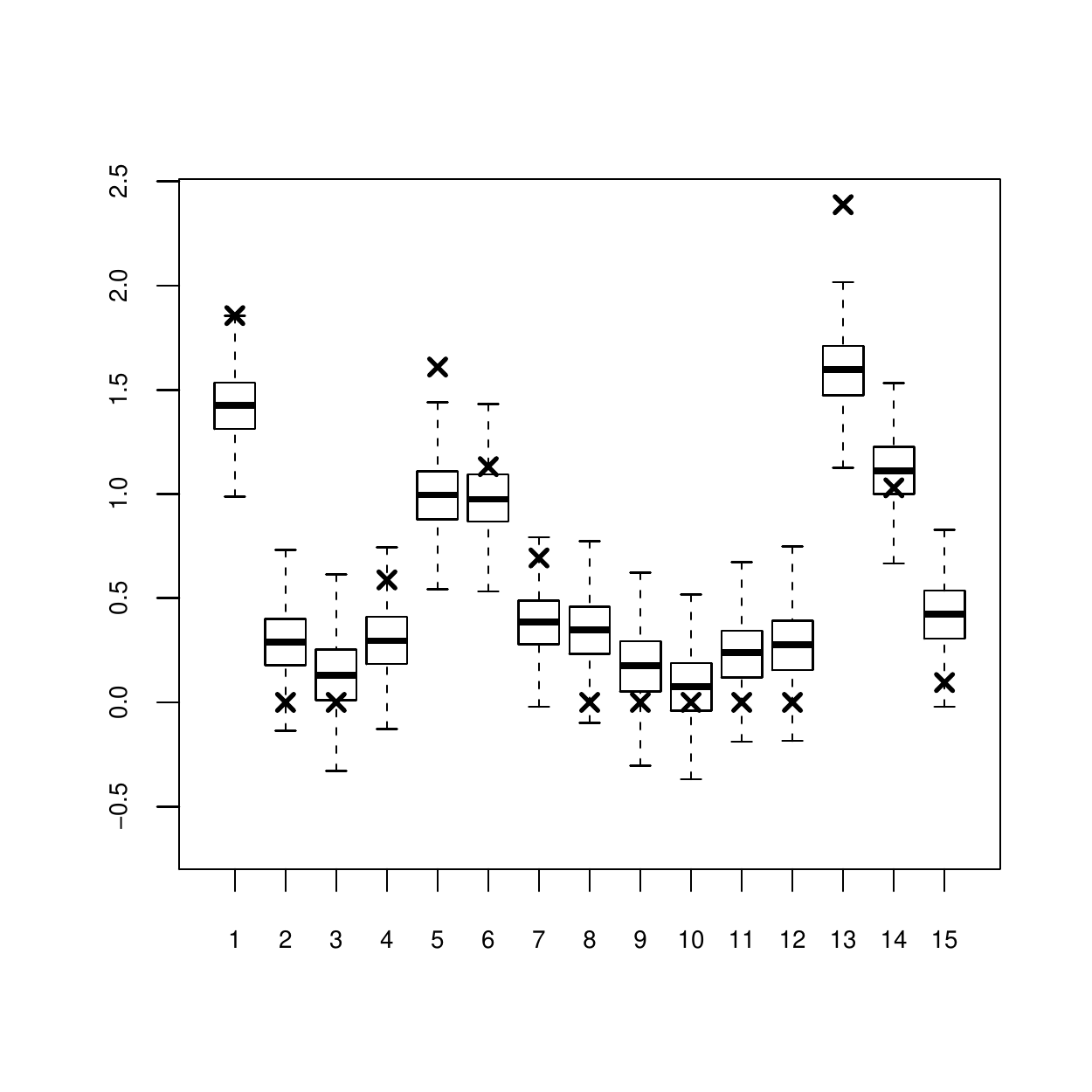}%
\includegraphics[width=0.15\linewidth, clip, trim= 35 35 20 50]{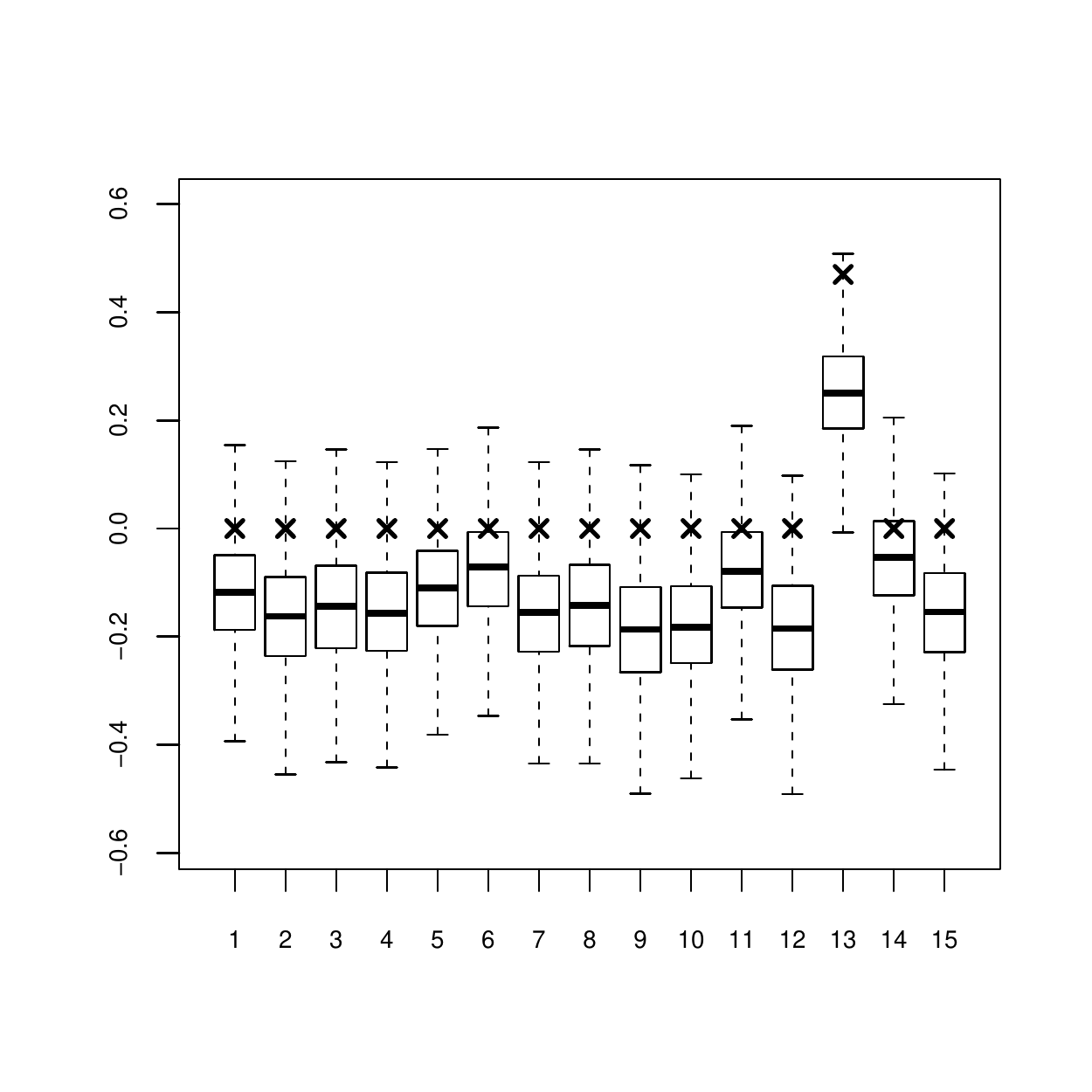}%
\includegraphics[width=0.15\linewidth, clip, trim= 35 35 20 50]{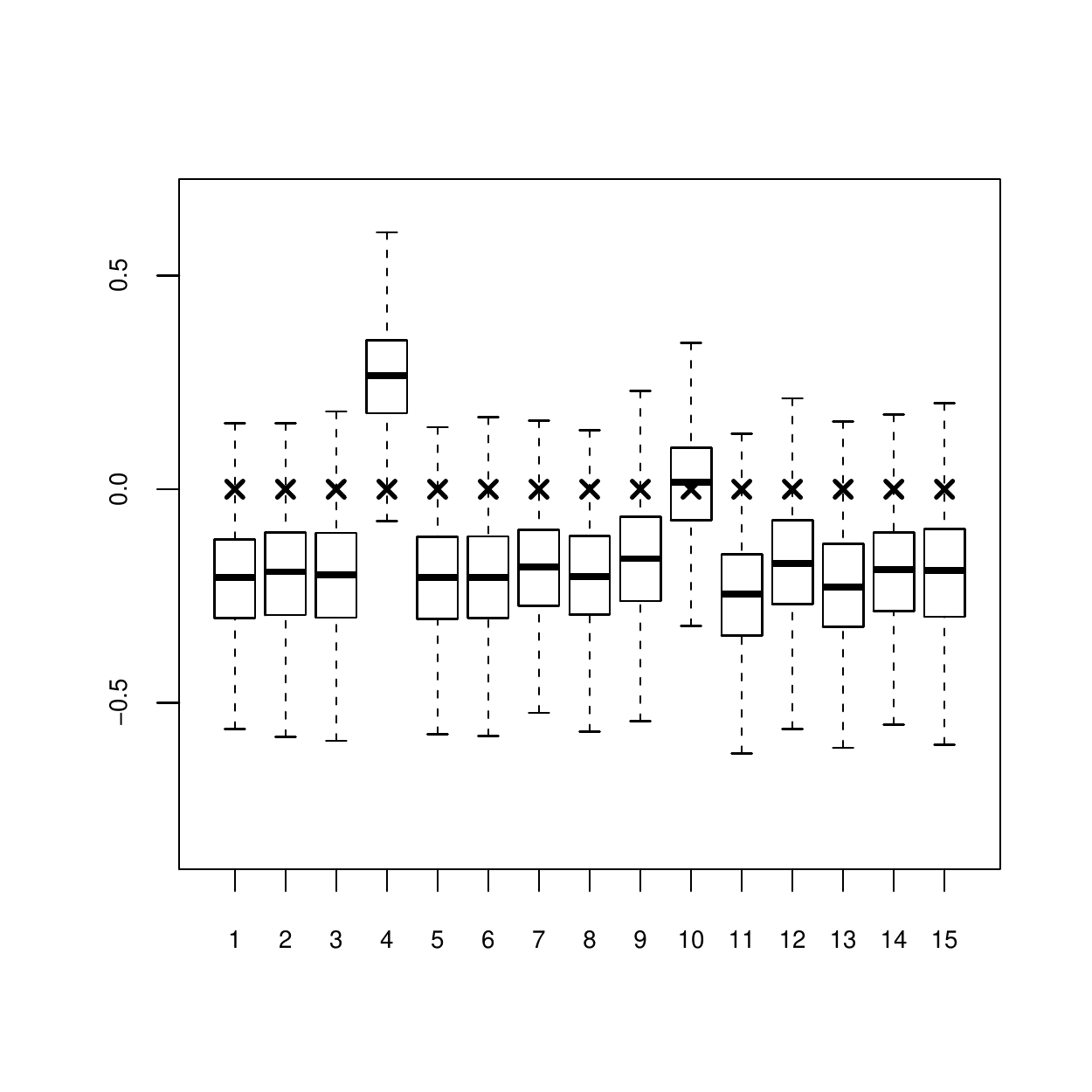}%
\includegraphics[width=0.15\linewidth, clip, trim= 35 35 20 50]{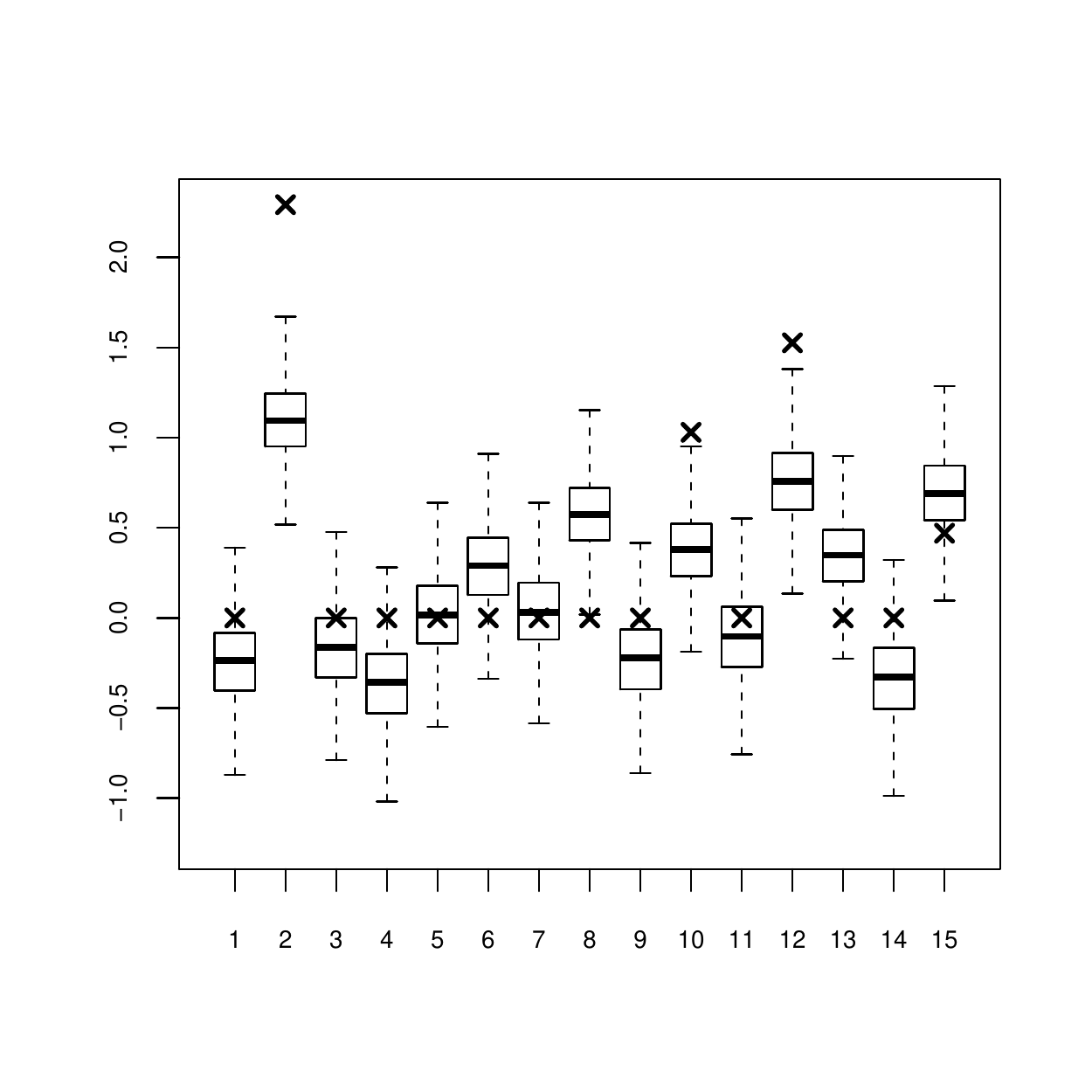}%
\includegraphics[width=0.15\linewidth, clip, trim= 35 35 20 50]{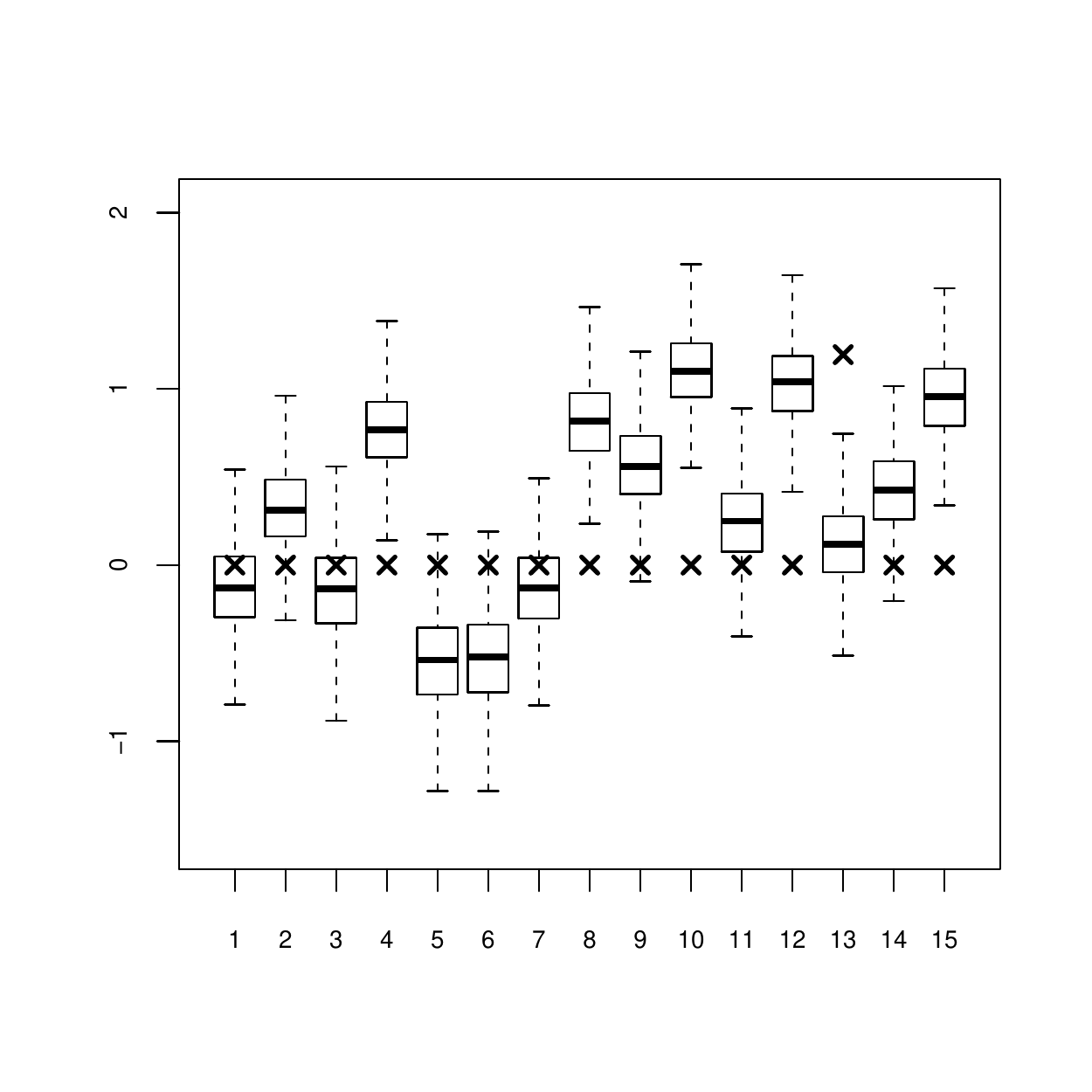}%
\phantom{\includegraphics[width=0.015\linewidth, clip, trim= 660 120 0 100 ]{legend_1.pdf}}
\caption{Image plots of the (transformed) radar observations~$\tilde{\vec{y}}^\tr_t$ (top row), marginal posterior means E$(\theta_{ti}|\D)$ and standard deviations SD$(\theta_{ti}|\D)$ of the precipitation field (second and third row, respectively). Posterior probabilities $\Pr(\theta_{ti}>0|\D)$ (fourth row) and  boxplots of the marginal posterior distributions of the precipitation field at the rain gauge locations; crosses show the (transformed) gauge observations~$\tilde{\vec{y}}^\tg_t$ (fifth row).}
\label{fig:real_images}
\end{center}
\end{figure}

Turning now to the dynamic states, Figure~\ref{fig:real_velocities} shows the marginal posterior means along with the~$95\%$ credible regions for the velocity components~$\nu_x$ (left) and $\nu_y$ (right) for $t=1,\dots,\tilde{T}$.
Inspection of the radar observations reveals that the rain clouds generally drift in a north-easterly direction and so it is pleasing to see that this feature is also observed within the inferred velocity components, with both $\nu_x$ and $\nu_y$ typically larger than zero.
Figure~\ref{fig:real_images}, top row, shows image plots of the (transformed) radar observations~$\tilde{\vec{y}}^\tr_t$ at observation times $t \in \{12,24,36,48,60,72\}$; the corresponding plots for all time-steps are given in Section~4 of the supplementary material.
The marginal posterior means (E$(\theta_{ti}|\D)$, $i=1,\dots,N$) of the precipitation field of interest are shown in the second row of Figure~\ref{fig:real_images} from which we can see that the areas in which we expect to observe rainfall at ground level (E$(\theta_{ti}|\D)>0$) correlate with the positive rainfall rates observed on the weather radar.
The third row of Figure~\ref{fig:real_images} shows the marginal standard deviations of the precipitation field (SD$(\theta_{ti}|\D)$, $i=1,\dots,N$) and from this we can see the reduced uncertainty around the rain gauge locations which is perhaps unsurprising given the increased information at these locations.
This analysis also reveals that our model is able to capture the locations/times where we are unlikely to observe any precipitation; the fourth row of Figure~\ref{fig:real_images} shows $\Pr(\theta_{ti} > 0|\D) $.
Finally, the bottom row of Figure~\ref{fig:real_images} contains boxplots of the marginal posterior distributions of the precipitation field at the rain gauge locations, $\pi(\theta_{t,\ell_g}|\D)$ for $g=1,\dots,15$; crosses show the (transformed) gauge observations~$\tilde{\vec{y}}^\tg_t$.

\begin{figure}[ht!]
\hspace{1cm} $+10$min \hspace{0.8cm} $+20$min \hspace{0.8cm} $+30$min \hspace{0.8cm} $+40$min \hspace{0.8cm} $+50$min \hspace{0.8cm} $+60$min
\vspace{-0.25cm}
\begin{center}
        \centering
\includegraphics[width=0.15\linewidth, clip, trim= 55 70 30 55]{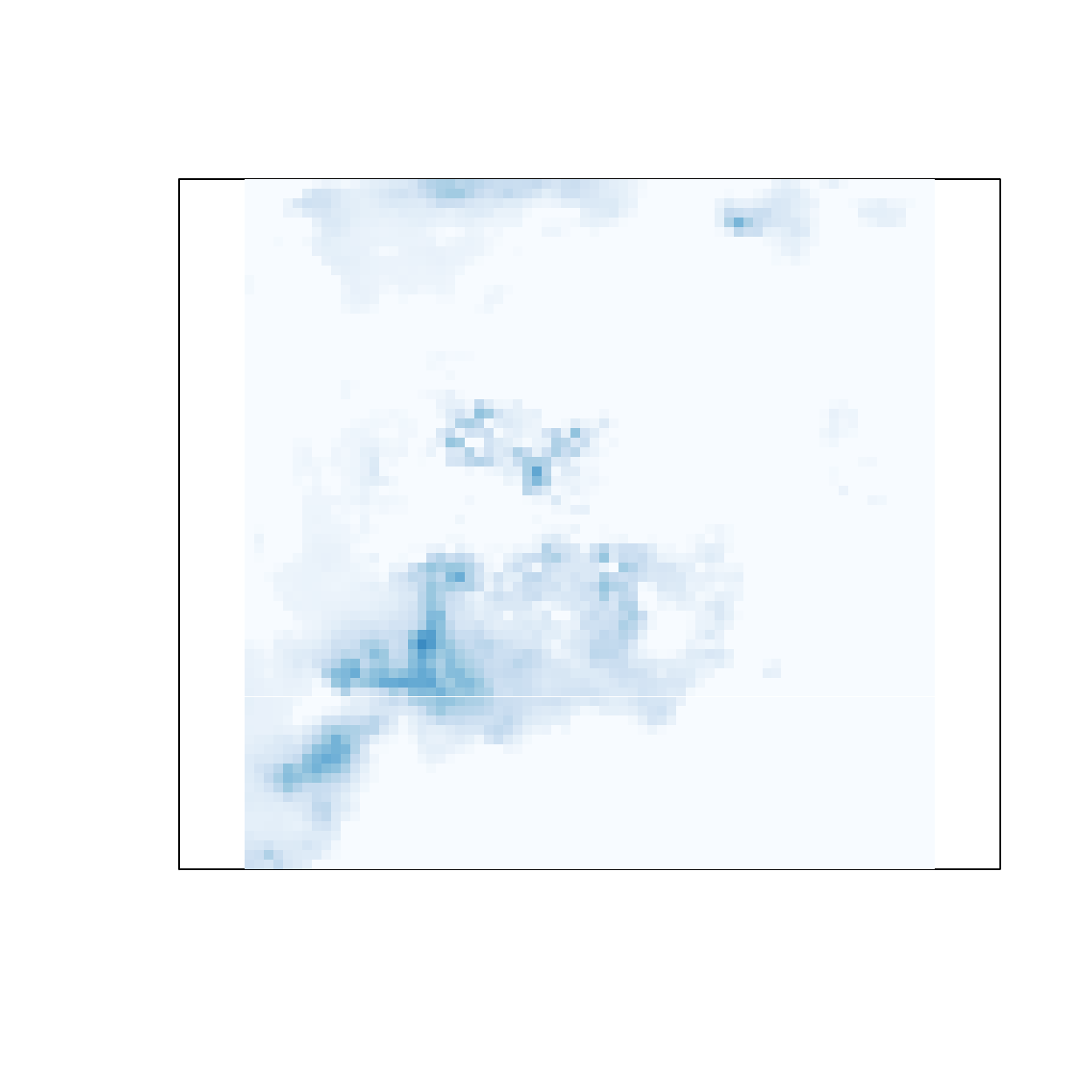}%
\includegraphics[width=0.15\linewidth, clip, trim= 55 70 30 55]{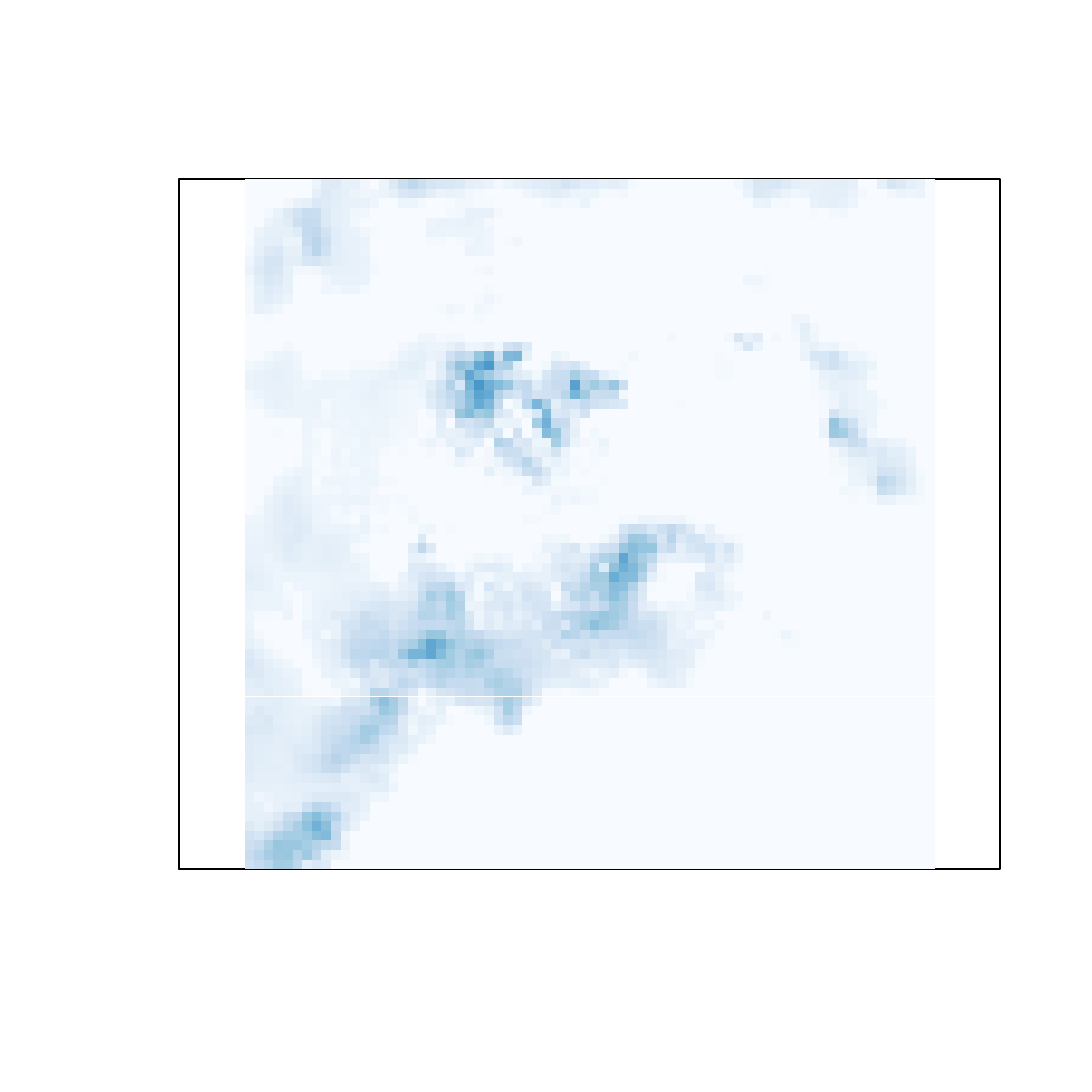}%
\includegraphics[width=0.15\linewidth, clip, trim= 55 70 30 55]{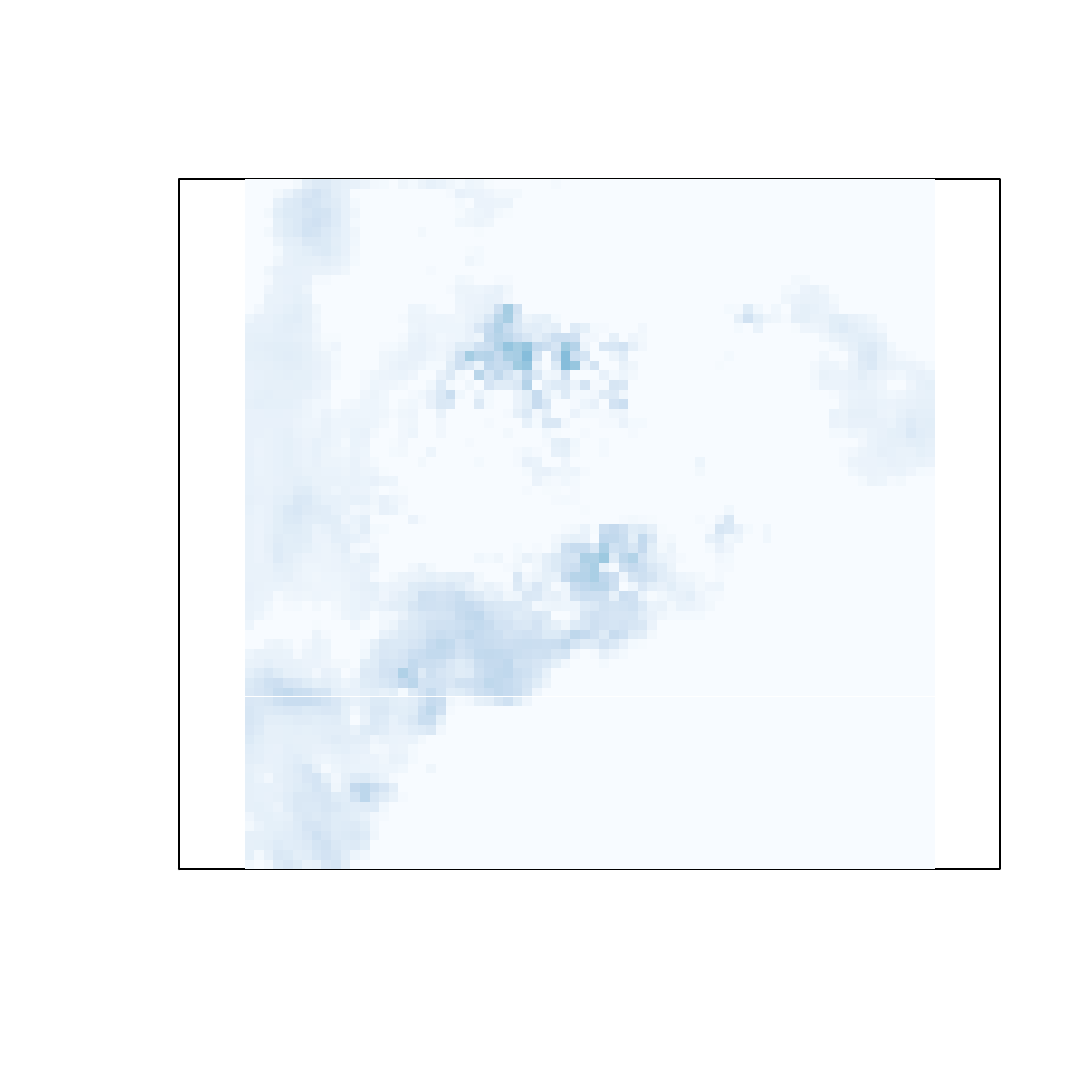}%
\includegraphics[width=0.15\linewidth, clip, trim= 55 70 30 55]{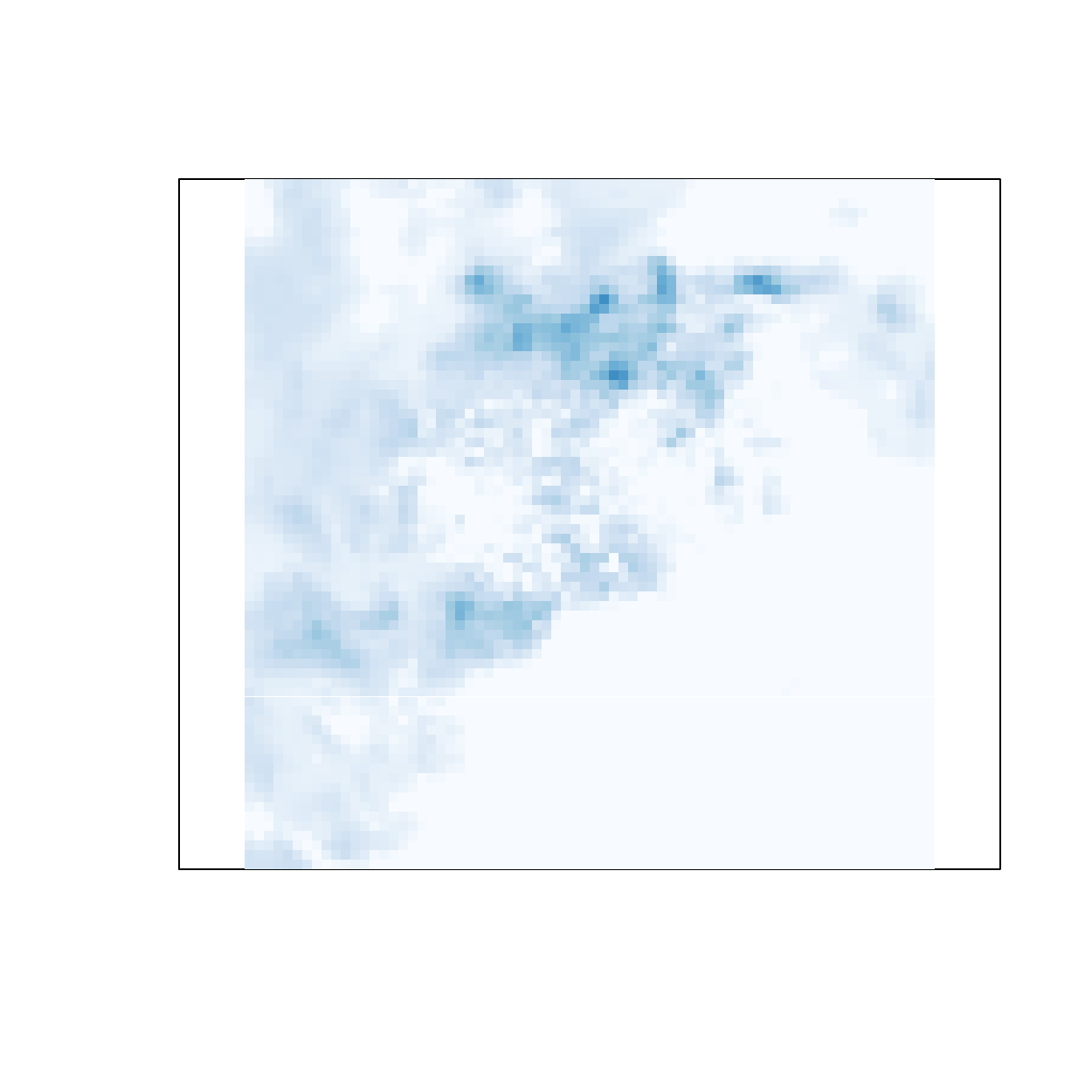}%
\includegraphics[width=0.15\linewidth, clip, trim= 55 70 30 55]{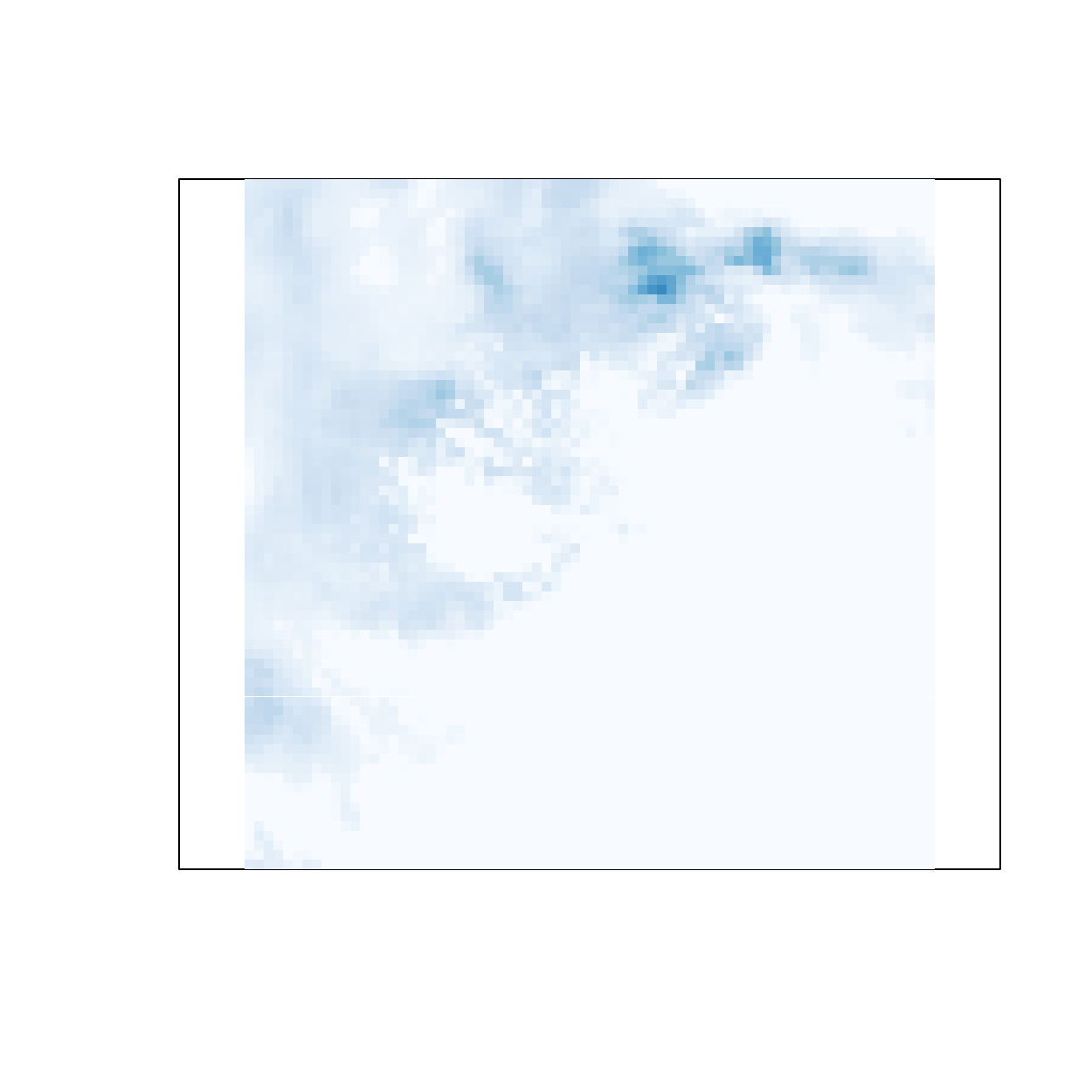}%
\includegraphics[width=0.15\linewidth, clip, trim= 55 70 30 55]{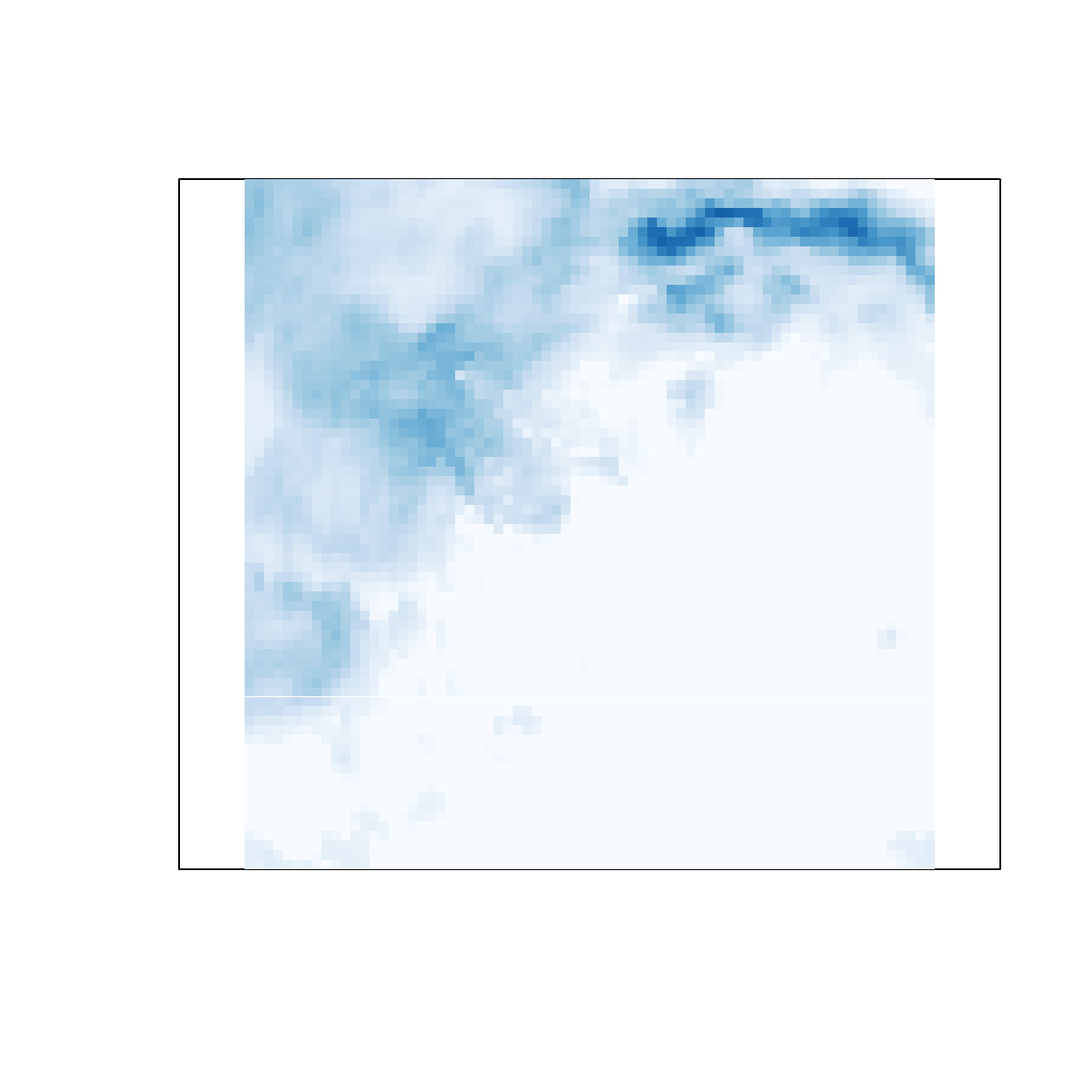}%
\includegraphics[width=0.015\linewidth, clip, trim= 660 120 0 100 ]{legend_4.pdf}
\includegraphics[width=0.15\linewidth, clip, trim= 55 70 30 55]{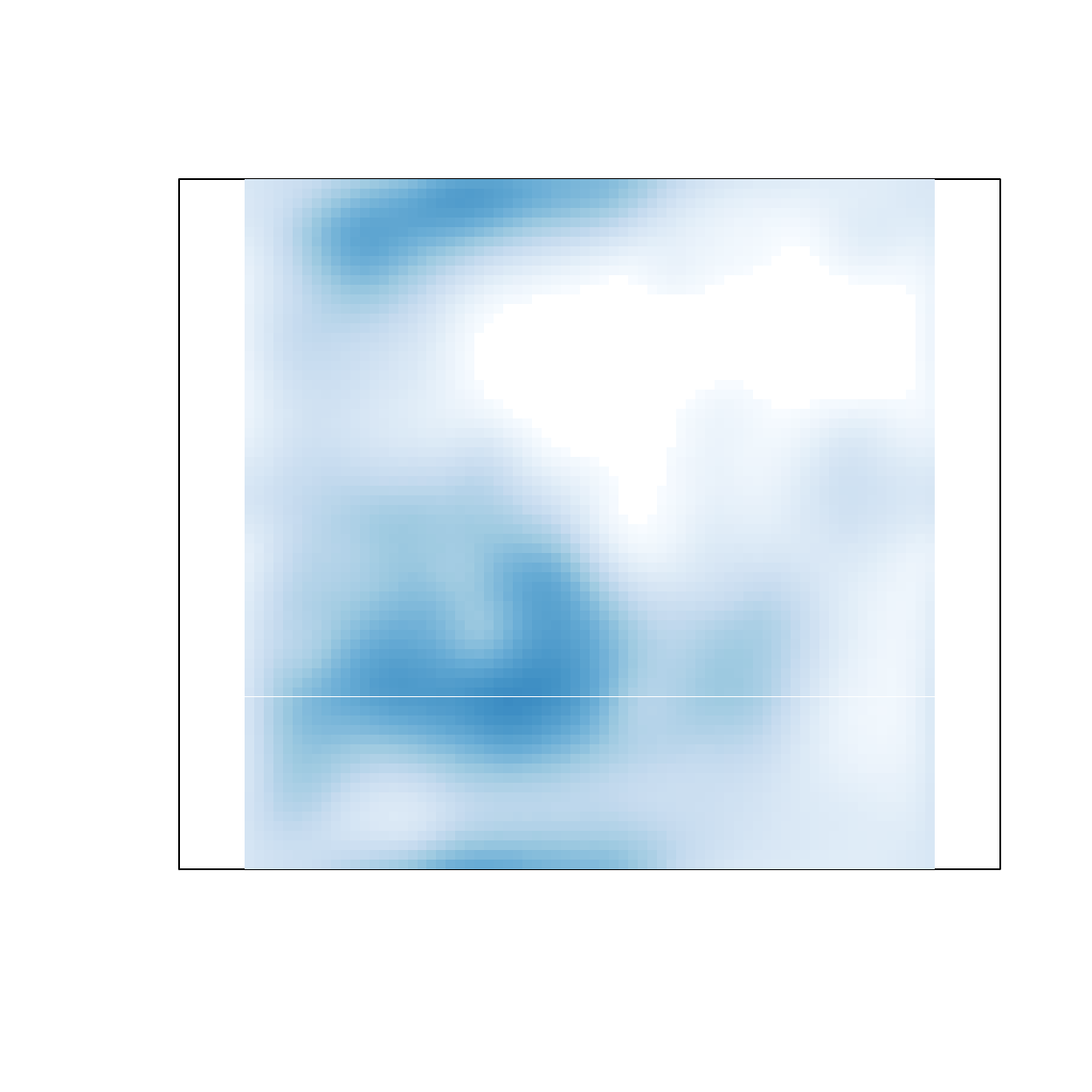}%
\includegraphics[width=0.15\linewidth, clip, trim= 55 70 30 55]{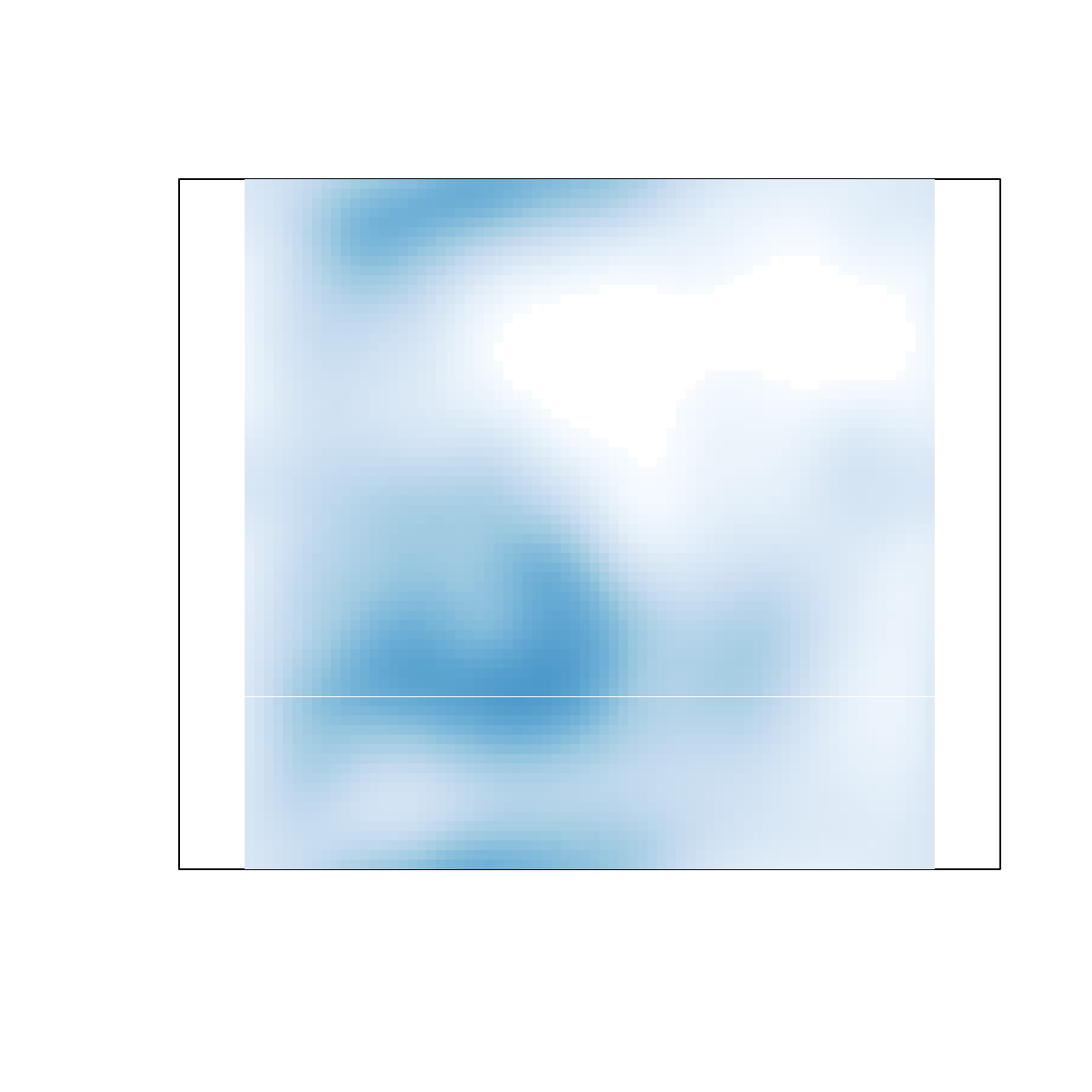}%
\includegraphics[width=0.15\linewidth, clip, trim= 55 70 30 55]{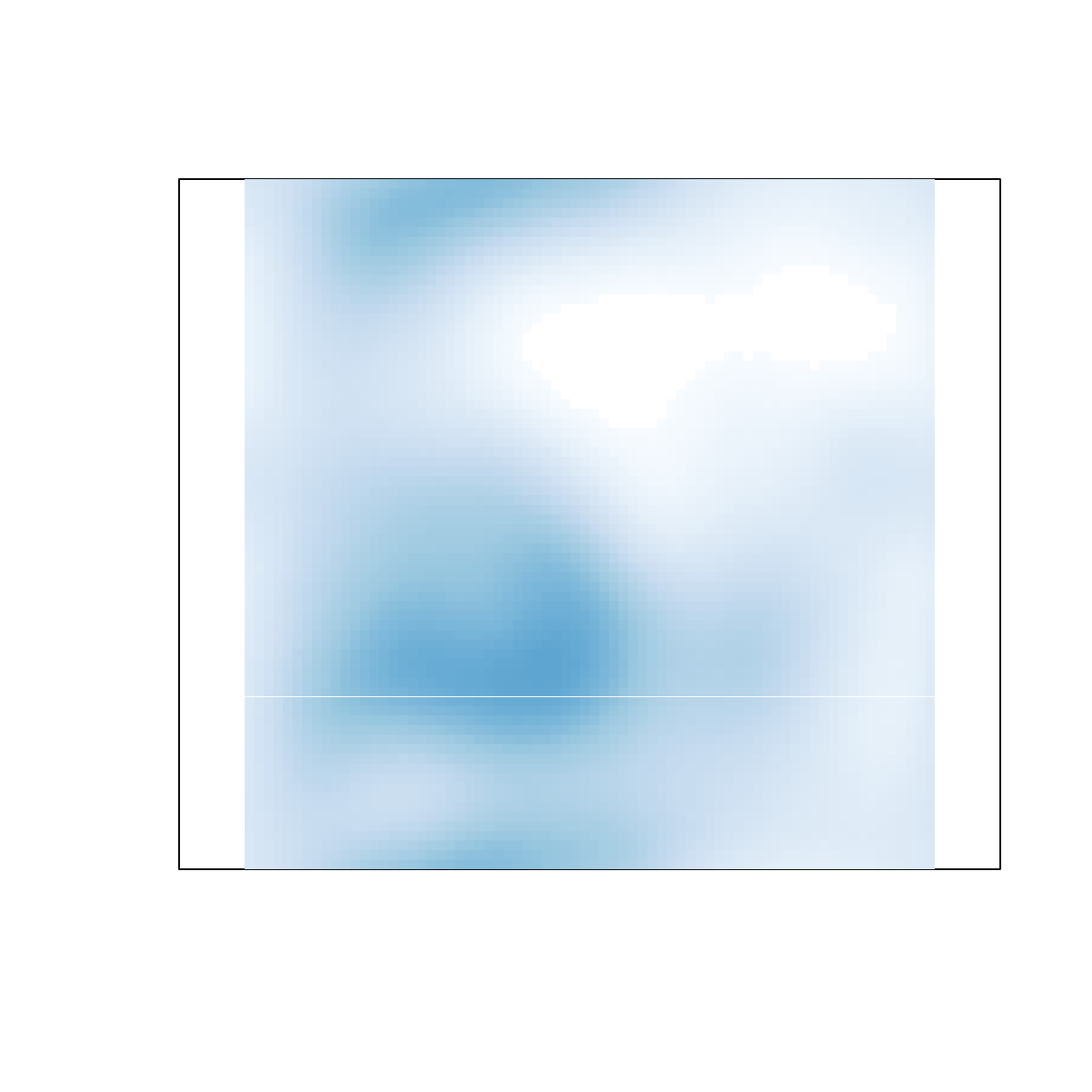}%
\includegraphics[width=0.15\linewidth, clip, trim= 55 70 30 55]{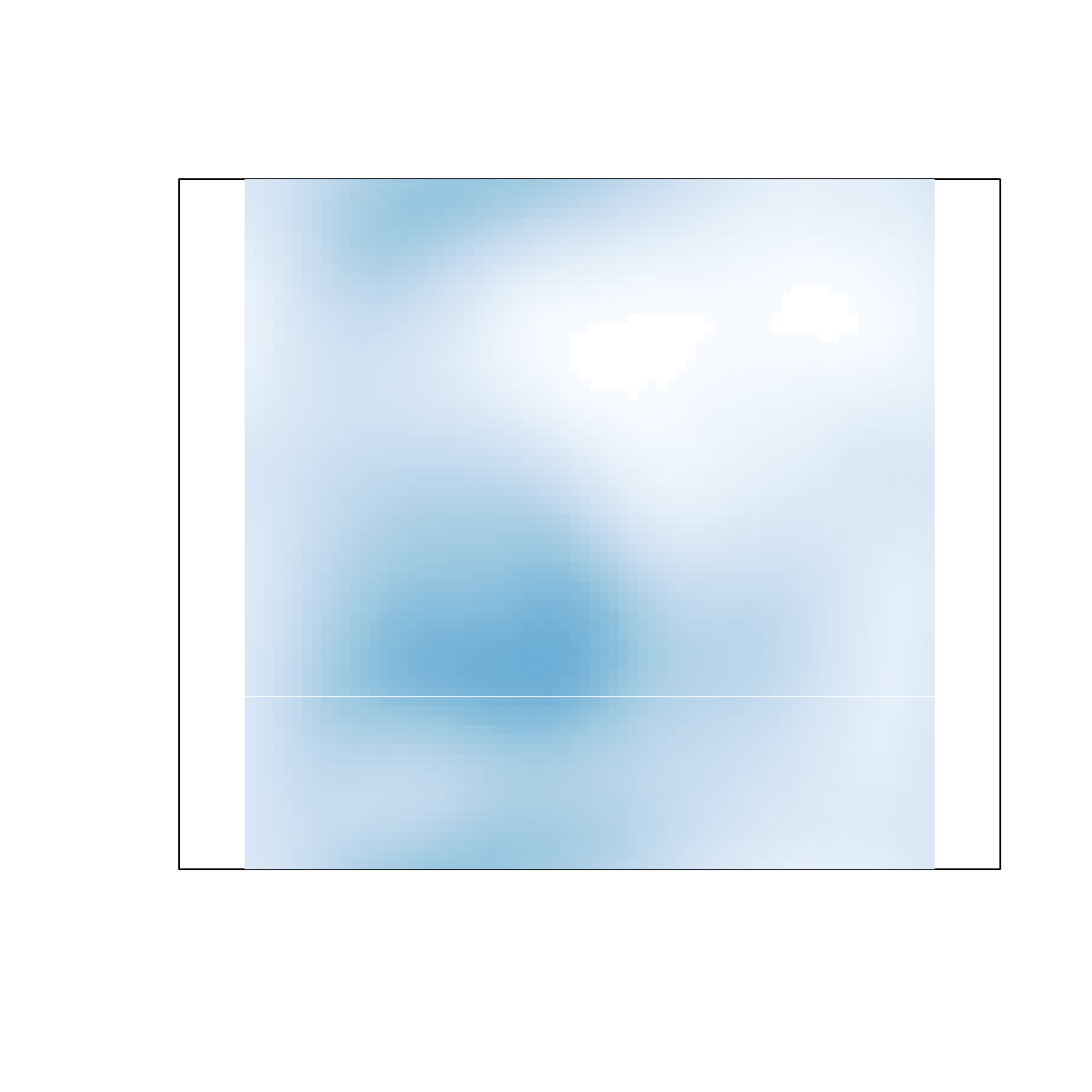}%
\includegraphics[width=0.15\linewidth, clip, trim= 55 70 30 55]{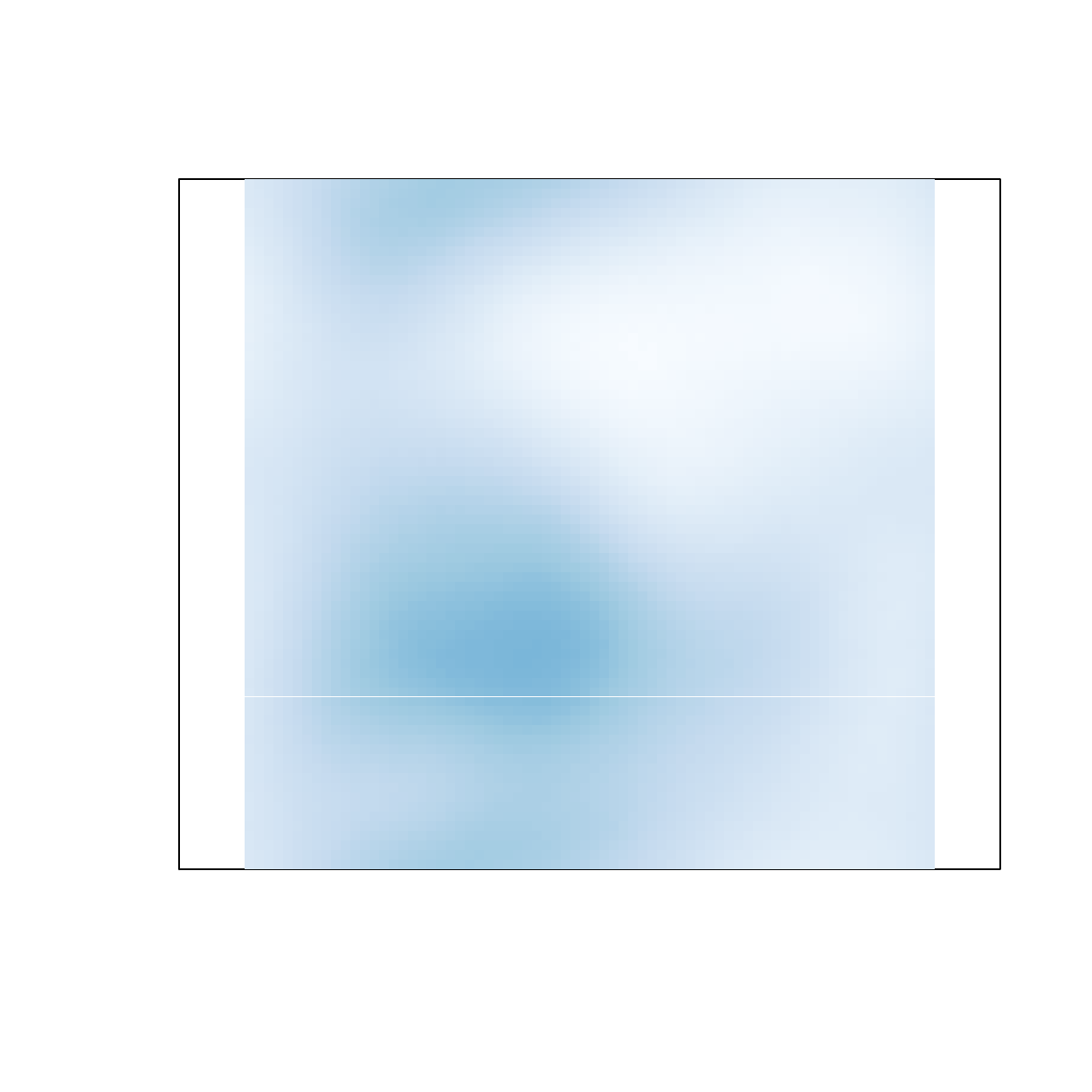}%
\includegraphics[width=0.15\linewidth, clip, trim= 55 70 30 55]{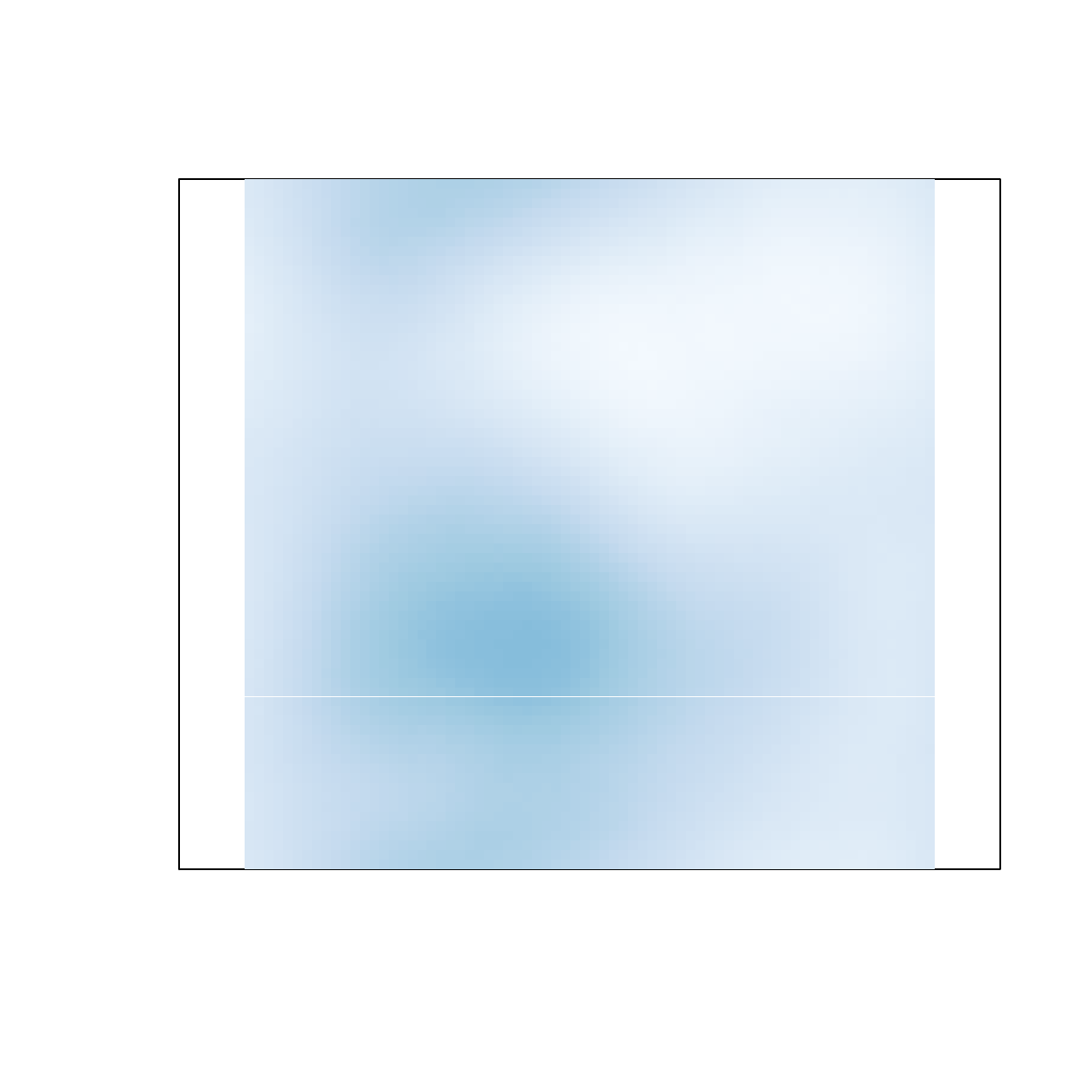}%
\includegraphics[width=0.015\linewidth, clip, trim= 660 120 0 100 ]{legend_4.pdf}
\includegraphics[width=0.15\linewidth, clip, trim= 55 70 30 55]{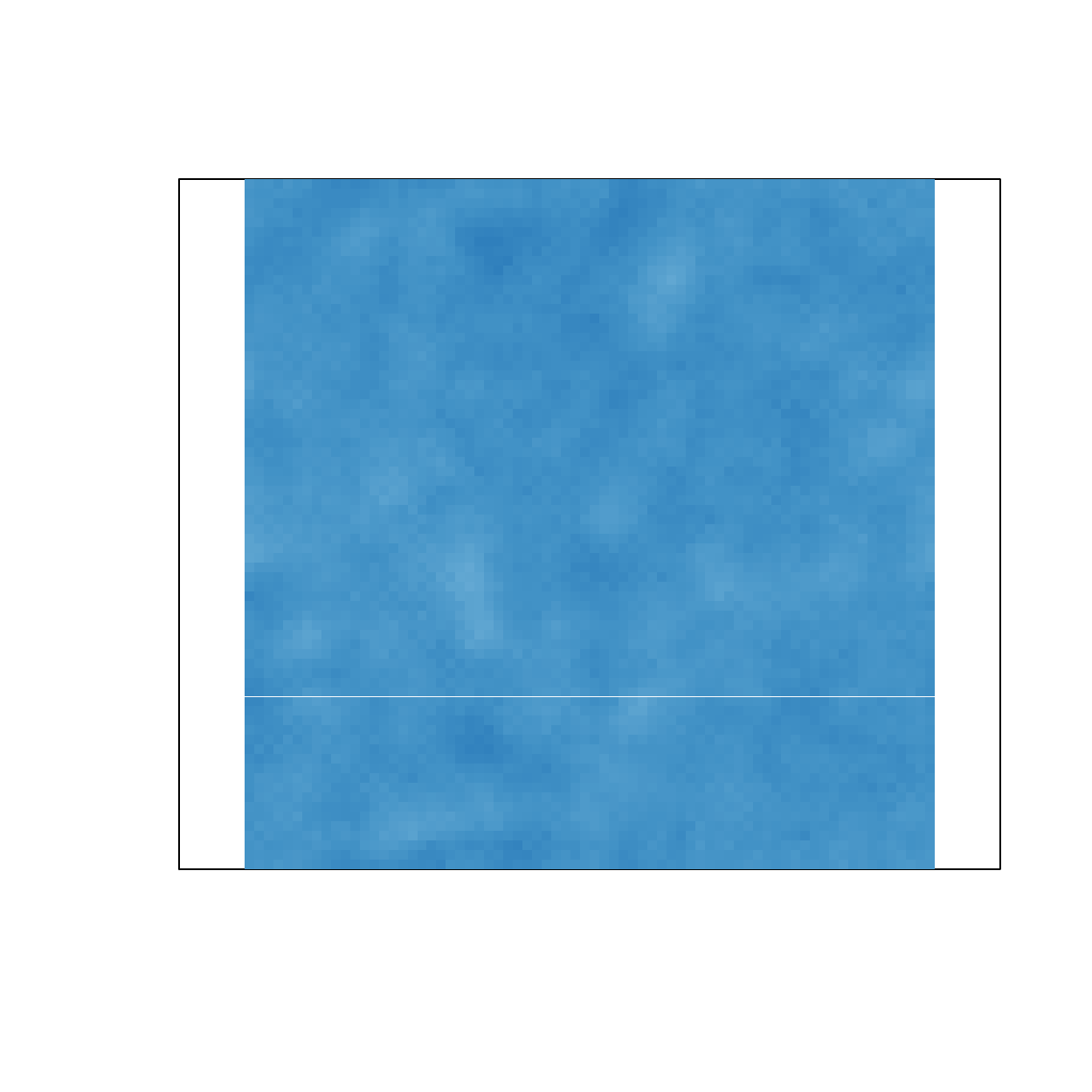}%
\includegraphics[width=0.15\linewidth, clip, trim= 55 70 30 55]{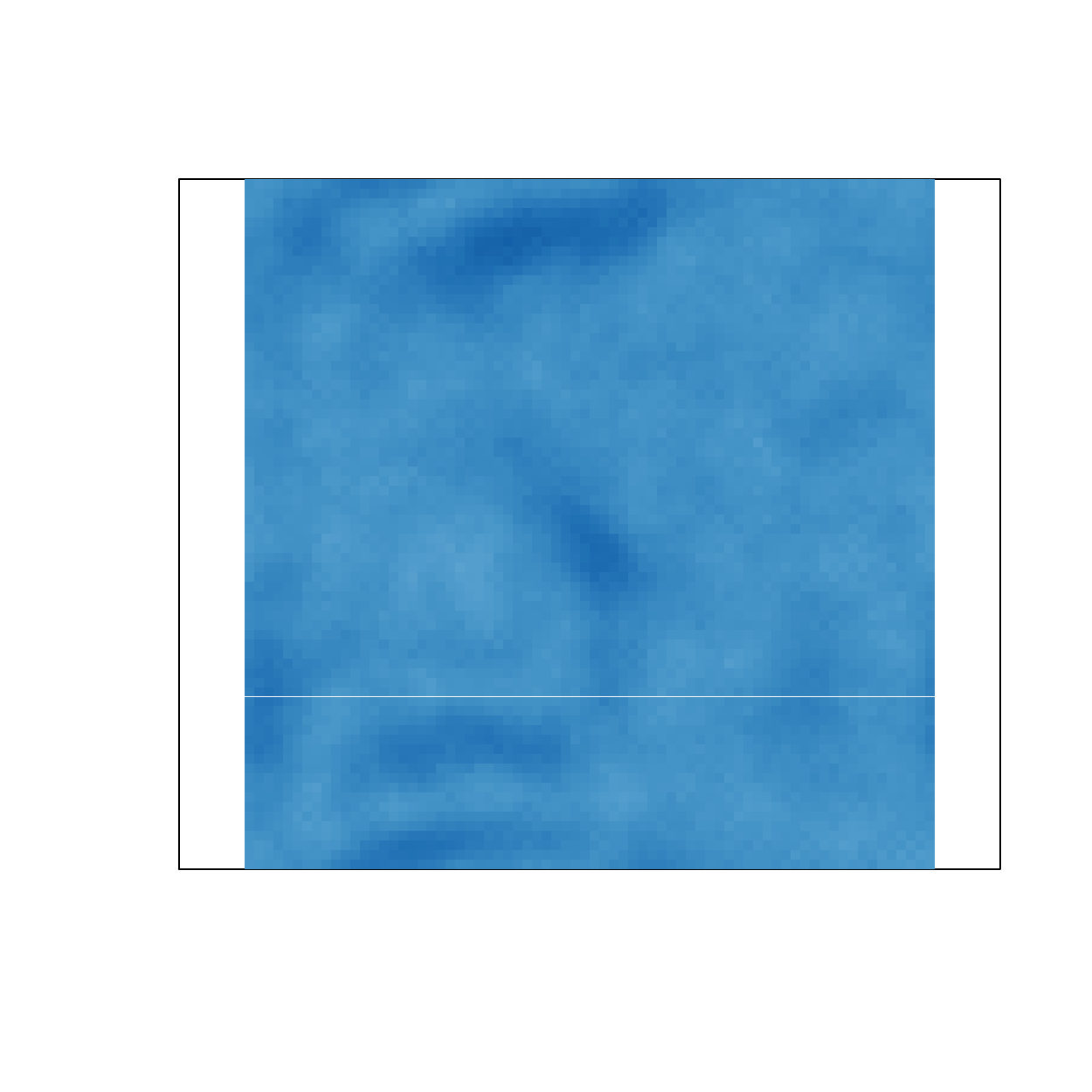}%
\includegraphics[width=0.15\linewidth, clip, trim= 55 70 30 55]{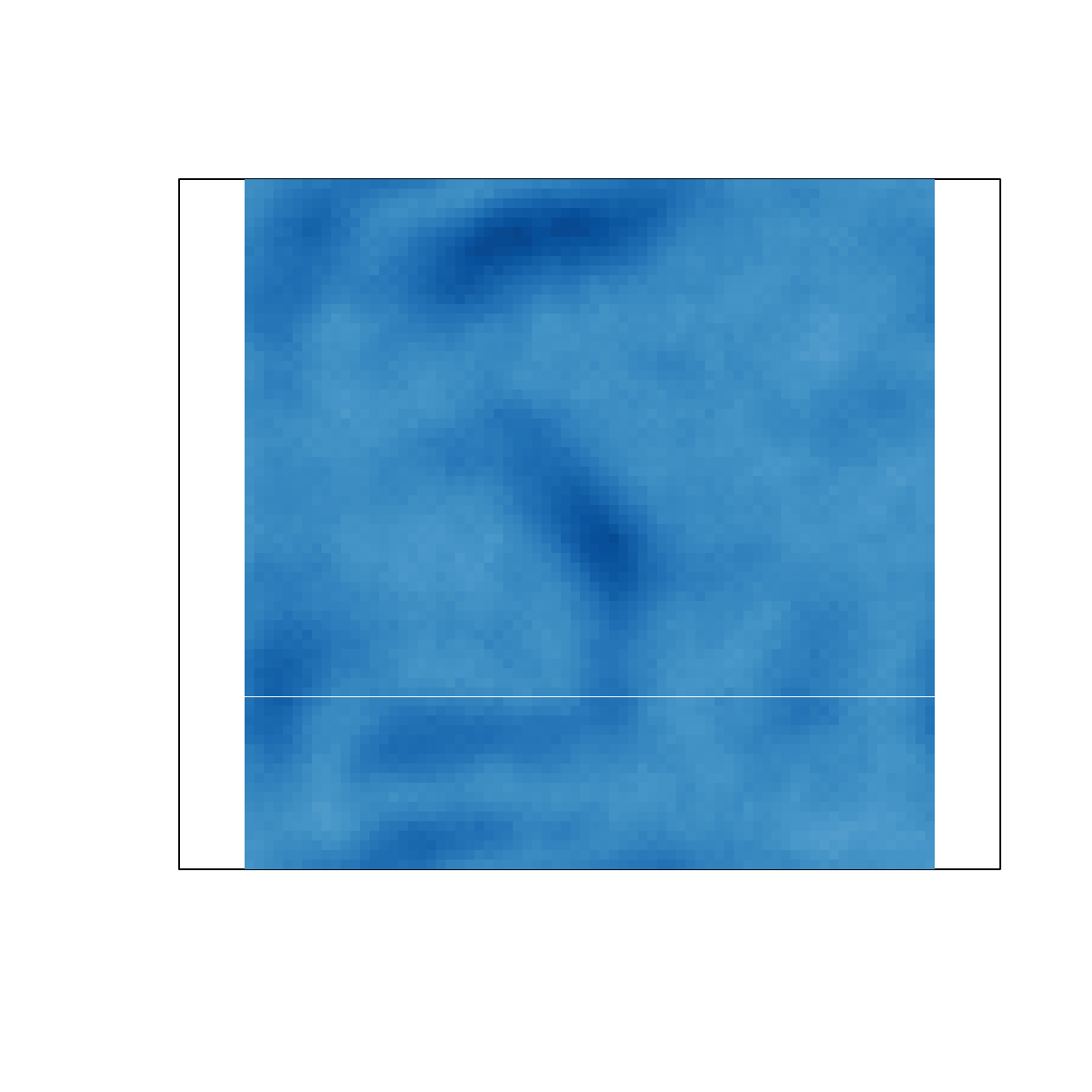}%
\includegraphics[width=0.15\linewidth, clip, trim= 55 70 30 55]{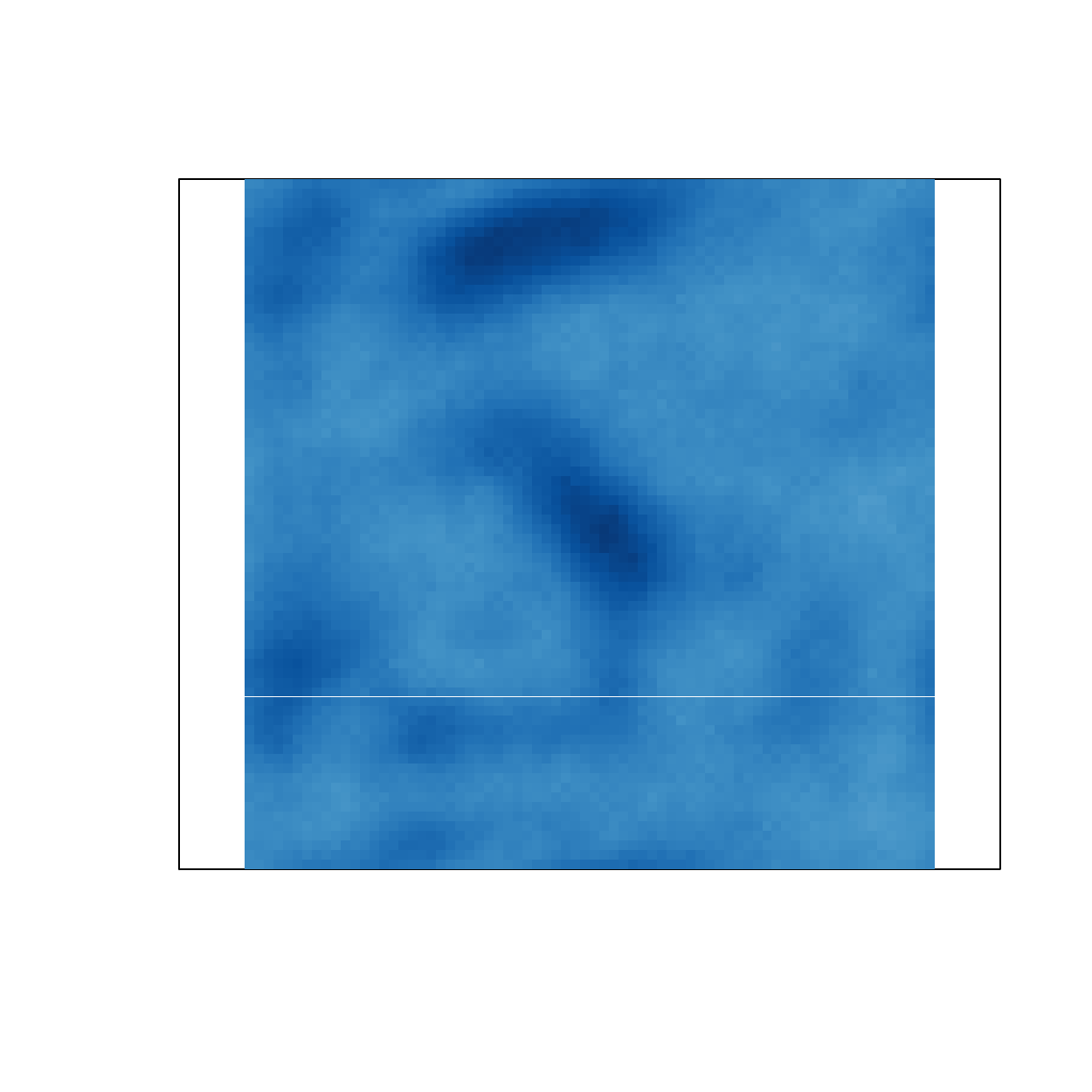}%
\includegraphics[width=0.15\linewidth, clip, trim= 55 70 30 55]{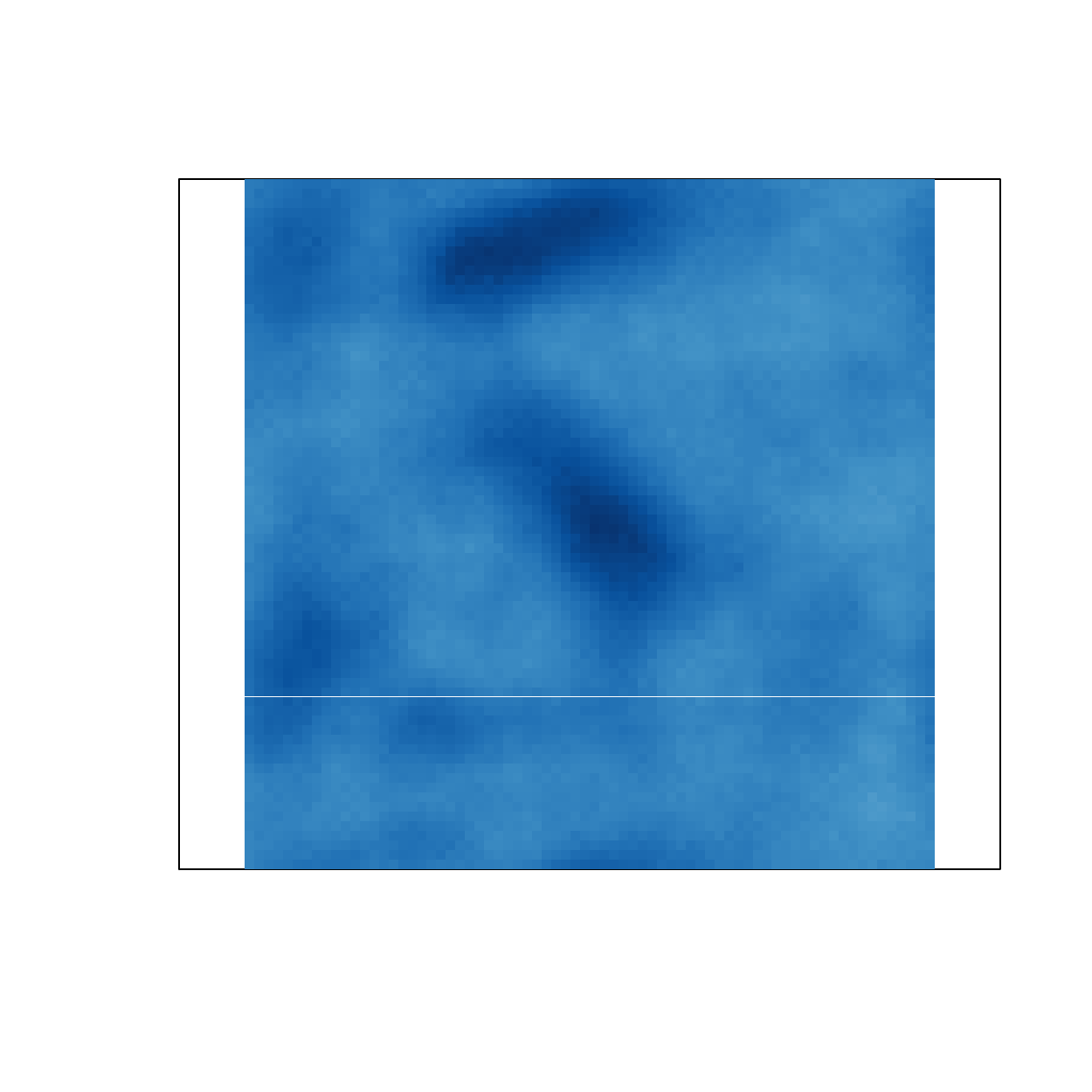}%
\includegraphics[width=0.15\linewidth, clip, trim= 55 70 30 55]{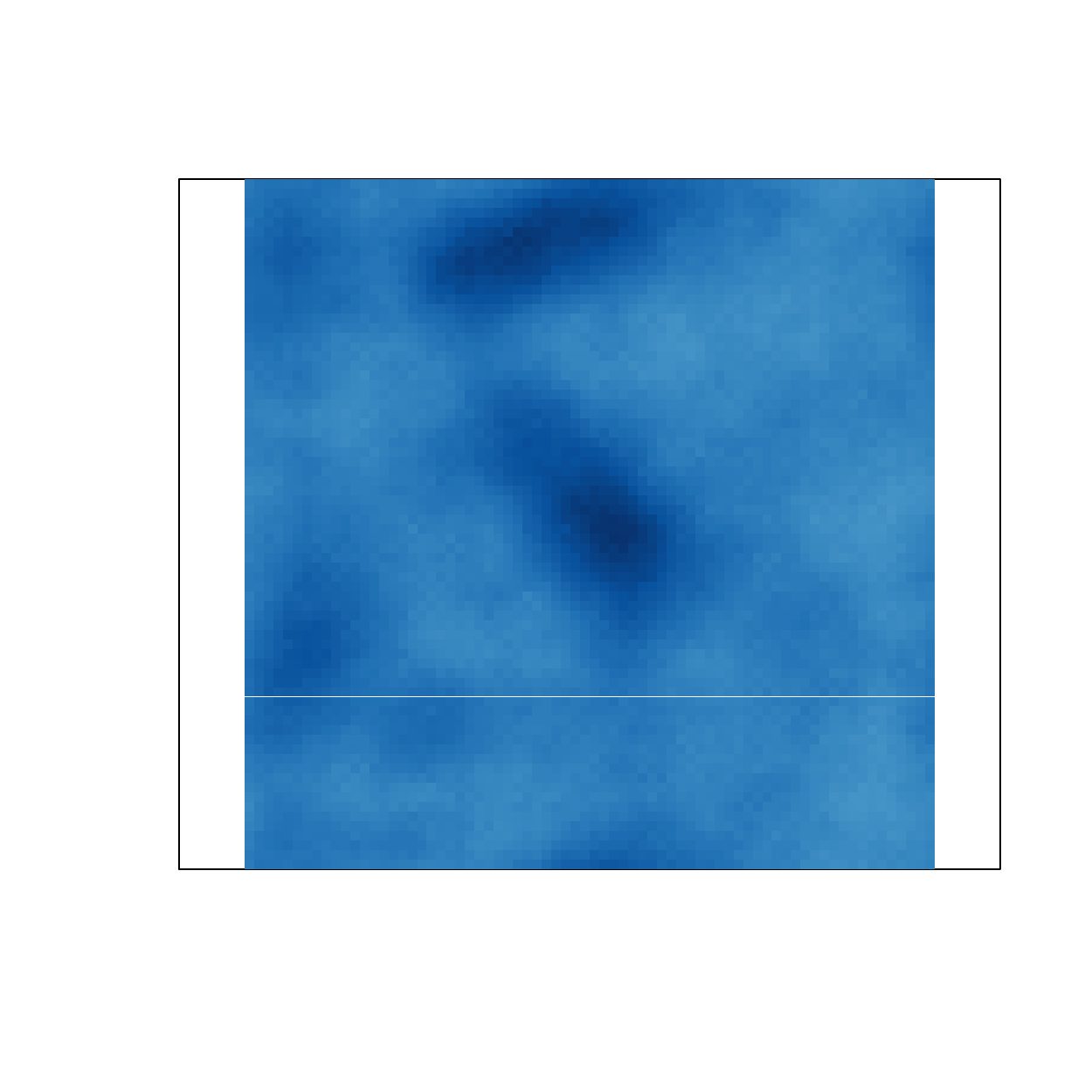}%
\includegraphics[width=0.015\linewidth, clip, trim= 660 120 0 100 ]{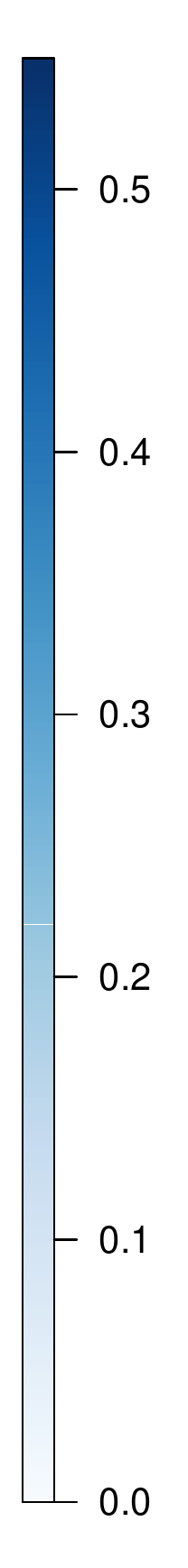}
\includegraphics[width=0.15\linewidth, clip, trim= 55 70 30 55]{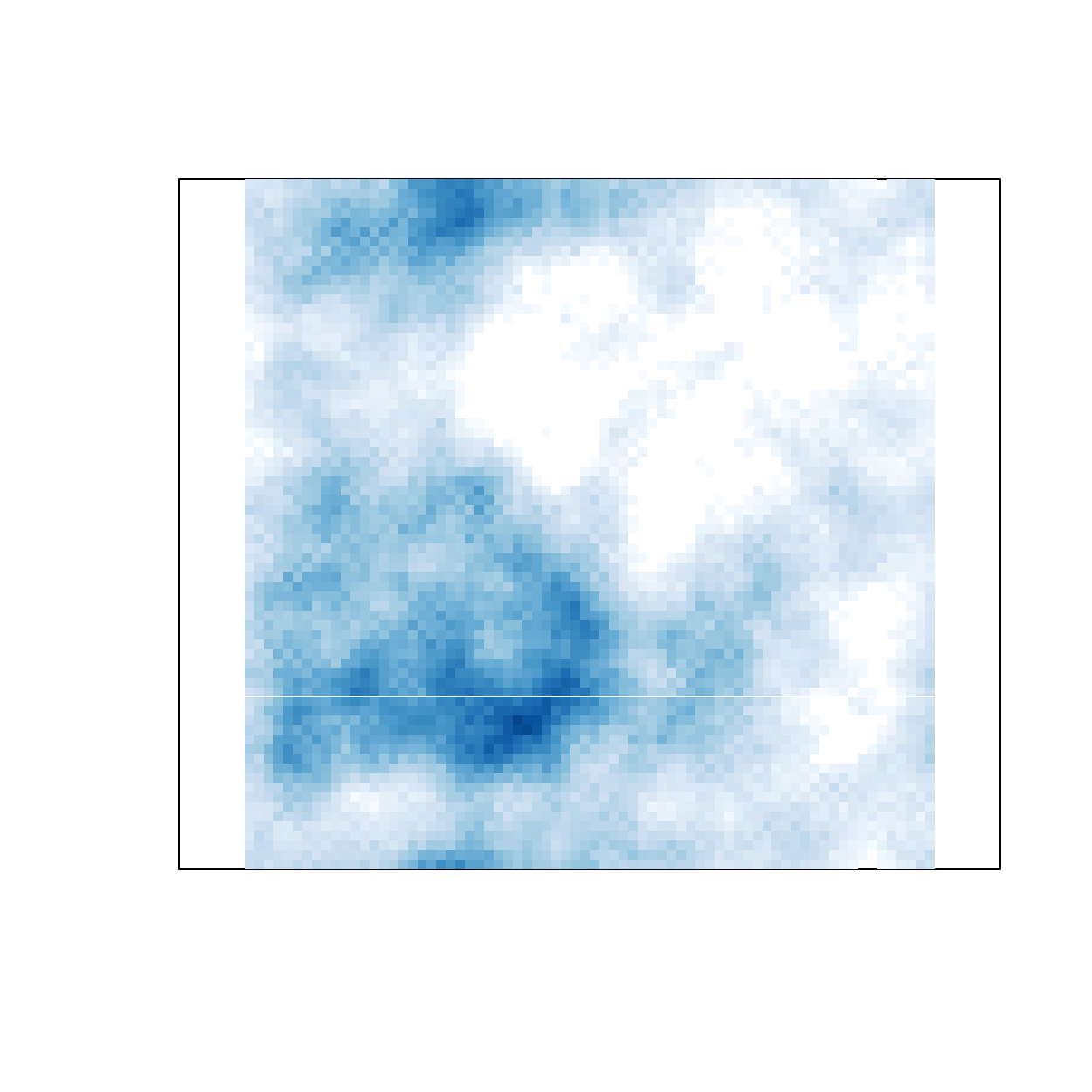}%
\includegraphics[width=0.15\linewidth, clip, trim= 55 70 30 55]{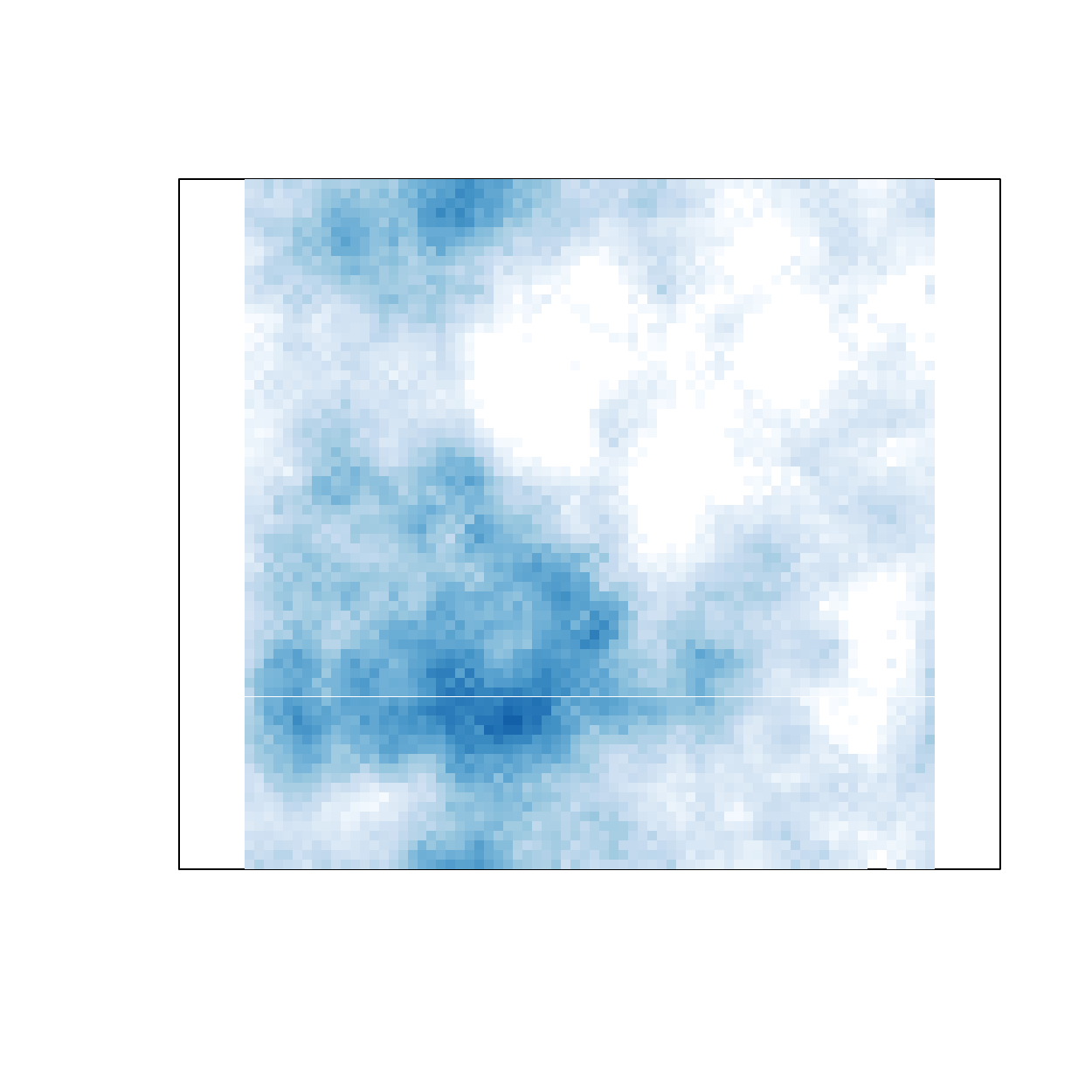}%
\includegraphics[width=0.15\linewidth, clip, trim= 55 70 30 55]{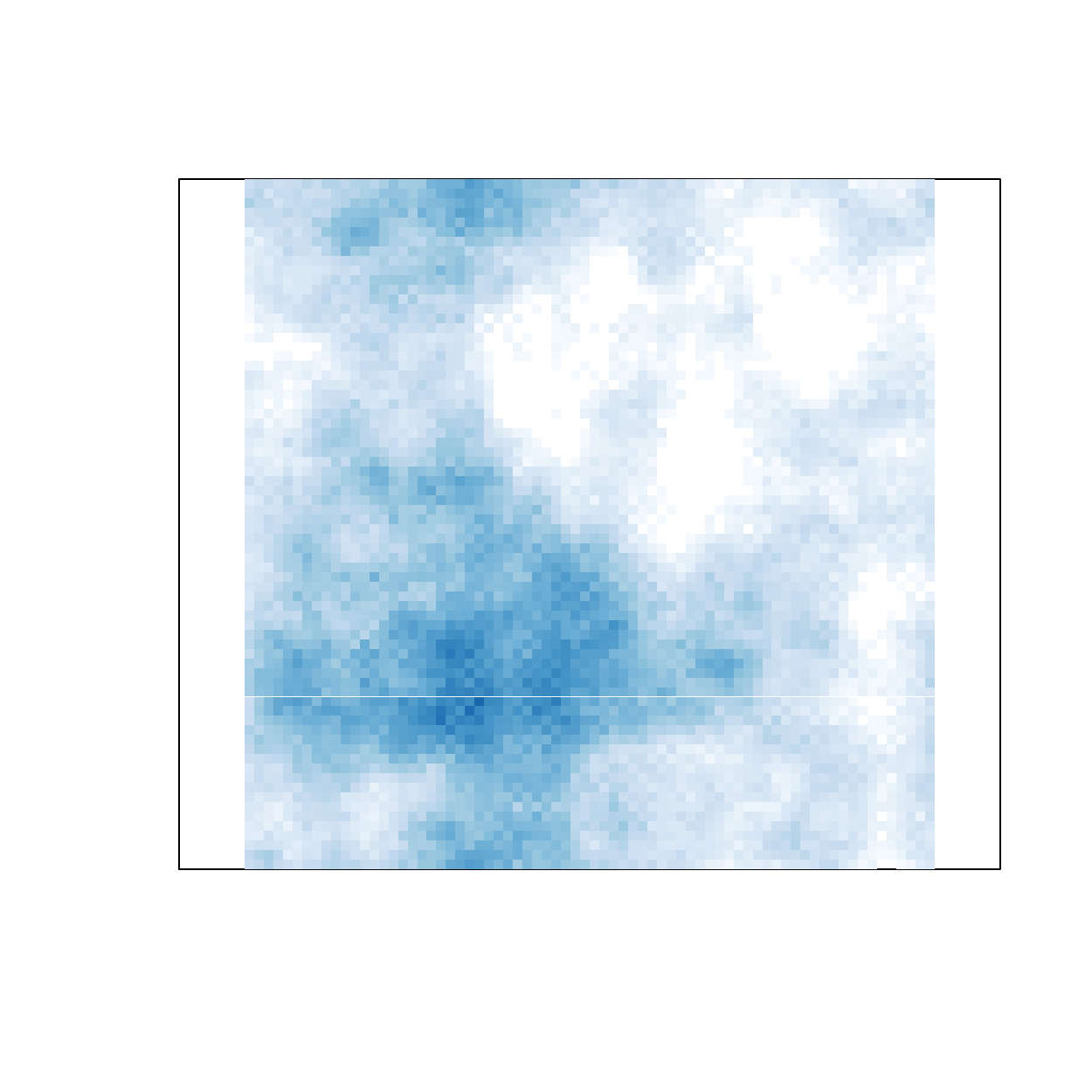}%
\includegraphics[width=0.15\linewidth, clip, trim= 55 70 30 55]{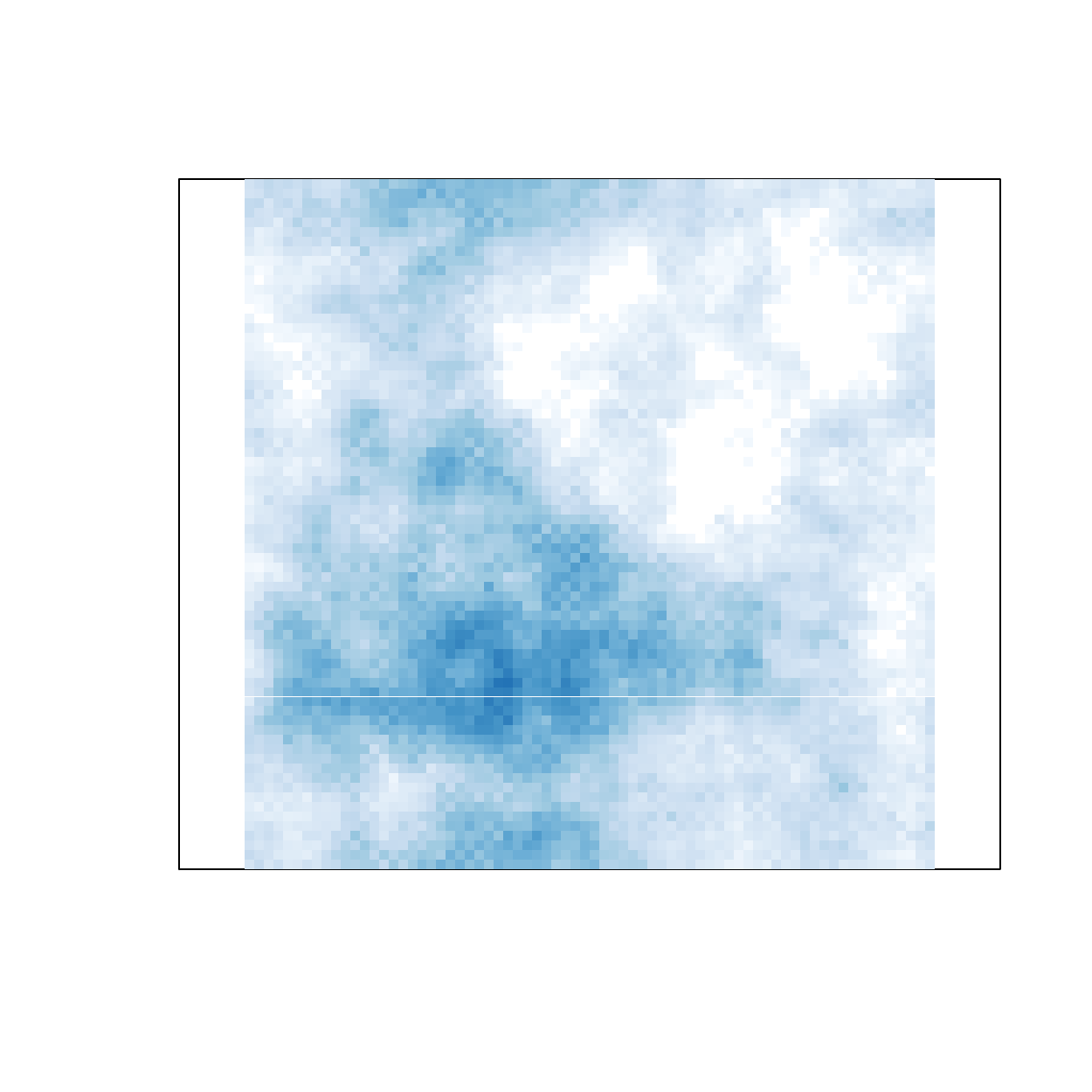}%
\includegraphics[width=0.15\linewidth, clip, trim= 55 70 30 55]{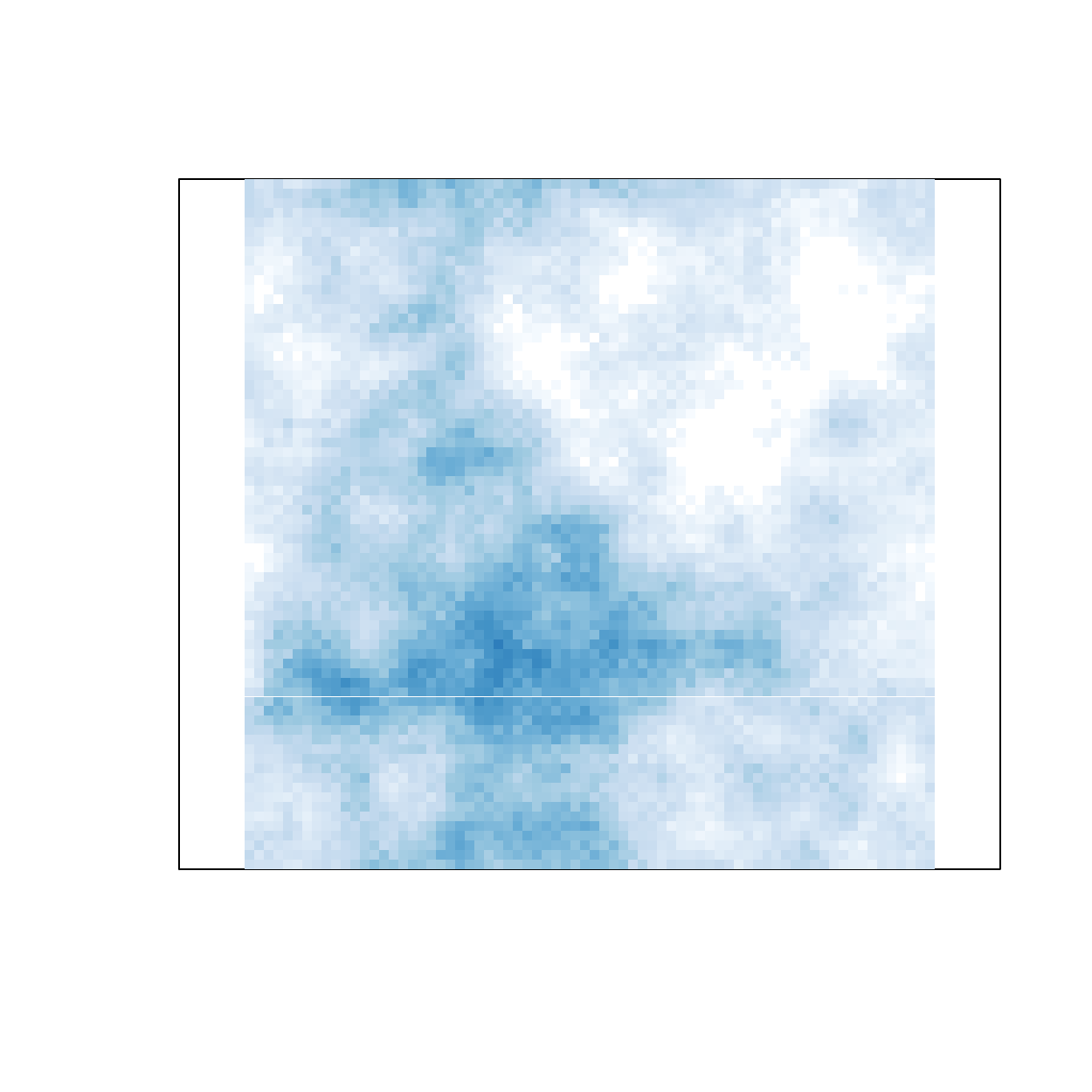}%
\includegraphics[width=0.15\linewidth, clip, trim= 55 70 30 55]{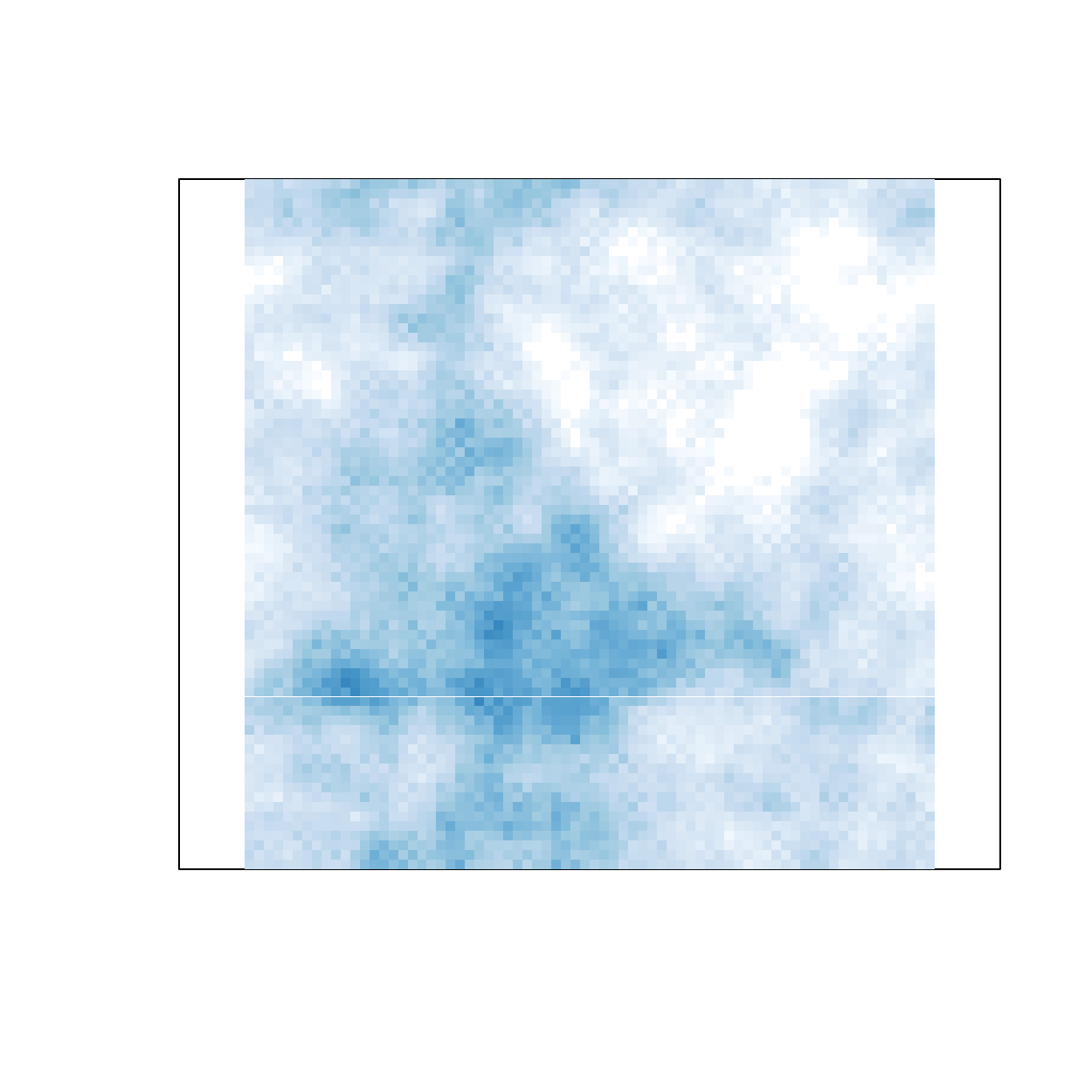}%
\includegraphics[width=0.015\linewidth, clip, trim= 660 120 0 100 ]{legend_4.pdf}
\includegraphics[width=0.15\linewidth, clip, trim= 55 70 30 55]{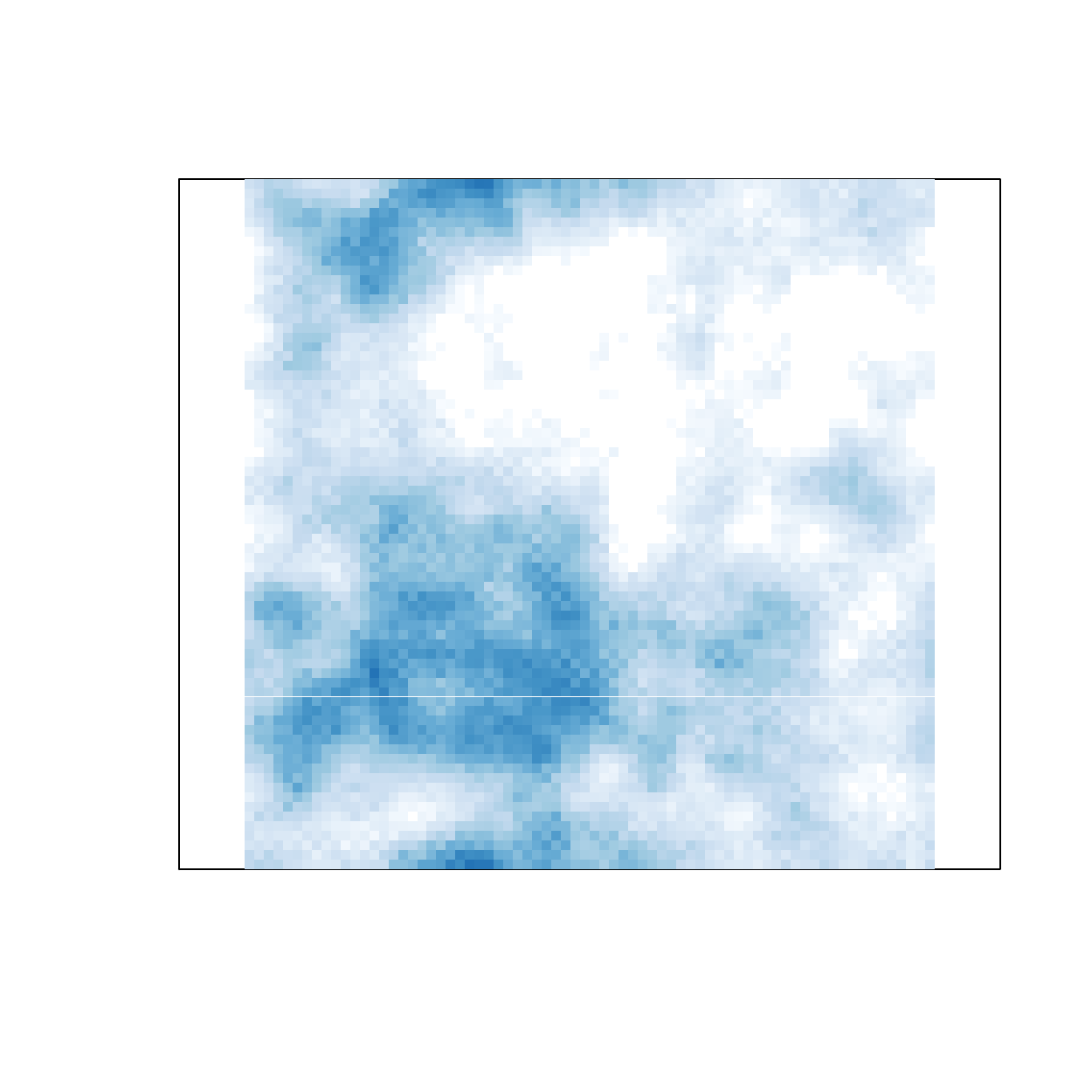}%
\includegraphics[width=0.15\linewidth, clip, trim= 55 70 30 55]{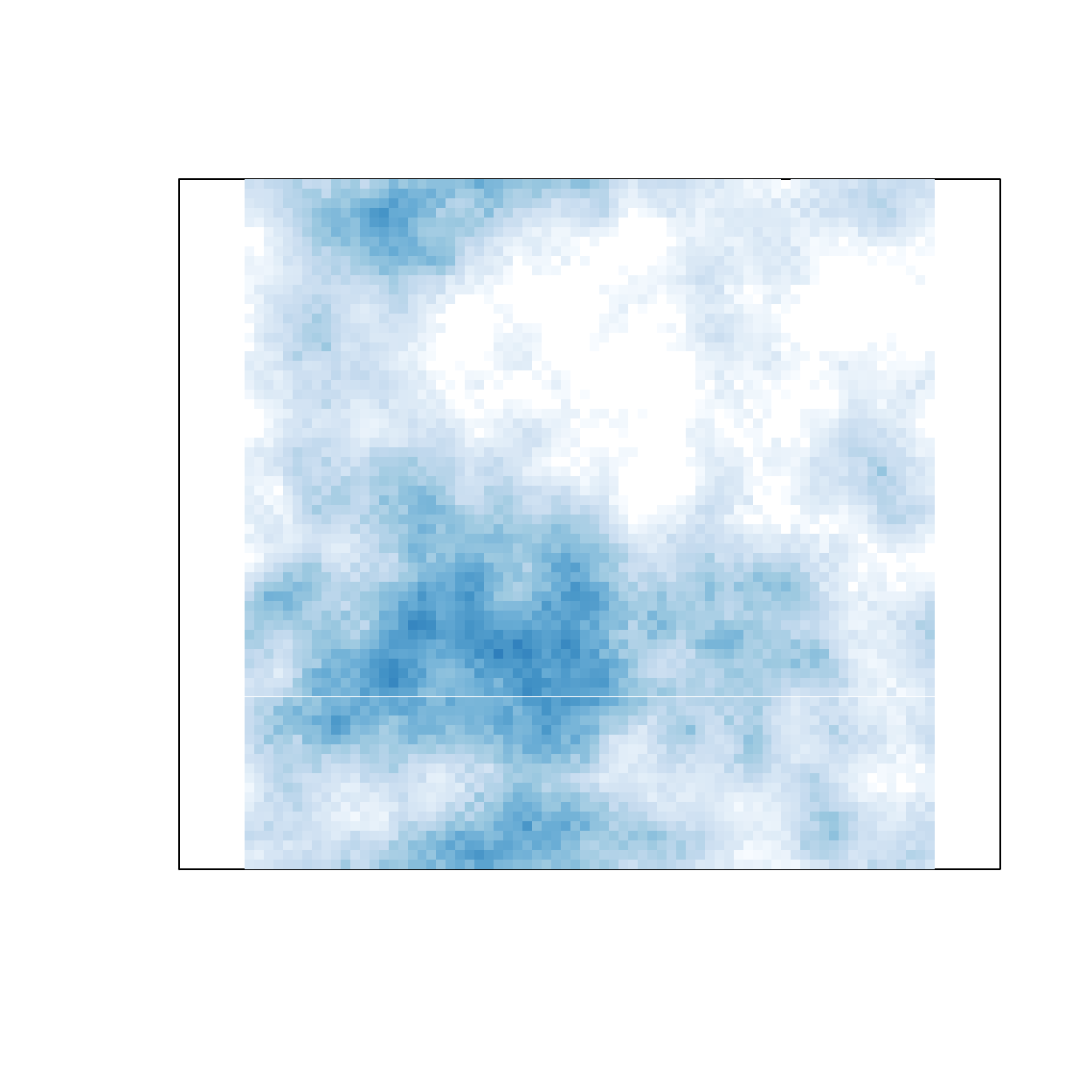}%
\includegraphics[width=0.15\linewidth, clip, trim= 55 70 30 55]{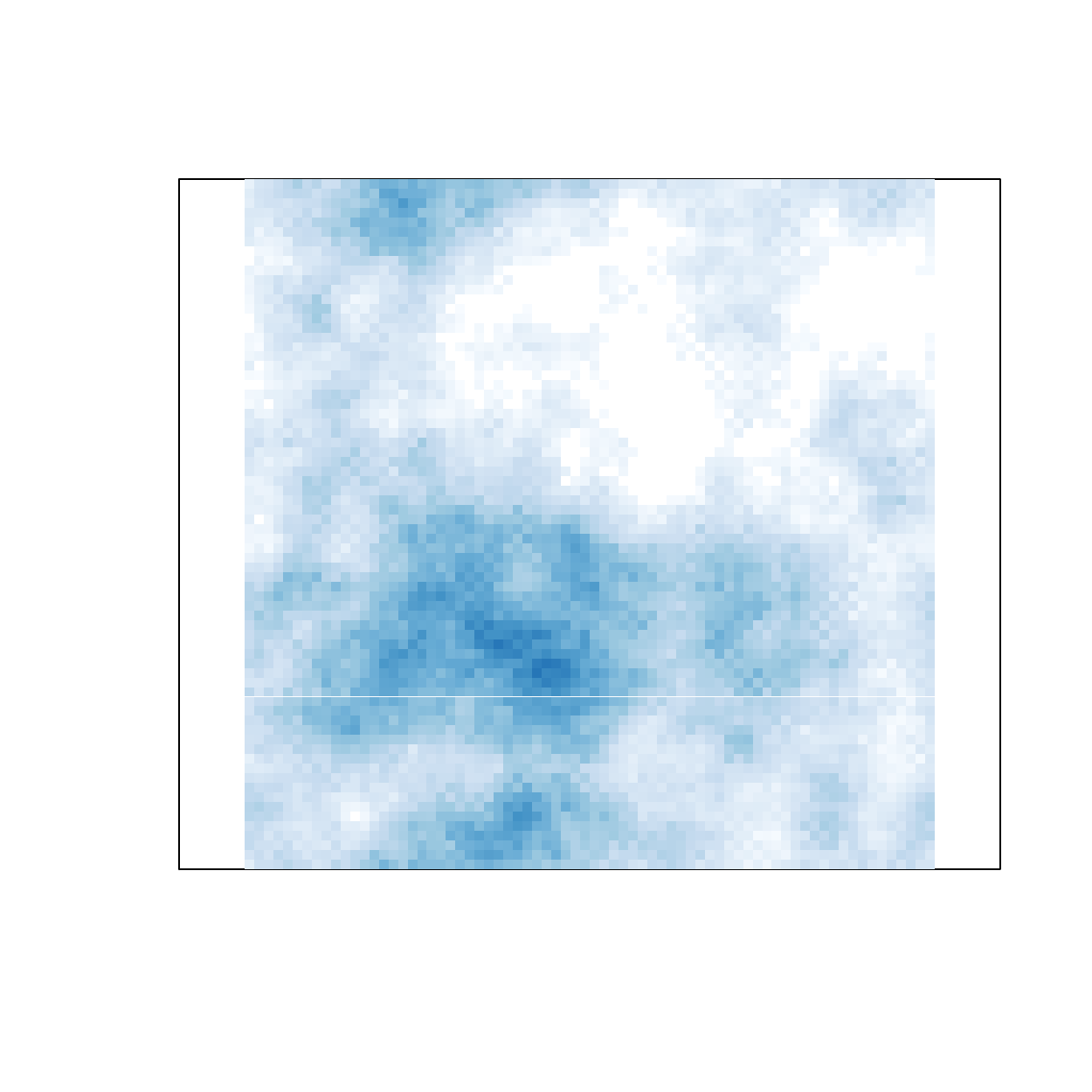}%
\includegraphics[width=0.15\linewidth, clip, trim= 55 70 30 55]{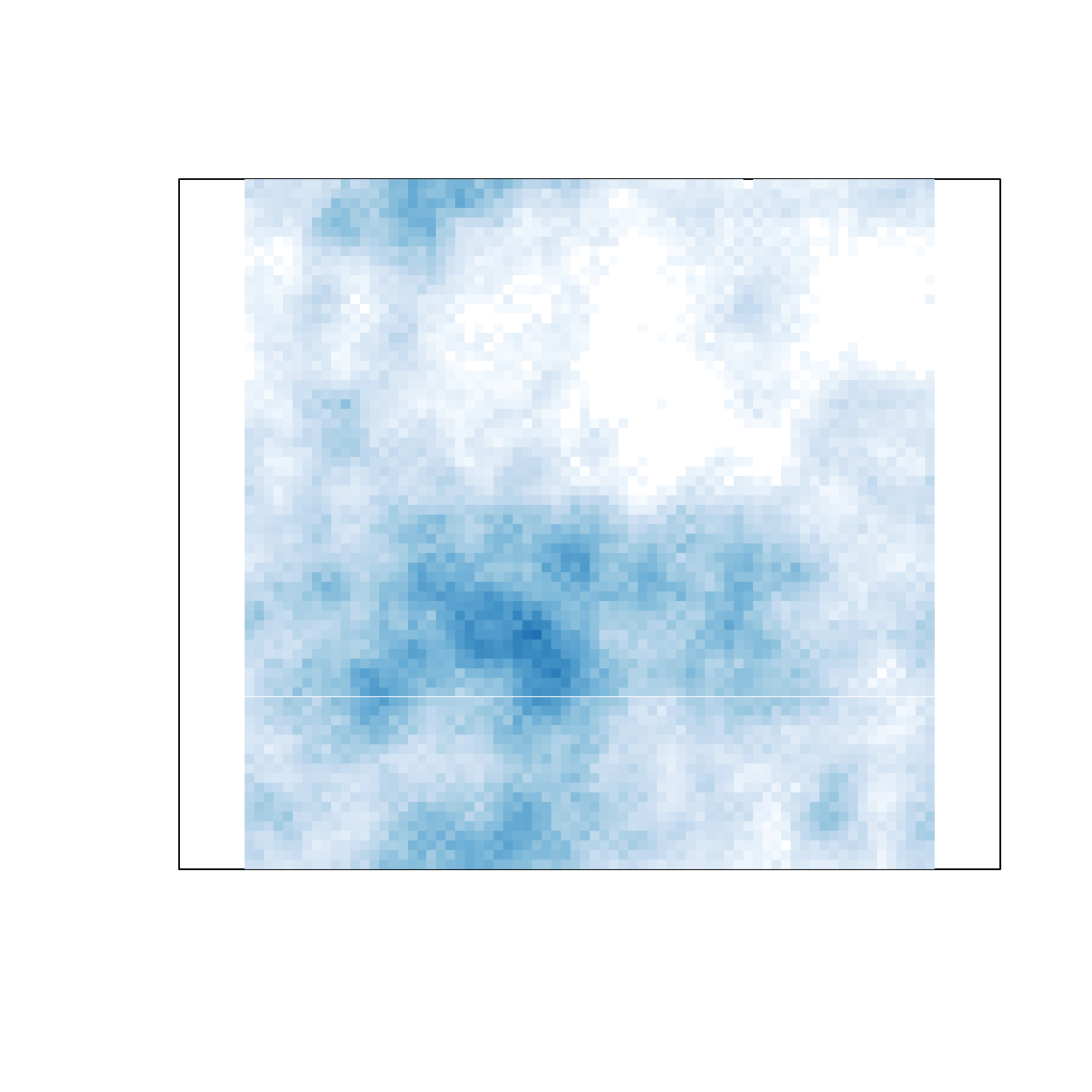}%
\includegraphics[width=0.15\linewidth, clip, trim= 55 70 30 55]{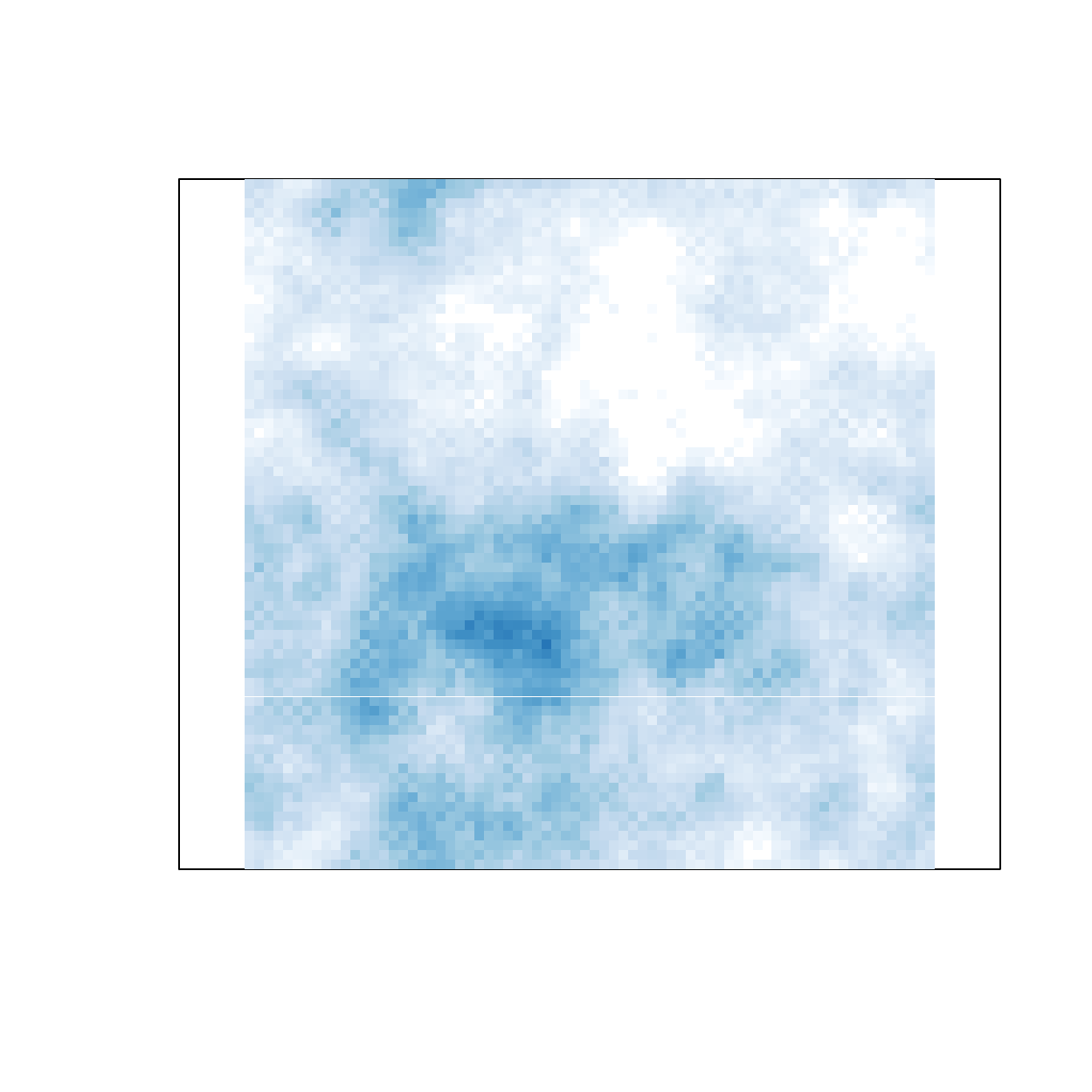}%
\includegraphics[width=0.15\linewidth, clip, trim= 55 70 30 55]{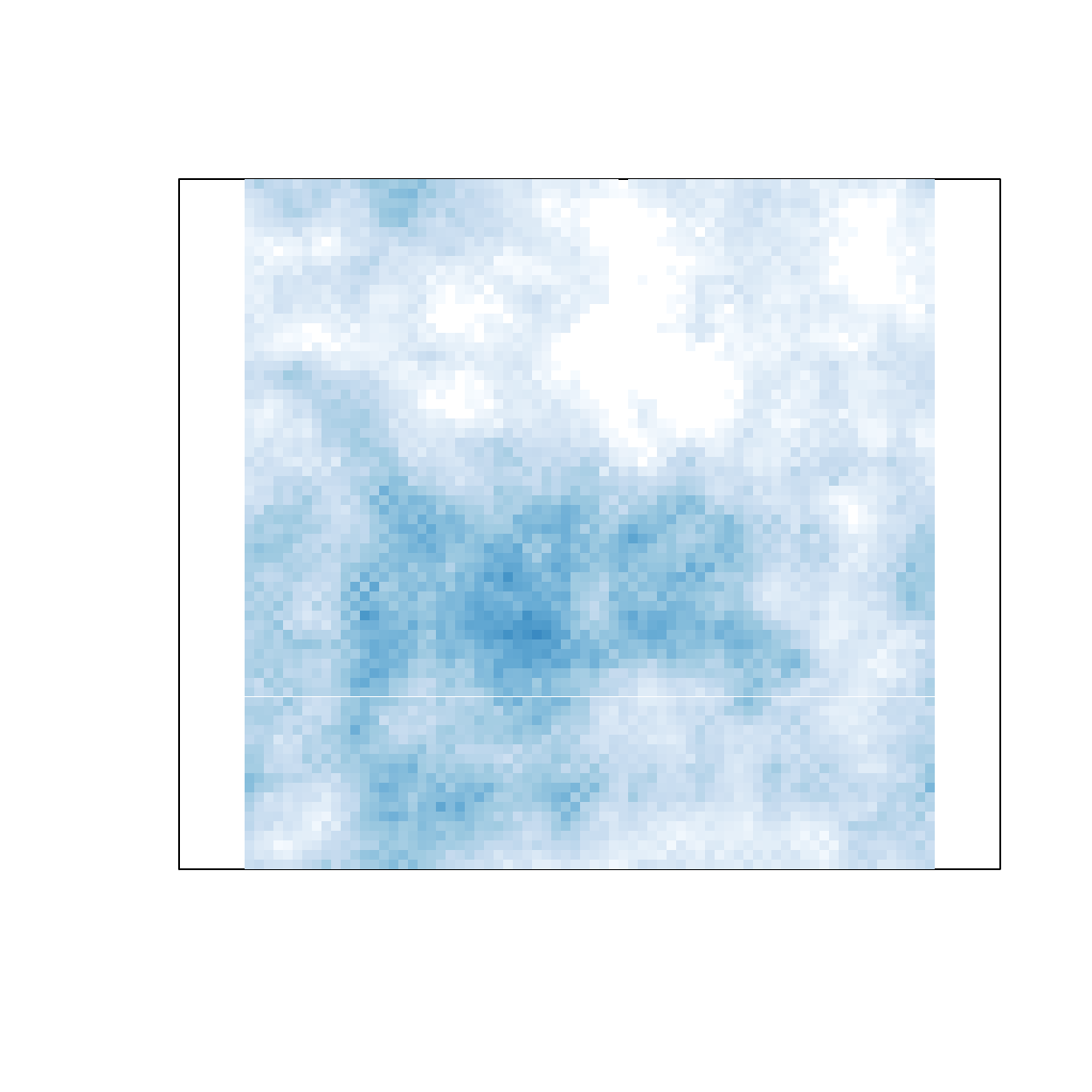}%
\includegraphics[width=0.015\linewidth, clip, trim= 660 120 0 100 ]{legend_4.pdf}
\includegraphics[width=0.15\linewidth, clip, trim= 55 70 30 55]{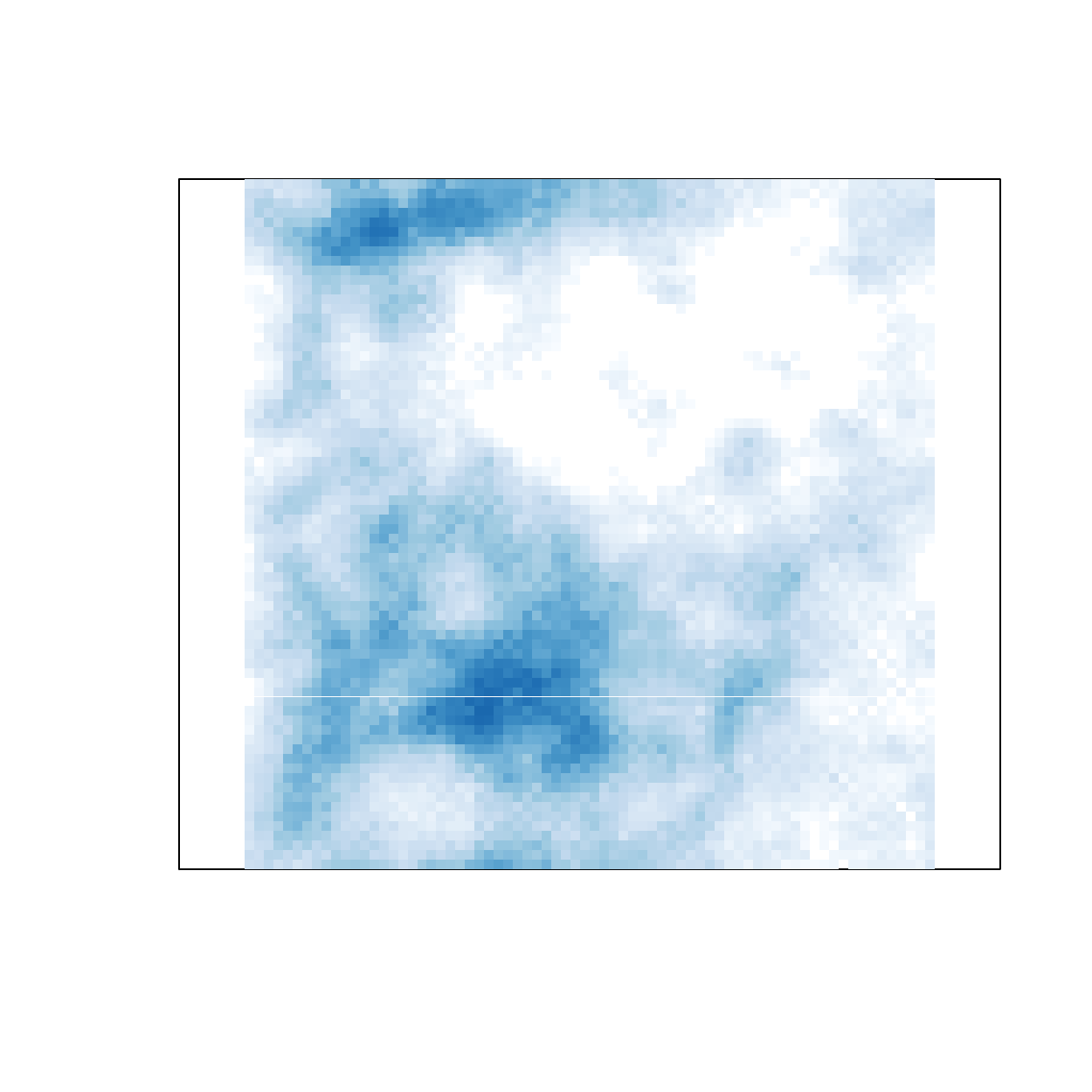}%
\includegraphics[width=0.15\linewidth, clip, trim= 55 70 30 55]{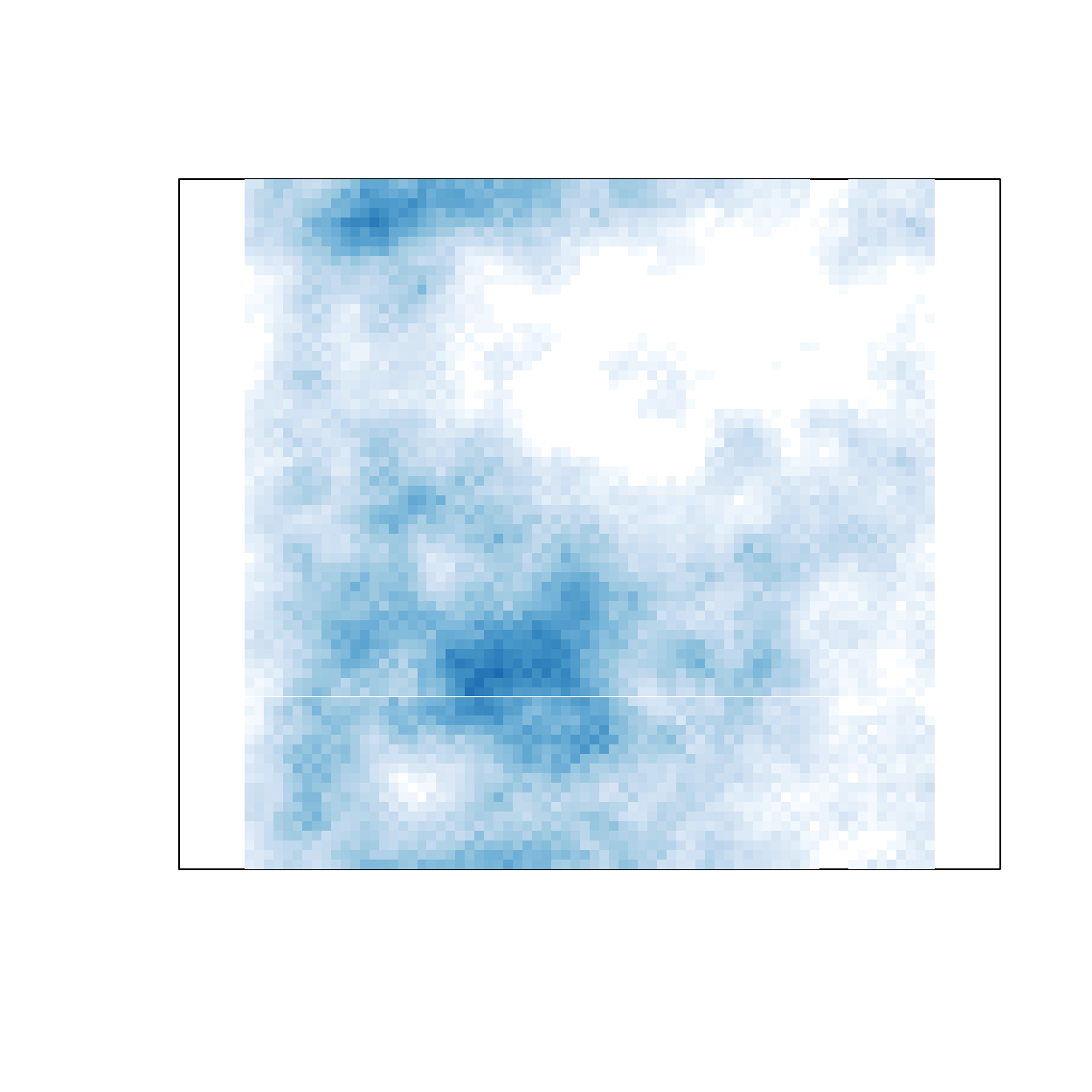}%
\includegraphics[width=0.15\linewidth, clip, trim= 55 70 30 55]{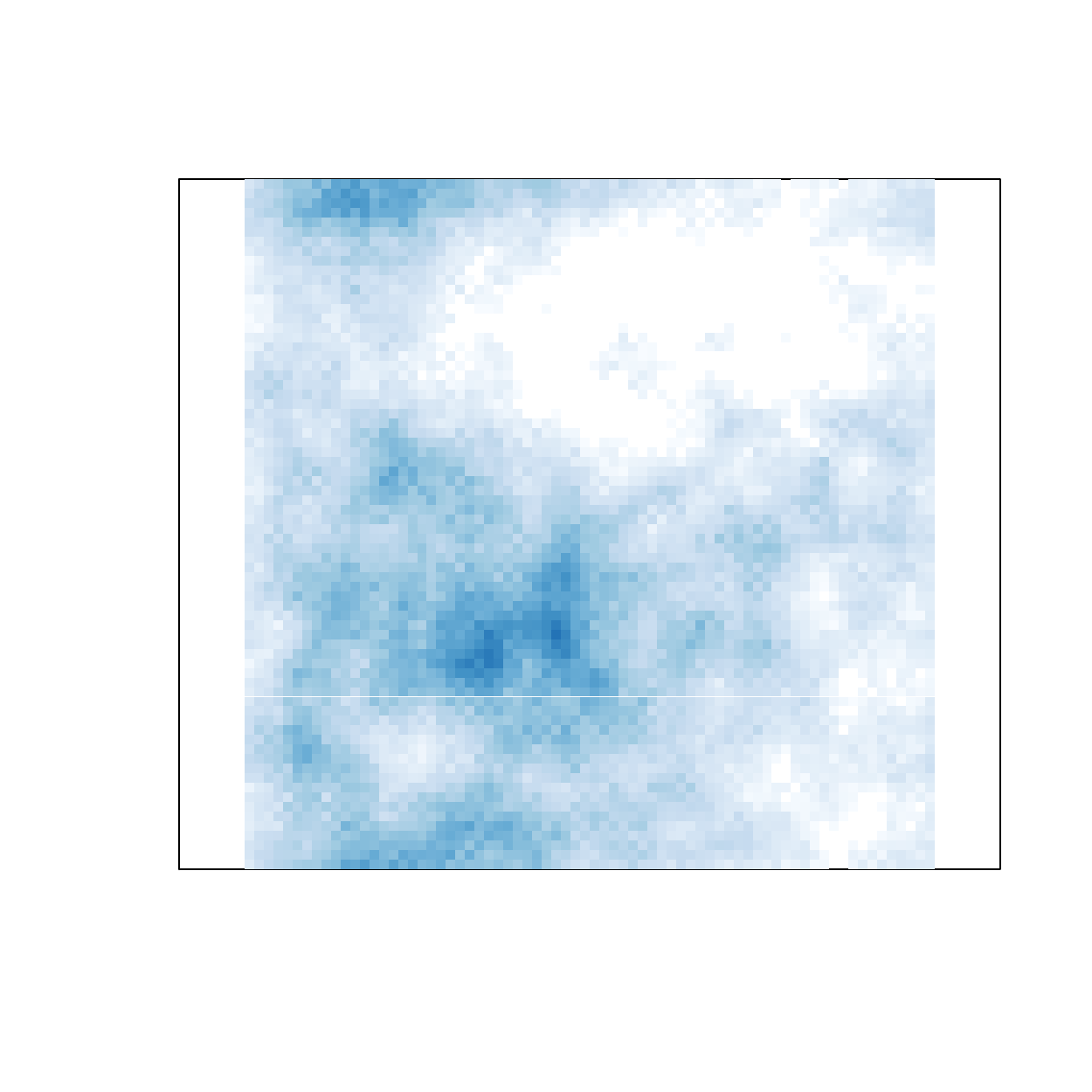}%
\includegraphics[width=0.15\linewidth, clip, trim= 55 70 30 55]{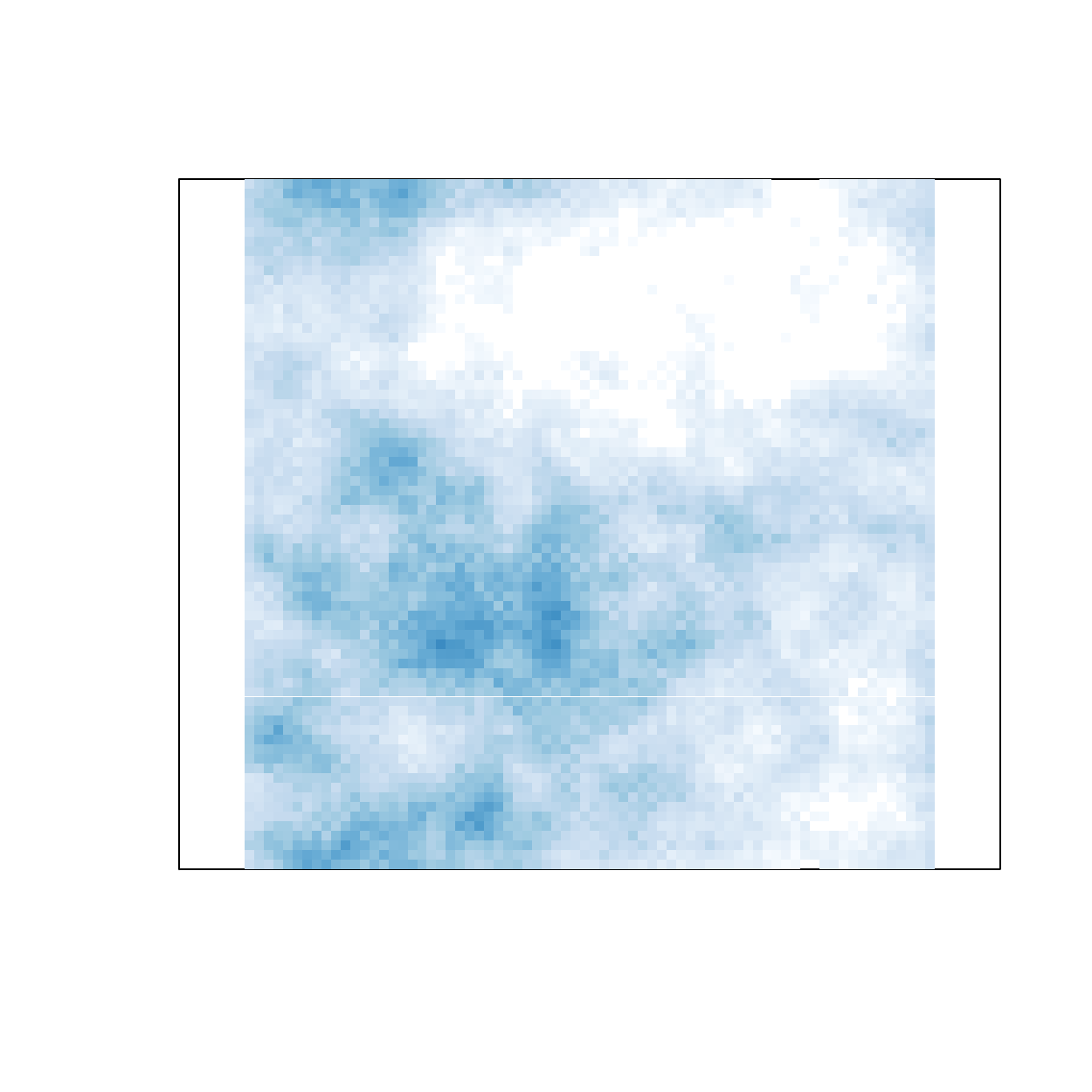}%
\includegraphics[width=0.15\linewidth, clip, trim= 55 70 30 55]{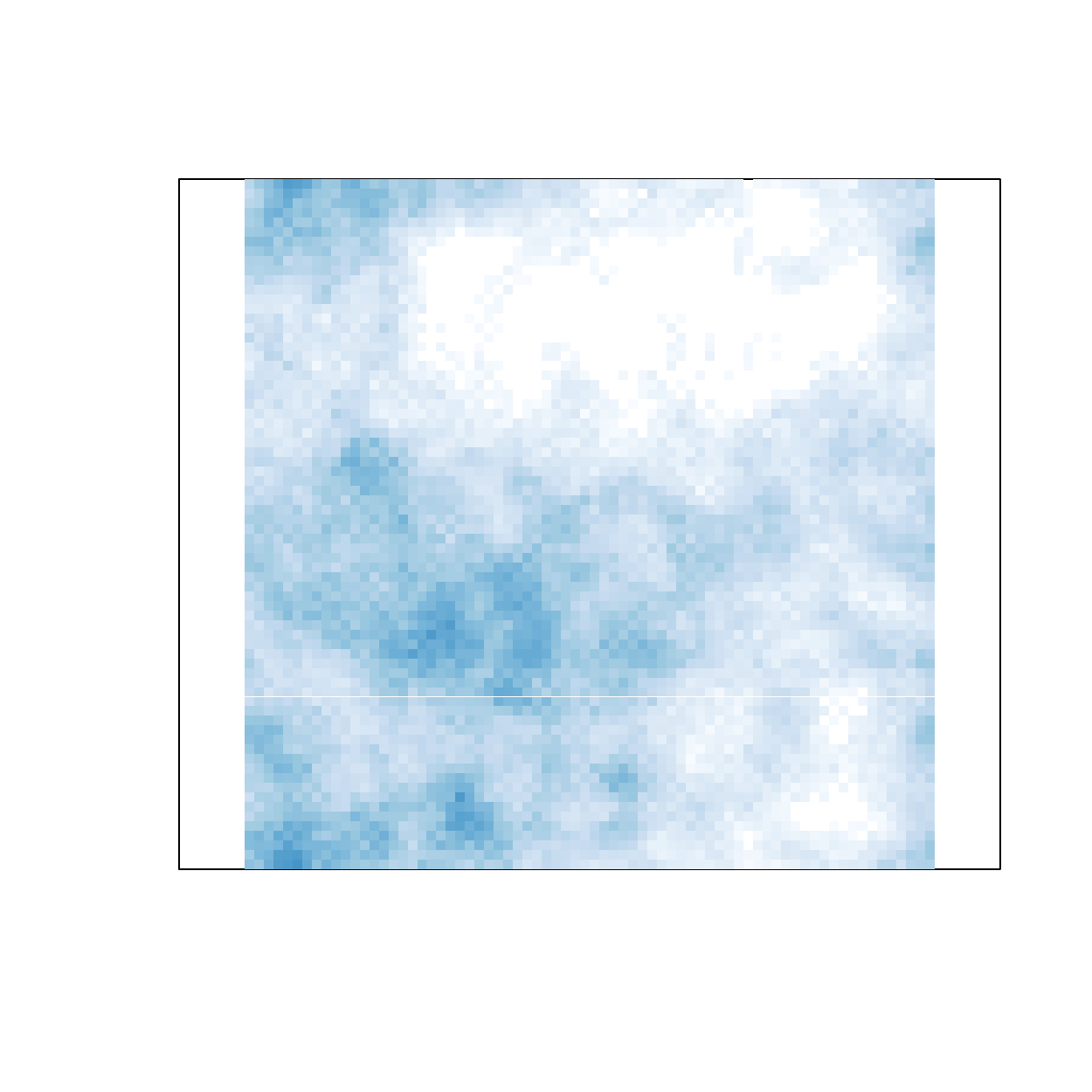}%
\includegraphics[width=0.15\linewidth, clip, trim= 55 70 30 55]{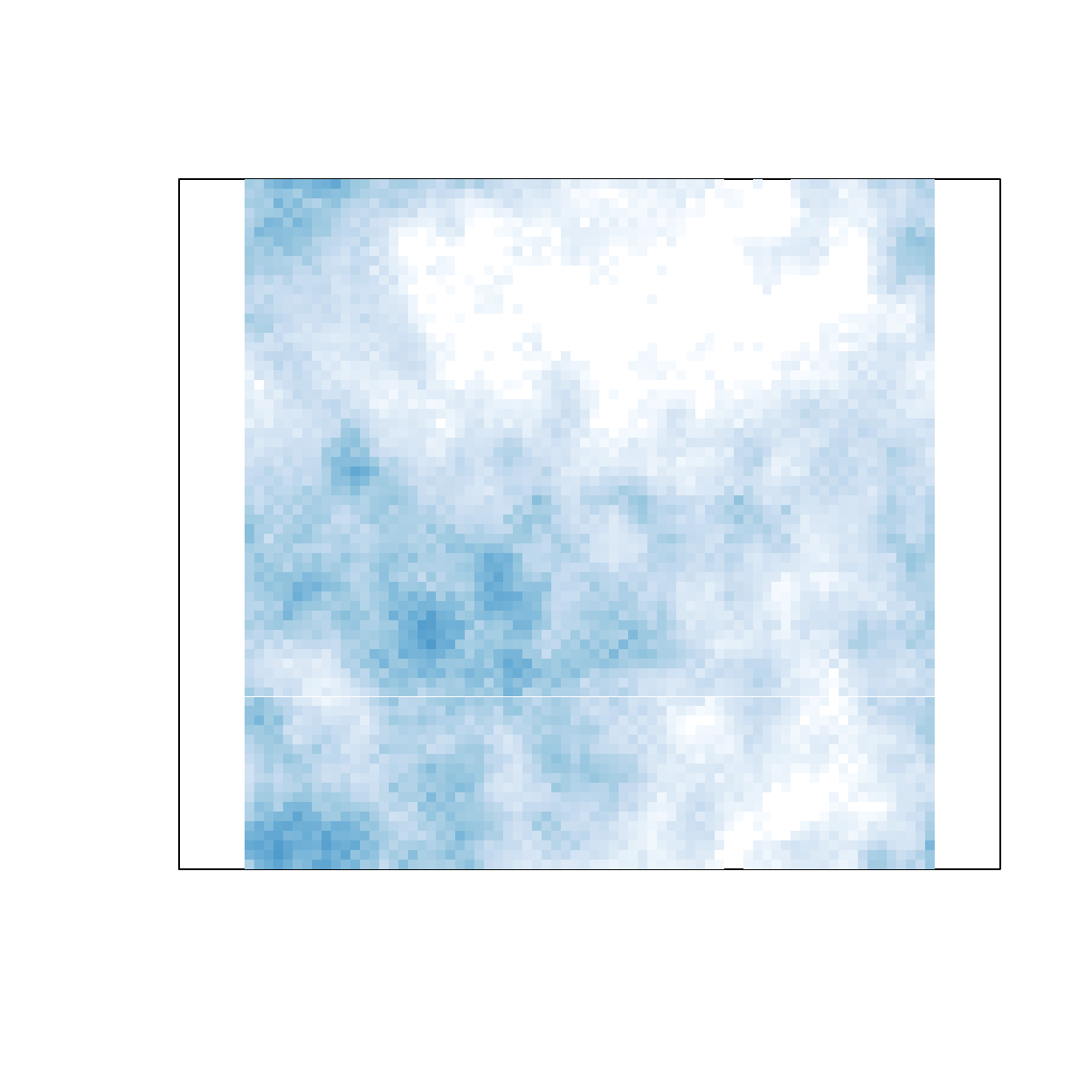}%
\includegraphics[width=0.015\linewidth, clip, trim= 660 120 0 100 ]{legend_4.pdf}
\caption{Image plots of the (transformed) radar observations at the subsequent~$6$ observation times (top row), marginal posterior predictive means and standard deviations of the corresponding precipitation field (second and third row, respectively). Rows four, five and six each show a typical predictive draw of the precipitation field.}
\label{fig:real_forecast_images}
\end{center}
\end{figure}

Our dynamical model~\eqref{eqn:full_bayes_model} allows us to straightforwardly generate predictions for future time-steps by applying the model recursions; Figure~\ref{fig:real_forecast_images}, top row, shows image plots of the (transformed) radar observations at the next~$6$ observation times.
The second row shows the marginal expectation of the precipitation field at each location and the corresponding standard deviations are shown in row~3.
The final three rows in Figure~\ref{fig:real_forecast_images} each show a typical predictive draw of the precipitation field.
These  suggest that our model is capable of producing plausible short-term forecasts (``nowcasts'') up to around time $t+30$min; recalling that the radar typically provides lower rainfall rates than those observed at ground level.
We note in passing that forecasts are also available on a finer temporal resolution than that of the  observed data as a result of introducing the imputed time-steps.
At time $t + 40$min a large cluster of rain clouds form in the northern region of our domain which highlights the unpredictability of weather systems even at modest forecast horizons.
That our model is not able to predict that event with much certainty is perhaps not surprising, although we note that we are able to capture an increased amount of  uncertainty in the region where that event is taking place; see Figure~\ref{fig:real_forecast_images}, third row.
Thus, as noted by other authors, although both mid-range and long-range forecasts are straightforward to obtain, these should be treated with caution in radar-based modelling applications as weather systems can develop (and decay)  quickly.

\section{Conclusion}
\label{sec:conc}
We have considered the challenging problem of modelling high-dimensional spatio-temporal precipitation fields.
In particular we have outlined a physically motivated statistical model obtained via a discretisation of a collection of stochastic partial differential equations with spatio-temporal dynamics also embedded within the source-sink process.
In contrast to other approaches in the literature our observation model enables information from both ground-based rain gauges and weather radar to be used to inferring the (latent) precipitation field at ground level which is of interest when modelling rainfall-induced surface water flows within urban areas.
Posterior sampling is achieved via an efficient MCMC scheme where the EnKS is embedded within a Gibbs sampling scheme that facilitates inference in reasonable time.
We considered a novel dataset from the Urban Observatory project (Newcastle upon Tyne, UK) that highlights how ground-based rain gauges can be effectively used to calibrate radar observations to infer likely rainfall rates at ground level.

\bibliographystyle{Chicago}
\bibliography{paper}

\begin{thebibliography}{}

\bibitem[\protect\citeauthoryear{Anderson and Anderson}{Anderson and
  Anderson}{1999}]{anderson1999monte}
Anderson, J.~L. and S.~L. Anderson (1999).
\newblock {A Monte Carlo implementation of the nonlinear filtering problem to
  produce ensemble assimilations and forecasts}.
\newblock {\em Monthly Weather Review\/}~{\em 127\/}(12), 2741--2758.

\bibitem[\protect\citeauthoryear{Barr, Johnson, Ming, Peppa, Dong, Wen, Robson,
  Smith, James, Wilkinson, et~al.}{Barr et~al.}{2020}]{barr2020flood}
Barr, S., S.~R. Johnson, X.~Ming, M.~Peppa, N.~Dong, Z.~Wen, C.~Robson,
  L.~Smith, P.~James, D.~J. Wilkinson, et~al. (2020).
\newblock {Flood-PREPARED}: A nowcasting system for real-time impact adaption
  to surface water flooding in cities.
\newblock {\em ISPRS Annals of Photogrammetry, Remote Sensing \& Spatial
  Information Sciences\/}~{\em 6}.

\bibitem[\protect\citeauthoryear{Berliner}{Berliner}{2003}]{berliner2003physical}
Berliner, L.~M. (2003).
\newblock Physical-statistical modeling in geophysics.
\newblock {\em Journal of Geophysical Research: Atmospheres\/}~{\em
  108\/}(D24).

\bibitem[\protect\citeauthoryear{Bernardo and Smith}{Bernardo and
  Smith}{1994}]{BernardoS94}
Bernardo, J.~M. and A.~F.~M. Smith (1994).
\newblock {\em Bayesian Theory}.
\newblock Chichester, U.K.: Wiley.

\bibitem[\protect\citeauthoryear{Bocquet and Sakov}{Bocquet and
  Sakov}{2014}]{bocquet2014iterative}
Bocquet, M. and P.~Sakov (2014).
\newblock An iterative ensemble {K}alman smoother.
\newblock {\em Quarterly Journal of the Royal Meteorological Society\/}~{\em
  140\/}(682), 1521--1535.

\bibitem[\protect\citeauthoryear{Carter and Kohn}{Carter and
  Kohn}{1994}]{carter1994gibbs}
Carter, C.~K. and R.~Kohn (1994).
\newblock {On Gibbs sampling for state space models}.
\newblock {\em Biometrika\/}~{\em 81\/}(3), 541--553.

\bibitem[\protect\citeauthoryear{Cressie}{Cressie}{1991}]{cressie1991stat}
Cressie, N. (1991).
\newblock {\em {Statistics for Spatial Data}}.
\newblock John Wiley \& Sons.

\bibitem[\protect\citeauthoryear{Cressie and Wikle}{Cressie and
  Wikle}{2015}]{cressiewikle2015}
Cressie, N. and C.~K. Wikle (2015).
\newblock {\em Statistics for spatio-temporal data}.
\newblock John Wiley \& Sons.

\bibitem[\protect\citeauthoryear{Damien and Walker}{Damien and
  Walker}{2001}]{damien2001sampling}
Damien, P. and S.~G. Walker (2001).
\newblock Sampling truncated normal, beta, and gamma densities.
\newblock {\em Journal of Computational and Graphical Statistics\/}~{\em
  10\/}(2), 206--215.

\bibitem[\protect\citeauthoryear{Evensen}{Evensen}{2003}]{evensen2003ensemble}
Evensen, G. (2003).
\newblock The ensemble {K}alman filter: Theoretical formulation and practical
  implementation.
\newblock {\em Ocean dynamics\/}~{\em 53\/}(4), 343--367.

\bibitem[\protect\citeauthoryear{Evensen and Van~Leeuwen}{Evensen and
  Van~Leeuwen}{2000}]{evensen2000ensemble}
Evensen, G. and P.~J. Van~Leeuwen (2000).
\newblock {An ensemble Kalman smoother for nonlinear dynamics}.
\newblock {\em Monthly Weather Review\/}~{\em 128\/}(6), 1852--1867.

\bibitem[\protect\citeauthoryear{Fr{\"u}hwirth-Schnatter}{Fr{\"u}hwirth-Schnatter}{1994}]{fruhwirth1994data}
Fr{\"u}hwirth-Schnatter, S. (1994).
\newblock Data augmentation and dynamic linear models.
\newblock {\em Journal of time series analysis\/}~{\em 15\/}(2), 183--202.

\bibitem[\protect\citeauthoryear{Garside and Wilkinson}{Garside and
  Wilkinson}{2003}]{garside2003dynamic}
Garside, L. and D.~J. Wilkinson (2003).
\newblock {Dynamic lattice-Markov spatio-temporal models for environmental
  data}.
\newblock {\em Bayesian Statistics\/}~{\em 7}, 535--542.

\bibitem[\protect\citeauthoryear{Gelman, Carlin, Stern, Dunson, Vehtari, and
  Rubin}{Gelman et~al.}{2014}]{gelman2014bayesian}
Gelman, A., J.~B. Carlin, H.~S. Stern, D.~B. Dunson, A.~Vehtari, and D.~B.
  Rubin (2014).
\newblock {\em {Bayesian Data Analysis}\/} (3rd ed.).
\newblock CRC press Boca Raton, FL.

\bibitem[\protect\citeauthoryear{Geman and Geman}{Geman and
  Geman}{1984}]{geman1987stochastic}
Geman, S. and D.~Geman (1984).
\newblock {Stochastic relaxation, Gibbs distributions, and the Bayesian
  restoration of images}.
\newblock {\em {IEEE Transactions on Pattern Analysis and Machine
  Intelligence}\/}~{\em 6}, 721--741.

\bibitem[\protect\citeauthoryear{Guennebaud, Jacob, et~al.}{Guennebaud
  et~al.}{2010}]{eigenwebv3}
Guennebaud, G., B.~Jacob, et~al. (2010).
\newblock {Eigen v3}.
\newblock http://eigen.tuxfamily.org.

\bibitem[\protect\citeauthoryear{Heaps, Boys, and Farrow}{Heaps
  et~al.}{2015}]{heaps2015bayesian}
Heaps, S.~E., R.~J. Boys, and M.~Farrow (2015).
\newblock {Bayesian modelling of rainfall data by using non-homogeneous hidden
  Markov models and latent Gaussian variables}.
\newblock {\em Journal of the Royal Statistical Society: Series C: Applied
  Statistics\/}, 543--568.

\bibitem[\protect\citeauthoryear{Hoffman and Frankel}{Hoffman and
  Frankel}{2018}]{hoffman2018numerical}
Hoffman, J.~D. and S.~Frankel (2018).
\newblock {\em Numerical methods for engineers and scientists}.
\newblock CRC press.

\bibitem[\protect\citeauthoryear{Houtekamer, He, and Mitchell}{Houtekamer
  et~al.}{2014}]{houtekamer2014parallel}
Houtekamer, P.~L., B.~He, and H.~L. Mitchell (2014).
\newblock {Parallel implementation of an ensemble Kalman filter}.
\newblock {\em Monthly Weather Review\/}~{\em 142\/}(3), 1163--1182.

\bibitem[\protect\citeauthoryear{Houtekamer and Mitchell}{Houtekamer and
  Mitchell}{2001}]{houtekamer2001sequential}
Houtekamer, P.~L. and H.~L. Mitchell (2001).
\newblock A sequential ensemble {K}alman filter for atmospheric data
  assimilation.
\newblock {\em Monthly Weather Review\/}~{\em 129\/}(1), 123--137.

\bibitem[\protect\citeauthoryear{James, Dawson, Harris, and Joncyzk}{James
  et~al.}{2014}]{james2014urban}
James, P.~M., R.~J. Dawson, N.~Harris, and J.~Joncyzk (2014).
\newblock Urban observatory environment.
\newblock http://dx.doi.org/10.17634/154300-19.

\bibitem[\protect\citeauthoryear{Katzfuss, Stroud, and Wikle}{Katzfuss
  et~al.}{2020}]{katzfuss2020ensemble}
Katzfuss, M., J.~R. Stroud, and C.~K. Wikle (2020).
\newblock {Ensemble Kalman methods for high-dimensional hierarchical dynamic
  space-time models}.
\newblock {\em Journal of the American Statistical Association\/}~{\em
  115\/}(530), 866--885.

\bibitem[\protect\citeauthoryear{Kendon, Roberts, Fowler, Roberts, Chan, and
  Senior}{Kendon et~al.}{2014}]{kendon2014heavier}
Kendon, E.~J., N.~M. Roberts, H.~J. Fowler, M.~J. Roberts, S.~C. Chan, and
  C.~A. Senior (2014).
\newblock Heavier summer downpours with climate change revealed by weather
  forecast resolution model.
\newblock {\em Nature Climate Change\/}~{\em 4\/}(7), 570--576.

\bibitem[\protect\citeauthoryear{Khare, Anderson, Hoar, and Nychka}{Khare
  et~al.}{2008}]{khare2008investigation}
Khare, S.~P., J.~L. Anderson, T.~J. Hoar, and D.~Nychka (2008).
\newblock {An investigation into the application of an ensemble Kalman smoother
  to high-dimensional geophysical systems}.
\newblock {\em Tellus A: Dynamic Meteorology and Oceanography\/}~{\em 60\/}(1),
  97--112.

\bibitem[\protect\citeauthoryear{Krainski, G{\'o}mez-Rubio, Bakka, Lenzi,
  Castro-Camilo, Simpson, Lindgren, and Rue}{Krainski
  et~al.}{2018}]{krainski2018advanced}
Krainski, E.~T., V.~G{\'o}mez-Rubio, H.~Bakka, A.~Lenzi, D.~Castro-Camilo,
  D.~Simpson, F.~Lindgren, and H.~Rue (2018).
\newblock {\em {Advanced spatial modeling with stochastic partial differential
  equations using R and INLA}}.
\newblock CRC Press.

\bibitem[\protect\citeauthoryear{Lawson, Biggeri, Dreassi, et~al.}{Lawson
  et~al.}{1999}]{lawson1999edge}
Lawson, A., A.~Biggeri, E.~Dreassi, et~al. (1999).
\newblock Edge effects in disease mapping.
\newblock {\em Disease Mapping and Risk Assessment for Public Health\/},
  85--97.

\bibitem[\protect\citeauthoryear{LeVeque}{LeVeque}{2007}]{leveque2007finite}
LeVeque, R.~J. (2007).
\newblock {\em {Finite difference methods for ordinary and partial differential
  equations: steady-state and time-dependent problems}}.
\newblock SIAM.

\bibitem[\protect\citeauthoryear{Liang and Smith}{Liang and
  Smith}{2015}]{liang2015high}
Liang, Q. and L.~S. Smith (2015).
\newblock A high-performance integrated hydrodynamic modelling system for urban
  flood simulations.
\newblock {\em Journal of Hydroinformatics\/}~{\em 17\/}(4), 518--533.

\bibitem[\protect\citeauthoryear{Liu, Yeo, and Lu}{Liu
  et~al.}{2021}]{liu2021statistical}
Liu, X., K.~Yeo, and S.~Lu (2021).
\newblock {Statistical modeling for spatio-temporal data from stochastic
  convection-diffusion processes}.
\newblock {\em Journal of the American Statistical Association\/}.

\bibitem[\protect\citeauthoryear{Marshall and Palmer}{Marshall and
  Palmer}{1948}]{marshall1948distribution}
Marshall, J.~S. and W.~M.~K. Palmer (1948).
\newblock The distribution of raindrops with size.
\newblock {\em Journal of meteorology\/}~{\em 5\/}(4), 165--166.

\bibitem[\protect\citeauthoryear{Milne}{Milne}{1953}]{milne1953numerical}
Milne, W.~E. (1953).
\newblock {\em Numerical solution of differential equations}, Volume~19.
\newblock Wiley New York.

\bibitem[\protect\citeauthoryear{Pfeifer and Deutsch}{Pfeifer and
  Deutsch}{1980}]{pfeifer1980three}
Pfeifer, P.~E. and S.~J. Deutsch (1980).
\newblock A three-stage iterative procedure for space-time modeling.
\newblock {\em Technometrics\/}~{\em 22\/}(1), 35--47.

\bibitem[\protect\citeauthoryear{Sans{\'o} and Guenni}{Sans{\'o} and
  Guenni}{2000}]{sanso2000nonstationary}
Sans{\'o}, B. and L.~Guenni (2000).
\newblock A nonstationary multisite model for rainfall.
\newblock {\em Journal of the American Statistical Association\/}~{\em
  95\/}(452), 1089--1100.

\bibitem[\protect\citeauthoryear{Sigrist, K{\"u}nsch, and Stahel}{Sigrist
  et~al.}{2015}]{sigrist2015stochastic}
Sigrist, F., H.~R. K{\"u}nsch, and W.~A. Stahel (2015).
\newblock Stochastic partial differential equation based modelling of large
  space--time data sets.
\newblock {\em Journal of the Royal Statistical Society: Series B (Statistical
  Methodology)\/}~{\em 77\/}(1), 3--33.

\bibitem[\protect\citeauthoryear{Smith and Roberts}{Smith and
  Roberts}{1993}]{smith1993bayesian}
Smith, A.~F. and G.~O. Roberts (1993).
\newblock {Bayesian computation via the Gibbs sampler and related Markov chain
  Monte Carlo methods}.
\newblock {\em Journal of the Royal Statistical Society: Series B
  (Methodological)\/}~{\em 55\/}(1), 3--23.

\bibitem[\protect\citeauthoryear{Spiegelhalter, Best, Carlin, and Van
  Der~Linde}{Spiegelhalter et~al.}{2002}]{spiegelhalter2002bayesian}
Spiegelhalter, D.~J., N.~G. Best, B.~P. Carlin, and A.~Van Der~Linde (2002).
\newblock Bayesian measures of model complexity and fit.
\newblock {\em Journal of the royal statistical society: Series B (statistical
  methodology)\/}~{\em 64\/}(4), 583--639.

\bibitem[\protect\citeauthoryear{Stroud, Stein, Lesht, Schwab, and
  Beletsky}{Stroud et~al.}{2010}]{stroud2010ensemble}
Stroud, J.~R., M.~L. Stein, B.~M. Lesht, D.~J. Schwab, and D.~Beletsky (2010).
\newblock An ensemble {K}alman filter and smoother for satellite data
  assimilation.
\newblock {\em Journal of the American Statistical Association\/}~{\em
  105\/}(491), 978--990.

\bibitem[\protect\citeauthoryear{Tanner and Wong}{Tanner and
  Wong}{1987}]{tanner1987calculation}
Tanner, M.~A. and W.~H. Wong (1987).
\newblock The calculation of posterior distributions by data augmentation.
\newblock {\em Journal of the American Statistical Association\/}~{\em
  82\/}(398), 528--540.

\bibitem[\protect\citeauthoryear{Tierney}{Tierney}{1994}]{tierney1994markov}
Tierney, L. (1994).
\newblock Markov chains for exploring posterior distributions.
\newblock {\em The Annals of Statistics\/}, 1701--1728.

\bibitem[\protect\citeauthoryear{Tobin}{Tobin}{1958}]{tobin1958estimation}
Tobin, J. (1958).
\newblock Estimation of relationships for limited dependent variables.
\newblock {\em Econometrica: journal of the Econometric Society\/}, 24--36.

\bibitem[\protect\citeauthoryear{Wikle, Berliner, and Cressie}{Wikle
  et~al.}{1998}]{wikle1998hierarchical}
Wikle, C.~K., L.~M. Berliner, and N.~Cressie (1998).
\newblock {Hierarchical Bayesian space-time models}.
\newblock {\em Environmental and Ecological Statistics\/}~{\em 5\/}(2),
  117--154.

\bibitem[\protect\citeauthoryear{Wikle and Cressie}{Wikle and
  Cressie}{1999}]{wikle1999dimension}
Wikle, C.~K. and N.~Cressie (1999).
\newblock {A dimension-reduced approach to space-time Kalman filtering}.
\newblock {\em Biometrika\/}~{\em 86\/}(4), 815--829.

\bibitem[\protect\citeauthoryear{Xia and Liang}{Xia and
  Liang}{2018}]{xia2018new}
Xia, X. and Q.~Liang (2018).
\newblock A new efficient implicit scheme for discretising the stiff friction
  terms in the shallow water equations.
\newblock {\em Advances in water resources\/}~{\em 117}, 87--97.

\bibitem[\protect\citeauthoryear{Xing, Liang, Wang, Ming, and Xia}{Xing
  et~al.}{2019}]{xing2019city}
Xing, Y., Q.~Liang, G.~Wang, X.~Ming, and X.~Xia (2019).
\newblock {City-scale hydrodynamic modelling of urban flash floods: the issues
  of scale and resolution}.
\newblock {\em Natural Hazards\/}~{\em 96\/}(1), 473--496.

\bibitem[\protect\citeauthoryear{Xu and Wikle}{Xu and
  Wikle}{2007}]{xu2007estimation}
Xu, K. and C.~K. Wikle (2007).
\newblock Estimation of parameterized spatio-temporal dynamic models.
\newblock {\em Journal of Statistical Planning and Inference\/}~{\em 137\/}(2),
  567--588.

\bibitem[\protect\citeauthoryear{Yeo and Johnson}{Yeo and
  Johnson}{2000}]{yeo2000new}
Yeo, I.-K. and R.~A. Johnson (2000).
\newblock A new family of power transformations to improve normality or
  symmetry.
\newblock {\em Biometrika\/}~{\em 87\/}(4), 954--959.

\end{thebibliography}

\end{document}